\documentclass[prd,amssymb,amsmath,amsfonts,nofootinbib,reprint,showpacs,longbibliography]{revtex4-1}

\usepackage{graphicx}
\usepackage{lmodern}
\usepackage{amsmath,amssymb}
\usepackage{mathrsfs}
\usepackage{amsfonts}
\usepackage[utf8]{inputenc}
\usepackage{url}
\usepackage[colorlinks]{hyperref}
\usepackage[table]{xcolor}
\usepackage{multirow}
\usepackage[normalem]{ulem}
\usepackage{lipsum}
\usepackage{marvosym}
\usepackage{enumerate}
\usepackage{enumitem}

\usepackage{color,soul}
\usepackage{acronym}

\usepackage{bm}
\usepackage{color}
\usepackage{commath}
\normalsize
\usepackage{booktabs}
\usepackage{tabularx}
\allowdisplaybreaks

\definecolor{cyan}{rgb}{0,0.9,0.9}
\definecolor{orange}{rgb}{0.9,0.5,0}
\definecolor{magenta}{rgb}{1,0,1}
\definecolor{purple}{rgb}{0.8,0.4,0.8}
\definecolor{gray}{rgb}{0.8242,0.8242,0.8242}
\definecolor{dodgerblue}{rgb}{0.12, 0.56, 1.0}

\newcommand{\red}[1]{\textcolor{black}{#1}}

\newcommand{\abh}[0]{\hat{a}}

\newcommand{\bmath}[1]{{\boldmath #1 \unboldmath}}

\def\minus{\hphantom{-}}

\newcolumntype{C}[1]{>{\centering\arraybackslash}m{#1}}

\newacro{adm}[ADM]{Arnowitt-Deser-Misner}
\newacro{bbh}[BBH]{binary black hole}
\newacro{bh}[BH]{black hole}
\newacroplural{bh}[BHs]{black holes}
\newacro{bhns}[BHNS]{black hole-neutron star}
\newacro{bhpt}[BHPT]{black hole perturbation theory}
\newacro{bns}[BNS]{binary neutron star}
\newacro{bf}[BF]{Bayes' factor}
\newacro{cbc}[CBC]{compact binary coalescence}
\newacro{ce}[CE]{Cosmic Explorer}
\newacro{da}[DA]{data analysis}
\newacro{et}[ET]{Einstein Telescope}
\newacro{emri}[EMRI]{extreme mass ratio inspiral}
\newacroplural{emri}[EMRIs]{extreme mass ratio inspirals}
\newacro{eob}[EOB]{Effective-One-Body}
\newacro{eom}[EOM]{equations of motion}
\newacro{fd}[FD]{frequency domain}
\newacro{fft}[FFT]{Fast Fourier transform}
\newacro{gw}[GW]{gravitational-wave}
\newacroplural{gw}[GWs]{Gravitational-waves}
\newacro{gr}[GR]{general relativity}
\newacro{grb}[GRB]{gamma-ray burst}
\newacro{grhd}[GRHD]{general-relativistic hydrodynamics}
\newacro{gwosc}[GWOSC]{Gravitational Wave Open Science Center}
\newacro{gwtc1}[GWTC-1]{the first gravitational-wave transients catalog}
\newacro{gsf}[GSF]{Gravitational Self Force}
\newacro{hm}[HM]{Higher mode}
\newacroplural{hm}[HMs]{Higher modes}
\newacro{ifo}[IFO]{interferometer}
\newacro{imr}[IMR]{inspiral-merger-ringdown}
\newacro{im}[IMR]{inspiral-to-merger}
\newacro{kagra}[KAGRA]{Kamioka Gravitational Wave Detector}
\newacro{ligo}[LIGO]{Laser Interferometer Gravitational-Wave Observatory}
\newacro{lisa}[LISA]{Laser Interferometer Space Antenna}
\newacro{lso}[LSO]{Last Stable Orbit}
\newacro{lvc}[LVC]{LIGO-Virgo Collaboration}
\newacro{lvk}[LVK]{LIGO-Virgo-Kagra Collaboration}
\newacro{lo}[LO]{leading order}
\newacro{ns}[NS]{neutron star}
\newacroplural{ns}[NSs]{neutron stars}
\newacro{nr}[NR]{numerical relativity}
\newacro{nqc}[NQCs]{Next-to-quasicircular corrections}
\newacro{nlo}[NLO]{next-to-leading order}
\newacro{nnlo}[NNLO]{next-to-next-to-leading order}
\newacro{n3lo}[N3LO]{next-to-next-to-next-to-leading order}
\newacro{n4lo}[N3LO]{next-to-next-to-next-to-next-to-leading order}
\newacro{ode}[ODE]{Ordinary Differential Equation}
\newacroplural{ode}[ODEs]{Ordinary Differential Equations}
\newacro{pn}[PN]{post-Newtonian}
\newacro{pm}[PM]{post-Minkowskian}
\newacro{pe}[PE]{parameter estimation}
\newacro{psd}[PSD]{power spectral density}
\newacro{pa}[PA]{post-adiabatic}
\newacro{qnm}[QNM]{quasi-normal mode}
\newacro{qc}[QC]{quasi-circular}
\newacro{snr}[SNR]{signal-to-noise ratio}
\newacro{spa}[SPA]{stationary-phase approximation}
\newacro{sxs}[SXS]{Simulating eXtreme Spacetimes}
\newacro{td}[TD]{time domain}
\newacro{ng}[NG]{Nect Generation}

\newcommand{\numCircular}{4}

\newcommand{\mineEccentric}{0.100}
\newcommand{\maxeEccentric}{0.900}
\newcommand{\minpEccentric}{6.000}
\newcommand{\maxpEccentric}{15.000}

\newcommand{\numEccentric}{69}

\newcommand{\minEHyperbolic}{1.001}
\newcommand{\maxEHyperbolic}{1.170}
\newcommand{\minPphiHyperbolic}{4.000}
\newcommand{\maxPphiHyperbolic}{20.000}

\newcommand{\numHyperbolic}{108}

\newcommand{\minaGlobal}{-0.800}
\newcommand{\maxaGlobal}{0.990}

\newcommand{\numTotalSimulations}{181}

\newcommand{\medianEccentricdE}{12.8}
\newcommand{\medianEccentricdJ}{7.2}
\newcommand{\EccentricrStrong}{10.0}
\newcommand{\EccentricNStrong}{65}
\newcommand{\medianEccentricdEStrong}{13.6}
\newcommand{\medianEccentricdJStrong}{7.4}
\newcommand{\fractionEccentricdETen}{43}
\newcommand{\fractionEccentricdJTen}{58}
\newcommand{\fractionEccentricdEFifty}{78}
\newcommand{\fractionEccentricdJFifty}{86}

\newcommand{\medianHyperbolicdE}{22.7}
\newcommand{\medianHyperbolicdJ}{21.5}

\newcommand{\HyperbolicNStrong}{17}
\newcommand{\medianHyperbolicdEStrong}{51.8}
\newcommand{\medianHyperbolicdJStrong}{27.2}
\newcommand{\fractionHyperbolicdETen}{27}
\newcommand{\fractionHyperbolicdJTen}{28}
\newcommand{\fractionHyperbolicdEFifty}{81}
\newcommand{\fractionHyperbolicdJFifty}{91}

\newcommand{\spearmanOmgmaxdE}{+0.33}
\newcommand{\spearmanOmgmaxdJ}{+0.24}
\newcommand{\spearmanRmindE}{-0.24}
\newcommand{\spearmanRmindJ}{-0.14}
\newcommand{\spearmanAdE}{-0.12}
\newcommand{\spearmanAdJ}{-0.05}

\newcommand{\spearmanOmgMaxEccentricdE}{+0.59}
\newcommand{\spearmanOmgMaxEccentricdJ}{+0.65}
\newcommand{\spearmanOmgMaxHyperbolicdE}{+0.69}
\newcommand{\spearmanOmgMaxHyperbolicdJ}{+0.68}

\newcommand{\srEccTotSDtInPct}{96}

\newcommand{\srEccTotSDtOutPct}{70}

\newcommand{\srEccTotMDtInPct}{26}

\newcommand{\srEccTotSNsInPct}{85}

\newcommand{\srEccTotSNsOutPct}{33}

\newcommand{\srEccTotMNsInPct}{96}

\newcommand{\srEccTotMNsOutPct}{52}

\newcommand{\srEccTotSNsdtInPct}{93}

\newcommand{\srEccTotSNsdtOutPct}{74}

\newcommand{\srEccTotMNsdtInPct}{96}

\newcommand{\srEccTotMNsdtOutPct}{90}

\newcommand{\srEccMtwoMDtInPct}{85}

\newcommand{\srEccMtwoMNsInPct}{37}

\newcommand{\srEccNsdtTotAllMinPct}{83}

\newcommand{\srHypTotSDtInFrac}{6/6}

\newcommand{\srHypTotMOutOfBandMinPct}{72}
\newcommand{\srHypTotMOutOfBandMaxPct}{93}

\newcommand{\srHypNsdtAllMinPct}{4}
\newcommand{\srHypNsdtAllMaxPct}{47}

\begin{document}

\title{Spin the black circle II: \\
tidal heating and torquing of a rotating black hole by a test mass on generic orbits}
\author{Rossella \surname{Gamba}${}^{1,2}$}
\author{Danilo \surname{Chiaramello}${}^{3,4}$}
\author{Estuti \surname{Shukla}${}^{2,5}$}
\author{Simone \surname{Albanesi}${}^{4,6}$}

\affiliation{${}^{1}$ Department of Physics, University of California, Berkeley, CA 94720, USA}
\affiliation{${}^{2}$ Institute for Gravitation \& the Cosmos, The Pennsylvania State University, University Park PA 16802, USA}
\affiliation{${}^{3}$ Department of Physics, Universit\'a degli Studi di Torino, Torino, 10125, Italy}
\affiliation{${}^{4}$ INFN sezione di Torino, Torino, 10125, Italy}
\affiliation{${}^{5}$ Department of Physics, The Pennsylvania State University, University Park, PA
16802, USA}
\affiliation{${}^{6}$ Friedrich-Schiller-Universit\"at Jena, Theoretisch-Physikalisches Institut, 07743 Jena, Germany}

\begin{abstract}
Horizon fluxes of energy and angular momentum
are a key strong-field effect in the dynamics of black holes,
encoding direct information about their nature.
In this work, we present a numerical study of these fluxes for a test
particle orbiting a Kerr black hole on equatorial geodesics, covering circular,
eccentric, and hyperbolic trajectories across a wide range of orbital parameters and black hole spins.
We reproduce known results for circular orbits and uncover a richer phenomenology for eccentric and hyperbolic ones: the instantaneous
fluxes can exhibit multiple peaks and sign changes, indicating a complex interplay between orbital dynamics and horizon perturbation.
We then compare these results against existing analytical post-Newtonian expressions, exploring resummation strategies to improve their
performance against numerical data. In particular, we propose a factorized and resummed
representation of the horizon fluxes that predicts the onset frequency of the
superradiant regime to within $10\%$ for $\gtrsim \srEccNsdtTotAllMinPct\%$ for both the energy and
angular momentum fluxes of bound configurations;
high-energy open orbits however remain a challenge.
This representation exactly reduces to the
circular limit by construction, independently of the perturbative
order of the remaining analytical terms.
For peak and orbit-averaged fluxes, the analytical models achieve acceptable
accuracy -- with relative errors at the $10\%$ level or below -- at large separations
and low eccentricities. However, they can exhibit deviations of $\sim \mathcal{O}(100\%)$
in the strong-field regime, motivating the need for improved flux prescriptions
and further investigations.
\end{abstract}

\date{\today}
\maketitle

\acresetall

\section{Introduction}
\label{sec:intro}

One of the defining features of \acp{bh} in general relativity is the presence of a \emph{horizon},
a surface that causally disconnects the interior of the \ac{bh} from the rest of the universe.
The notion of \ac{bh} horizons has a long history, and has evolved from the global
definition of the event horizon to more local or quasi-local concepts, such as apparent, trapping, isolated, and dynamical
horizons~\citep{Hayward:1994yy, Thornburg:1995cp, Ashtekar:2003hk, Ashtekar:2025wnu}.
A key physical consequence of the presence of a horizon is the Hawking area theorem~\citep{Christodoulou:1970wf, Hawking:1971tu},
which states that the total area of \ac{bh} event horizons cannot decrease in classical general relativity
under the assumptions of cosmic censorship and the null energy condition.
This result, reminiscent of the second law of thermodynamics~\citep{Bardeen:1973gs}, has been tested following the observation of
\acp{gw} from \ac{bbh} coalescences~\citep{DelPozzo:2016kmd, Isi:2020tac} by the \ac{lvk}~\citep{KAGRA:2013rdx, LIGOScientific:2014pky, VIRGO:2014yos}. 
The first such test was performed using data from GW150914~\citep{LIGOScientific:2016vlm,LIGOScientific:2016wkq},
finding agreement with the theorem's prediction with $97\%$ probability. More recently, Refs.~\cite{LIGOScientific:2025rid,LIGOScientific:2025obp,Prasad:2026imj}
carried out a similar analysis using data from GW250114~\citep{LIGOScientific:2025rid}, again confirming the law to
more than $3 \sigma$ credibility, the most stringent test of this kind to date, thanks to the impressive signal-to-noise ratio of this event.

Beyond the area theorem, the analogy between \ac{bh} mechanics and thermodynamics has been further
developed over the years~\citep{Bardeen:1973gs}, leading to the formulation of the four laws of \ac{bh} mechanics.
Of particular relevance is the first law, which relates variations in the mass, spin, and area of perturbed stationary \acp{bh}.
Originally formulated for isolated \acp{bh}, the first law has since been extended to binary systems of comparable-mass \acp{bh} in circular orbits
within the \ac{pn} framework~\citep{LeTiec:2011ab, Blanchet:2012at},
multiple \acp{bh} with generic distributions of perfect fluid matter~\citep{Friedman:2001pf, Uryu:2010su}, and scatterings
in the test-mass limit including all radiation effects~\citep{Gonzo:2024xjk}. This law has also been put to the test with
\ac{gw} observations~\citep{Wang:2023jah}, providing further confirmation of general relativity in the strong-field regime.

The growing interest in exploiting changes in mass, spin, and area of \acp{bh}
as fundamental probes of general relativity motivates a detailed investigation
of the various mechanisms that can induce such variations.
Tidal heating and torquing are among the most interesting of these processes, whereby fluxes of energy and angular momentum
are absorbed by (or extracted from) the \ac{bh} horizon~\citep{Hartle:1973zz, Alvi:2001mx}.
These effects were initially studied in the context of \ac{bh} perturbation theory~\citep{Poisson:1994yf,Tagoshi:1997jy,Mino:1997bx},
where a point particle orbits a much larger central \ac{bh}, using the Teukolsky
formalism~\citep{Teukolsky:1972my,Teukolsky:1973ha,Bernuzzi:2012ku,Taracchini:2013wfa,Fujita:2014eta,Shah:2014tka,OSullivan:2015lni}.
Such results were later extended to systems of comparable-mass \acp{bh}~\citep{Alvi:2001mx, Poisson:2004cw}.
In the framework developed by Poisson and collaborators~\citep{Poisson:2004cw,Taylor:2008xy,Comeau:2009bz,Poisson:2009qj,Poisson:2014gka,Poisson:2018qqd},
horizon fluxes can be computed in the ``slow-motion'' approximation, where each \ac{bh} is treated as an isolated, tidally deformed object,
characterized by a set of even (``electric'') and odd (``magnetic'') parity tidal multipole moments.
For \ac{bbh} systems, the tidal moments have been computed up to 1.5\ac{pn}~\citep{Taylor:2008xy, Poisson:2014gka},
enabling the calculation of horizon fluxes up to
\ac{nnlo} on quasi-circular~\citep{Chatziioannou:2012gq, Chatziioannou:2016kem, Saketh:2022xjb}
and generic (planar) orbits~\citep{Datta:2023wsn, Chiaramello:2024unv}.
In parallel, full 3+1 \ac{nr} simulations of merging \acp{bh} have been used to compute
horizon fluxes for comparable-mass systems on quasicircular~\citep{Scheel:2014ina} and
hyperbolic~\citep{Nelson:2019czq,Jaraba:2021ces,Rodriguez-Monteverde:2024tnt,Kogan:2025vml} orbits.
In these works, the fluxes were extracted using apparent horizons that track
the evolution of the masses and spins of each \ac{bh} during the simulation.
These results have demonstrated that horizon fluxes
can have a measurable impact on the \acp{bh} dynamics and the emitted \ac{gw} signal
when the \acp{bh} are moving on eccentric or hyperbolic orbits~\citep{Nelson:2019czq,Jaraba:2021ces,Rodriguez-Monteverde:2024tnt,Kogan:2025vml,Datta:2023wsn, Chiaramello:2024unv} or when the
system evolves through many orbits, as in the case of \acp{emri}~\citep{Bernuzzi:2012ku, Datta:2024vll}.

A striking manifestation of tidal heating and torquing is superradiance,
a phenomenon through which energy and angular momentum are \textit{extracted} from a rotating \ac{bh}.
Superradiance can be understood as a Penrose-like process in which a rotating \ac{bh} transfers angular momentum ---
and with it rotational energy --- to its companion, while simultaneously absorbing gravitational-wave energy that increases its irreducible mass;
the latter can never decrease, as it is simply the square root of the horizon surface area.
The direction of the total exchange of energy is determined by the balance
between these two effects: the extraction of rotational energy when the \ac{bh} sheds angular momentum,
and the concurrent increase of its irreducible mass.
For a \ac{bh} in a binary system on a quasicircular orbit, the onset of superradiance is determined by the frequency of the tidal perturbation relative to
the angular velocity of the \ac{bh} horizon, $\Omega_{\rm H}$. In the test mass limit,
this condition emerges naturally from frequency-domain solutions of the Teukolsky equation~\citep{Taracchini:2013wfa,Fujita:2014eta},
leading to a global prefactor proportional to $(\Omega - \Omega_{\rm H})$ in the analytical expressions
for both the energy and angular momentum horizon fluxes.
For generic orbits and comparable mass \acp{bh}, the situation is considerably more intricate.
\citet{Chiaramello:2024unv} (hereafter Paper~I) proposed a factorized form of the 
analytical expressions for the horizon fluxes associated with generic orbital motion that
includes a similar prefactor.
This correctly reproduces the quasicircular limit, and predicts
a more complex dependence of the superradiance effect on the orbital parameters for eccentric or hyperbolic orbits.
Moreover, its form implies that energy and angular momentum variations can decouple.
At present, this prediction has not been thoroughly tested against numerical calculations.

In this work, we aim to carry out such tests by numerically computing the horizon fluxes of energy and angular momentum
induced by test particles moving on generic orbits around a Kerr \ac{bh}, considering circular, eccentric, and hyperbolic orbits.
In doing so, we will also assess the validity of the analytical expressions derived in Paper~I,
and explore possible resummation strategies to improve their agreement with numerical data and extend their range of validity.
The structure of the manuscript is as follows.
In Sec.~\ref{sec:numerics} we summarize the numerical framework
employed to compute the horizon fluxes for test particles on Kerr spacetime, present the catalog of
configurations considered and discuss some of the salient features of our numerical results.
In Sec.~\ref{sec:framework} we review the analytical expressions
for the horizon fluxes derived in Paper I, and introduce possible resummation strategies.
In Sec.~\ref{sec:results} we present the comparison between numerical and analytical results,
with a focus on the performance of the resummed expressions and predictivity of the superradiance prefactor.
Finally, in Sec.~\ref{sec:conclusions} we summarize our findings and discuss future directions.

\textit{Conventions:} Unless otherwise stated we use geometric units with \(G = c = 1\);
occasionally we will keep explicit dependence on $c$ in our expressions to mark \ac{pn} orders.

\section{Numerical setup}
\label{sec:numerics}

In this section we briefly summarize the numerical framework we adopt to evolve the dynamics of a test particle
in Kerr spacetime and to compute the associated horizon fluxes of energy and angular momentum, as well as
the orbital configurations that we consider. For more details, we refer the reader to
Refs.~\citet{Harms:2013ib, Harms:2014dqa}.

\subsection{Dynamics and fluxes}

We evolve the motion of a test particle of mass $\mu$ in a fixed Kerr spacetime background
of mass \(M\) and dimensionless spin $\abh = a/M = S/M^2$. 
In this work, we restrict our attention to planar geodesic trajectories -- specifically circular, eccentric, and hyperbolic orbits, neglecting radiation-reaction effects.
This approximation is adequate here for two reasons:
first, in the test-mass limit, dissipation is not expected to significantly affect the orbital motion
over the timescales of interest in this work~\footnote{This approximation may not hold for highly energetic
hyperbolic encounters close to the transition between scattering and capture.
We do not consider such configurations in this work.};
second, from a more practical standpoint and as will become clear below, the prescription we use
to compute horizon fluxes becomes unreliable once the test particle falls into the central \ac{bh}~\citep{Harms:2014dqa, Poisson:2004cw},
reducing the insight gained from studying fully plunging orbits.

We solve Hamilton's equations for a test particle
in Kerr spacetime in (dimensionless) Boyer-Lindquist coordinates:
\begin{align}
    \dot{r} =& \Bigl(\frac{A}{B}\Bigr)^{1/2} \frac{\partial \hat{H}_{\rm Kerr}}{\partial p_{r*}} \, , \\
    \dot{\varphi} =& \frac{\partial \hat{H}_{\rm Kerr}}{\partial p_{\varphi}} \equiv \Omega \, , \\
    \dot{p}_{r*} =& -\Bigl(\frac{A}{B}\Bigr)^{1/2} \frac{\partial \hat{H}_{\rm Kerr}}{\partial r} \, , \\
    \dot{p}_{\varphi} =& 0 \, ,
\end{align}
where \( \hat{H}_{\rm Kerr} \) is the (equatorial) $\mu$-normalized Kerr Hamiltonian written in terms of the
centrifugal radius \( r_c \)~\citep{Damour:2014sva}:
\begin{align}
    r_c^2 =& r^2 + \abh^2 + \frac{2 \abh^2}{r} \, , \\
    \hat{H}_{\rm Kerr} =& \sqrt{A \biggl(1 + \frac{p_{\varphi}^2}{r_c^2}\biggr) + p_{r_*}^2} + 2 \frac{\abh p_\varphi}{r r_c^2}  \, ,
\end{align}
and \( A \) and \( B \) are the Kerr metric potentials:
\begin{align}
    A =& \frac{1 + 2 u_c}{1 + 2 u} (1 - 2 u_c) \, , \\
    B =& \frac{1}{1 - 2 u_c + \abh^2 u^2} \, ,
\end{align}
with \( u = 1/r \) and \( u_c = 1/r_c \), while $p_{r_*} = \sqrt{A/B} p_r$ is the canonical momentum associated with the tortoise coordinate $r_*$\citep{Damour:2014sva}.

\begin{figure*}
    \includegraphics[width=0.8\textwidth]{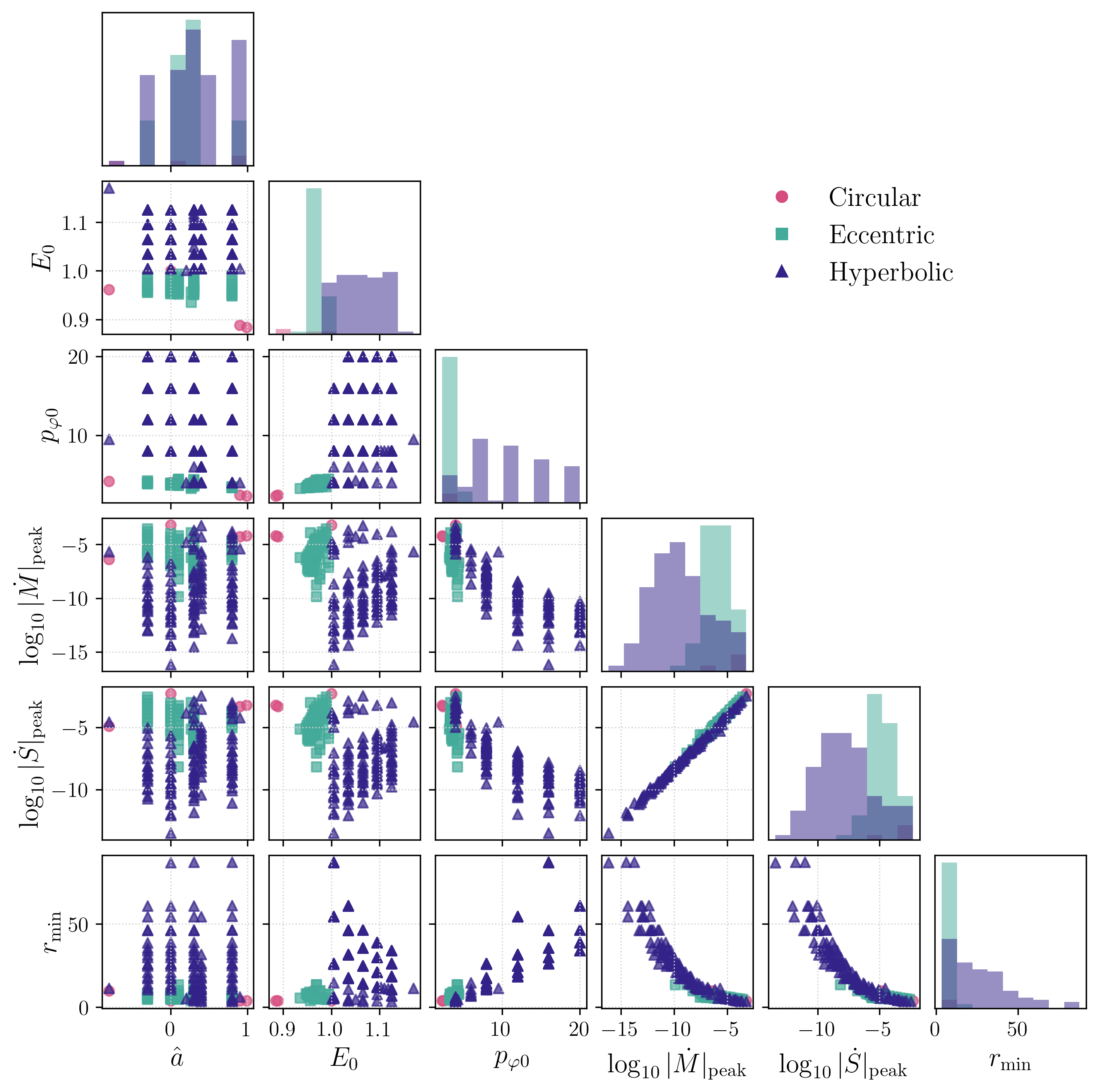}
    \caption{Summary corner plot of the initial parameters and maximum horizon fluxes for all simulated systems,
    divided by orbital configuration: circular (pink), eccentric (green) and hyperbolic (blue). Our entire dataset
    comprises \numTotalSimulations~simulations, with \numCircular~circular, \numEccentric~eccentric, and \numHyperbolic~hyperbolic ones.
    The maximum fluxes are obtained for the lowest approach distances $r_{\rm min}$, as expected.
    }
    \label{fig:summary_corner}
\end{figure*}

We explore the relevant parameter space in terms of \( \abh \) and different dynamical initial conditions, depending
on the shape of the orbit. For hyperbolic trajectories, we characterize the system using
the test particle's initial energy \(E_0 \) and angular momentum \(p_{\varphi 0} \) at a fixed initial separation $r_0$.
For eccentric ones we employ the eccentricity \(e_0 \) and semilatus rectum \(p_0 / M \) of the orbit at a fixed initial
anomaly $\zeta_0 = \pi$, such that the particle evolution begins at the apocenter, $r_0 =  p_0/(1+e_0\cos\zeta_0) = p_0/(1-e_0)$.
Finally, as usual, for circular orbits the initial radial separation \(r_0\) fully determines the trajectory.

\begin{figure*}
    \includegraphics[width=0.45\textwidth]{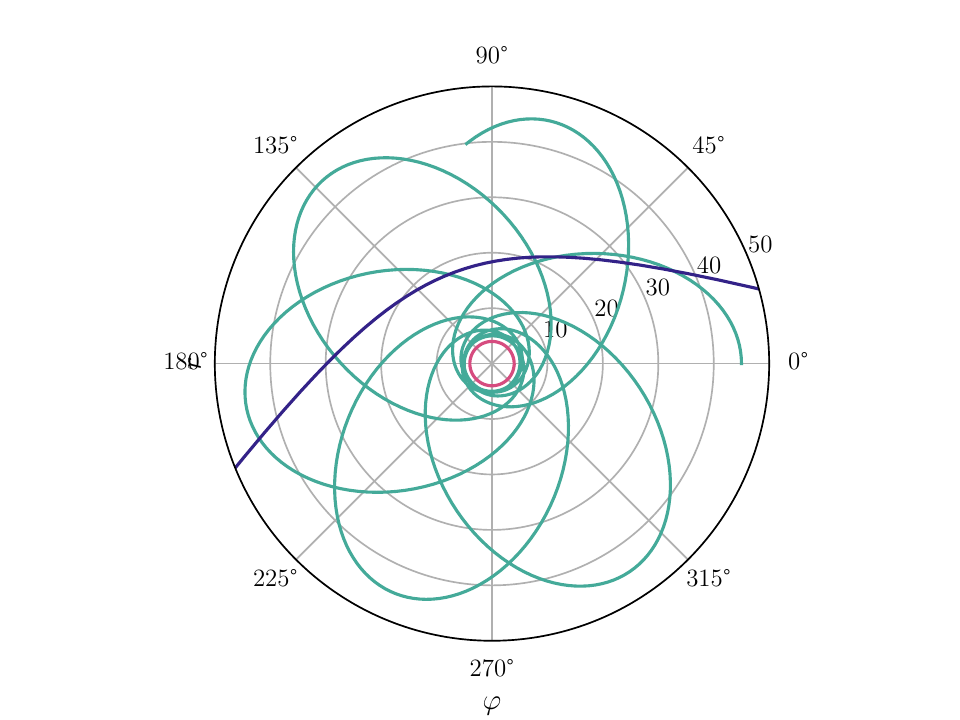}
    \includegraphics[width=0.45\textwidth]{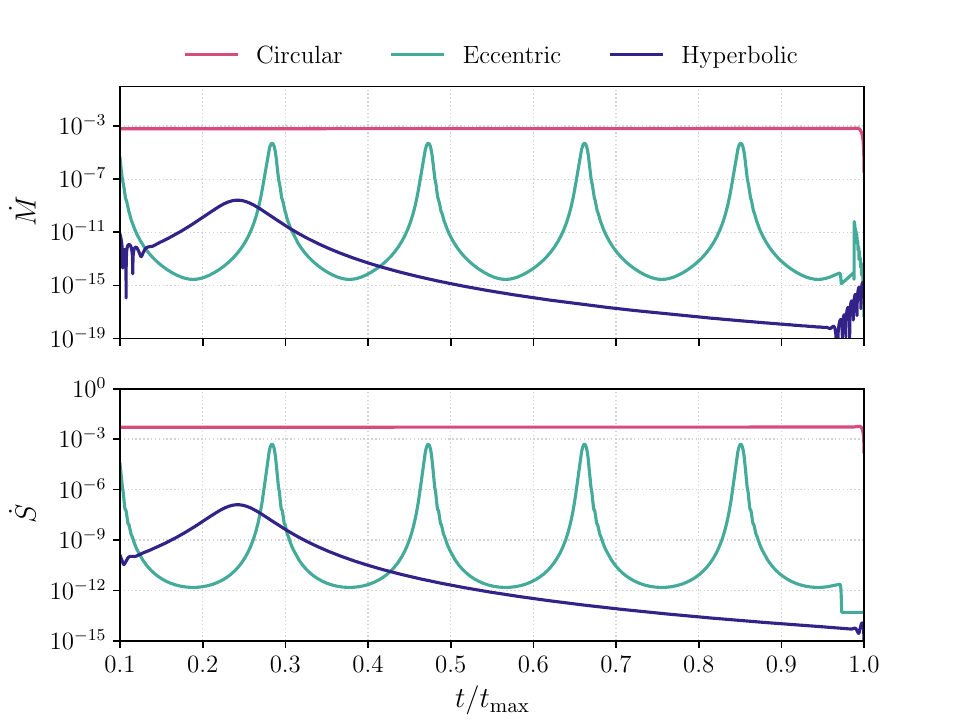}
    \caption{Three representative configurations considered in this work. Left panel: trajectories of the test
    particle around the Kerr black hole for circular (pink), eccentric (green), and hyperbolic (blue) orbits. Right panel:
    corresponding horizon fluxes of energy (top panel) and angular momentum (bottom panel) as functions of (normalized) time.
    The eccentric and hyperbolic configurations exhibit peaks in the fluxes at closest approach, while the circular orbit
    shows constant fluxes, as expected. 
    Some numerical noise can be observed at both the beginning and end of the dynamics.
    The former can be readily traced to junk radiation ($t\sim 0$); the latter is due to the fact that the evaluation of Eq.~\eqref{eq:fplus}
    involves an integral over future times. As one approaches the end of the dynamics $(t \sim t_{\rm max})$, unphysical behaviors appear.
    }
    \label{fig:example_configurations}
\end{figure*}

The geodesic trajectories provide the particle perturbation source term of the Teukolsky equation
(see Eq.~18 of \citet{Harms:2014dqa}). To solve it we employ \texttt{Teukode}, a time-domain solver for the \(2+1\) Teukolsky equation,
which uses an eighth-order finite-difference scheme for the spatial derivatives and a 
fourth-order Runge--Kutta scheme for the time evolution.
This is carried out on a horizon-penetrating, hyperboloidal foliation~\citep{Zenginoglu:2008wc, Bernuzzi:2011aj, Harms:2014dqa} of the spacetime,
which enables the computation of gravitational perturbations in a background Kerr spacetime and
the extraction of waveforms at future null infinity without the need for extrapolation~\citep{Fontbute:2025ixd, Bernuzzi:2025hhu}.
To compute the mass and angular momentum fluxes at the horizon, we solve for the gravitational perturbation ($s=+2$)
Newman-Penrose scalar $\psi_{0}$. The equations in \texttt{Teukode} are solved for a fixed azimuthal $m$-mode of the Weyl scalar ($\Psi_{0m}$),
with the total fluxes computed as:
\begin{align}
\dot{M}
&= \frac{r_+^{2}+a^{2}}{4\kappa}
   \sum_m \biggl[
      2\kappa \int_{-1}^{1}\! d\xi\, \lvert f^{+}_{Hm}\rvert^{2} \notag\\
&\qquad\qquad
      -\, i m \Omega_{\rm H}
      \int_{-1}^{1}\! d\xi\,
      \bigl(f^{+*}_{Hm}f^{-}_{Hm}-f^{+}_{Hm}f^{-*}_{Hm}\bigr)
   \biggr] \, , \label{eq:mdot}\\[1em]
\dot{S}
&= -\,\frac{r_+^{2}+a^{2}}{4\kappa}
   \sum_m i m
   \biggl[
      \int_{-1}^{1}\! d\xi\,
      \bigl(f^{+*}_{Hm}f^{-}_{Hm}-f^{+}_{Hm}f^{-*}_{Hm}\bigr)
   \biggr] \, . \label{eq:jdot}
\end{align}
where $\Omega_{\rm H}=a/(2Mr_+)$ is the angular velocity of the horizon, $\kappa = (r_+-M)/(r_+^{2}-a^{2})$ is the surface gravity,
$r_+$ is the horizon radius and the complex quantities $f^{\pm}_{\rm{H} m}$ are defined as, 
\begin{align}
f^{+}_{\rm{H}m}(v,\theta)
&= -\,e^{\kappa v}
   \int_{v}^{\infty}\! dv'\,
   e^{-(\kappa - i m \Omega_{\rm H})v'}\,
   \Psi_{0m}(v',r_+,\theta), \label{eq:fplus}\\[0.5em]
f^{-}_{\rm{H} m}(v,\theta)
&= -\!\int_{-\infty}^{v}\! dv'\,
   e^{i m \Omega_{\rm H} v'}\,
   \Psi_{0m}(v',r_+,\theta). \label{eq:fminus}
\end{align}

For further details on the formalism developed by Poisson, we refer the interested reader to~\citet{Poisson:2004cw},
and~\citet{Harms:2014dqa} for its implementation in \texttt{Teukode}.
To numerically evaluate Eqs.~\eqref{eq:mdot}--\eqref{eq:fminus}, \texttt{Teukode} employs Simpson's rule for the 
angular integrations and the trapezoidal rule for the time integration. 
These integrals are evaluated in post-processing, using a timestep of \( dt = 0.01 \) M.

Note that the fluxes at an advanced time $v$ also depend on the future evolution of the curvature perturbation, via $f^+_{\rm{H} m}$.
Although the contribution of far future times is exponentially suppressed by the factor $e^{-\kappa v}$ in the integrand, this feature would propagate unphysical behavior in $\Psi_{0m}$ on plunging orbits caused by the disappearing particle source term to earlier times in the horizon fluxes, making them unreliable already around the time of the light ring crossing.

For circular, hyperbolic, and eccentric trajectories, we evolve the system for
\(\geq 1000\,M\), \( \geq 1400\,M\), and \( \geq 2000\,M\) respectively, to ensure that at least
one full orbit is completed for bound orbits, and that the particle has sufficiently receded from
the black hole after the encounter for hyperbolic orbits.

\subsection{Code tests and convergence}
\label{sec:tests}

Before further analyzing our numerical results, we summarize here the tests performed to  assess their numerical accuracy
and validate our implementation. All details are available in App.~\ref{sec:code_tests}.

We perform three main sets of tests: self-convergence studies, varying both the radial and angular grid resolutions;
mode truncation tests, where we estimate the impact of including different numbers of \(m\)-modes in the flux calculation;
and comparisons with earlier results available in the literature for circular orbits.
Given the large number of configurations considered in this work, we focus our first two tests on three representative
systems, one for each orbital type (circular, eccentric, hyperbolic).

Based on the results of the tests listed above, by default we perform simulations including \( m = 0, 1, 2 \) modes
and choose a grid resolution of \( n_r \times n_{\theta} = 3601 \times 161 \), finding a convergence order of $2$ when varying either grid spacing.
Combining the errors from $m$-mode truncation and grid discretization, we estimate the overall (relative) numerical uncertainty
to be of the order of $10^{-3} - 10^{-2}$
for both eccentric and hyperbolic orbits. These numbers are configuration-dependent;
we treat them as indicative of the overall accuracy of our results and use them when comparing
against the analytical predictions described in Sec.~\ref{sec:framework}.

As a final consistency check of our numerical results, we also verify that we correctly reproduce the expected
superradiance behavior for circular orbits, which is well-known in the literature and can be derived from frequency-domain solutions
of the Teukolsky equation.
Specifically, following the notation of~\citet{Fujita:2014eta}, the flux of energy in the case of a circular orbit of frequency $\Omega$ is given by:
\begin{align}
    \label{eq:testmass_fluxes}
    \langle \dot{M} \rangle &= \sum_{\ell, m} \dfrac{\alpha_{\ell m \omega} |\tilde{Z}^{\rm H}_{\ell m \omega}|^2}{4 \pi m^2 \Omega^2} \nonumber \\
    &\equiv \Omega \left(\Omega - \Omega_{\rm H}\right) \sum_{\ell, m} \dfrac{\tilde{\alpha}_{\ell m \omega} |\tilde{Z}^{\rm H}_{\ell m \omega}|^2}{4 \pi m^2 \Omega^2} \, ,
\end{align}
where in the second equality we have extracted from the $\alpha_{\ell m \omega}$ coefficients a common factor
featuring the orbital frequency (see Eqs.~(17) and~(18) of~\citet{Fujita:2014eta}). The angular momentum flux
in this case is just found from $\langle \dot{M} \rangle = \Omega \langle \dot{S} \rangle$.
Since the leftover $\tilde{\alpha}_{\ell m \omega}$ are nonnegative, the overall sign
of the fluxes is determined by the prefactor $\Omega - \Omega_{\rm H}$: if the orbital frequency is lower than
the \ac{bh}'s horizon frequency, $\langle \dot{M} \rangle < 0$, and the particle is extracting rotational
energy from the central \ac{bh}, rather than losing energy through its horizon.
This factorization is not explicit in the time-domain method we use to compute the fluxes;
verifying it is therefore a non-trivial consistency check
of our results. We consider the case of a circular orbit satisfying $\Omega = \Omega_{\rm H}$, and confirm
that both fluxes vanish (to numerical precision), in accordance with Eq.~\eqref{eq:testmass_fluxes}.

\subsection{Parameter space and overview of results}
\label{subsec:overview}

We survey black-hole spins in the range \( \abh \in [\minaGlobal, \red{\maxaGlobal}] \).
For unbound trajectories, we vary the orbital energy \( E_0 \in [\minEHyperbolic, \maxEHyperbolic] \)
and angular momentum \( p_{\varphi 0} \in [\minPphiHyperbolic, \maxPphiHyperbolic] \).
For bound systems, the eccentricity spans \red{\( e_0 \in [\mineEccentric, \maxeEccentric] \)}  with the
semilatus rectum \( p_0 \in [\red{\minpEccentric}, \maxpEccentric] \).
All the configurations considered in this work are depicted in Fig.~\ref{fig:summary_corner}. In total,
we simulate \numTotalSimulations~systems, of which \numCircular~are circular, \red{\numEccentric}~eccentric, and \red{\numHyperbolic}~on hyperbolic orbits.

\begin{figure*}
\includegraphics[width=0.49\textwidth]{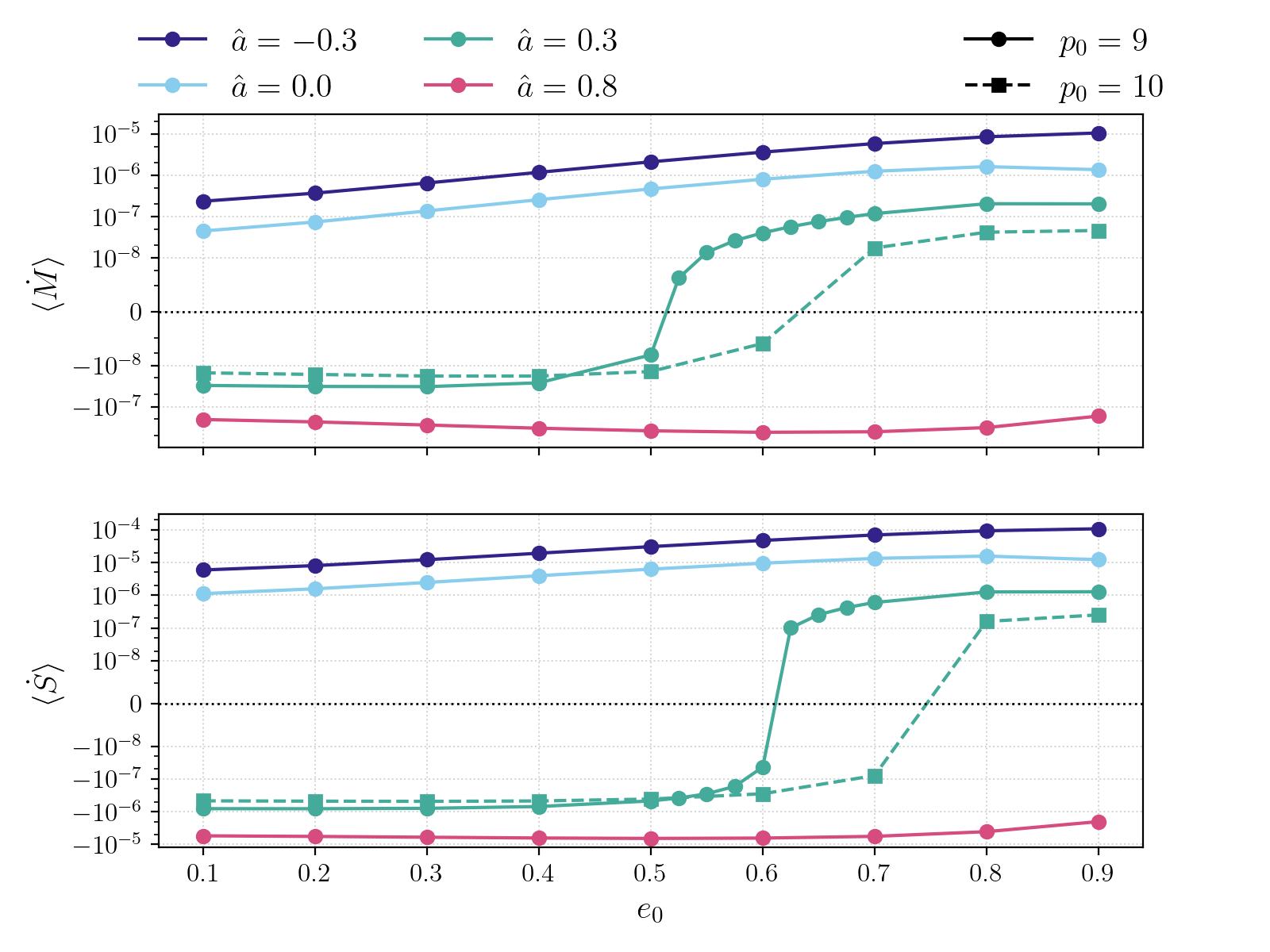}
\includegraphics[width=0.49\textwidth]{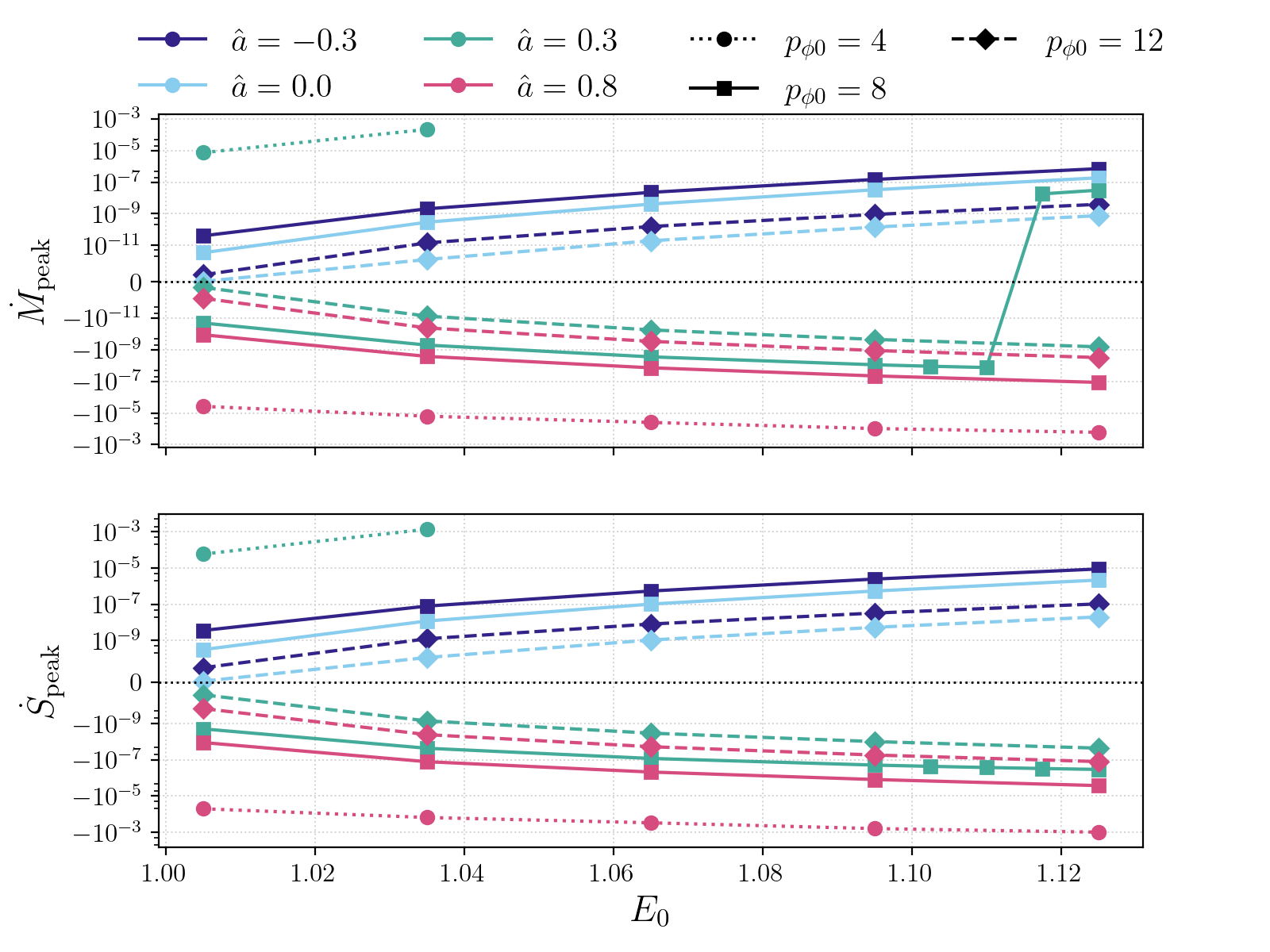}
    \caption{Left: orbit-averaged energy (top) and angular momentum (bottom) fluxes as a function of the eccentricity $e_0$ of the
    orbit at fixed semilatus rectum $p_0$ and spin $a$ for a representative sample of simulations.
    The orbit-average procedure simplifies the complex behavior of the
    instantaneous fluxes for positive $\abh$. In this case, we observe that for fixed values of ($p_0, \abh$) there 
    exists a critical value of eccentricity beyond which the fluxes change sign from negative to positive. Notably, this eccentricity
    value is not the same for energy and momentum fluxes. Right: peak energy (top) and angular momentum (bottom) fluxes during the close encounter
    as a function of the orbital energy \( E_0 \) at fixed angular momentum \( p_{\varphi} \) and spin \( \abh \).
    Similar to the eccentric case, for positive spins there exists a critical value of \( E_0 \)
    beyond which the peak fluxes change sign from negative to positive. Again, this value is not the same for energy
    and angular momentum fluxes.}
    \label{fig:fluxes_orbavg_peak}
\end{figure*}

The left panel of Fig.~\ref{fig:example_configurations} shows three representative systems among those simulated in this work.
Depending on the orbital configuration, the horizon fluxes exhibit starkly different behaviors.
Circular orbits produce constant fluxes, while eccentric and hyperbolic orbits show pronounced peaks
at periastron passage and close encounter, respectively.
Globally, we observe that: (i) the peak value of the fluxes is rather strongly correlated with the distance of minimum approach (see Fig.~\ref{fig:summary_corner});
(ii) energy and angular momentum fluxes are positively correlated;
(iii) regardless of the orbital configuration, the instantaneous $\dot{M}, \dot{S} > 0$ for  $ \abh \leq 0$;
(iv) when $\abh > 0$, the behavior of the instantaneous fluxes is at times complicated by one or more sign changes,
in the case of eccentric and hyperbolic orbits.
To further elucidate the last point, we consider integrated quantities
rather than instantaneous ones for eccentric orbits, and peak values for hyperbolic ones.
The left panel of Fig.~\ref{fig:fluxes_orbavg_peak} shows the orbit-averaged fluxes $\langle \dot{M} \rangle$
and $\langle \dot{S} \rangle$ computed on eccentric orbits, defined as:
\begin{equation}
    \langle X \rangle = \frac{1}{T_r}\int_0^{T_r} X dt \, ,
\end{equation}
where $T_r$ is the time period between two subsequent periastron passages.
This procedure reveals that typically the averaged fluxes' magnitudes increase with larger $e_0$
at fixed $p_0$ (i.e., they increase with smaller periastron distance)
and larger $|\abh|$. The sign of the fluxes is negative up to a critical value of $e_0$, beyond which the non-superradiant regime
dominates over the superradiant one, leading to positive averaged fluxes.
Interestingly, this critical value is not the same for $\langle \dot{M} \rangle$
and $\langle \dot{S} \rangle$, already indicating that the two are not trivially related for generic orbits.

Similar observations hold for hyperbolic orbits when considering the peak energy and momentum fluxes during the encounter,
as shown in the right panel of Fig.~\ref{fig:fluxes_orbavg_peak}. Larger $|\abh|$ values lead to larger flux magnitudes,
and for $\abh = 0.3$ there exists a critical value of $E_0$ (at fixed $p_{\varphi,0}$) beyond which
$\dot{M}_{\rm peak}$ changes sign from negative to positive. Again, this critical value is not the same
for energy and angular momentum fluxes.

\section{Analytical framework}
\label{sec:framework}

In this section we build on the results of Paper I and introduce possible factorization
and resummation strategies for the analytical expressions of the horizon fluxes on generic orbits.
These will be compared in Sec.~\ref{sec:results} with our numerical results.

\subsection{Summary of Paper I}
In Paper I we derived analytical \ac{pn} expressions for the fluxes of energy and angular momentum exchanged by a
\ac{bbh} system and each of its component \acp{bh}. Focusing without loss of generality on the primary \ac{bh},
expressions for the rates of change of its mass, $\dot{m}_1$, and spin, $\dot{S}_1$, were computed for generic planar
orbits up to the relative 1.5\ac{pn} order, which corresponds to 4\ac{pn} order beyond the leading infinity flux.
Working in \ac{eob} coordinates, we report here only the general structure of the flux expressions
(see Eqs.~(18) of Paper~I for the complete forms):
\begin{subequations}
\label{eq:pn_flux_raw}
\begin{align}
\dfrac{\dot{m}_1}{M} =& -\dfrac{8}{5}\nu^2 \biggl(\dfrac{m_1}{M}\biggr)^3 \dfrac{\chi_1}{r^6} \biggl\{\bigl(1 + 3\chi_1^2\bigr) \dfrac{p_\varphi}{r^2} \nonumber \\
                      +& \dfrac{1}{c^2} \dot{m}_1^{\rm NLO} (r, p_\varphi, p_r) + \dfrac{1}{c^3} \dot{m}_1^{\rm NNLO}(r, p_\varphi, p_r)\biggr\} \, , \\
\dfrac{\dot{S}_1}{M^2} =& -\dfrac{8}{5}\nu^2 \biggl(\dfrac{m_1}{M}\biggr)^3 \dfrac{\chi_1}{r^6} \biggl\{1 + 3\chi_1^2 \nonumber \\
                        +& \dfrac{1}{c^2} \dot{S}_1^{\rm NLO} (r, p_\varphi, p_r) + \dfrac{1}{c^3} \dot{S}_1^{\rm NNLO}(r, p_\varphi, p_r)\biggr\} \, ,
\end{align}
\end{subequations}
where $\nu = m_1 m_2 / M^2$ is the symmetric mass ratio and $\chi_1 = S_1/m_1^2$ is the dimensionless spin of the primary \ac{bh}.
In the absence of the radial momentum, these expressions reduce to the known quasicircular results~\citep{Saketh:2022xjb}.
Inspired by earlier works focused on the test-mass and quasicircular limit, we also proposed an alternative form of the
analytical fluxes that isolates the overall sign behavior by incorporating what we called the ``superradiance prefactor":
\begin{widetext}
\begin{subequations}
\label{eq:pn_flux_hfact}
\begin{align}
\dfrac{\dot{m}_1}{M} =& -\dfrac{16}{5}\nu^2 \biggl(\dfrac{m_1}{M}\biggr)^4 \dfrac{1+\sigma_1}{r^6} \biggl[\Omega_{\rm H}^1 - \dfrac{1}{c^3} \biggl(\dfrac{p_\varphi}{r^2} + 3 \dfrac{p_r^2}{p_\varphi}\biggr)\biggr]
                      \biggl\{\bigl(1 + 3\chi_1^2\bigr) \dfrac{p_\varphi}{r^2}
                          + \dfrac{1}{c^2} \dot{m}_{1, \rm{fact.}}^{\rm NLO} (r, p_\varphi, p_r) + \dfrac{1}{c^3} \dot{m}_{1, \rm{fact.}}^{\rm NNLO}(r, p_\varphi, p_r)\biggr\} \, , \\
\dfrac{\dot{S}_1}{M^2} =& -\dfrac{16}{5}\nu^2 \biggl(\dfrac{m_1}{M}\biggr)^4 \dfrac{1 + \sigma_1}{r^6} \biggl(\Omega_{\rm H}^1 - \dfrac{1}{c^3} \dfrac{p_\varphi}{r^2}\biggr)
                      \biggl\{1 + 3\chi_1^2
                            + \dfrac{1}{c^2} \dot{S}_{1, \rm{fact.}}^{\rm NLO} (r, p_\varphi, p_r) + \dfrac{1}{c^3} \dot{S}_{1, \rm{fact.}}^{\rm NNLO}(r, p_\varphi, p_r)\biggr\}\, ,
\end{align}
\end{subequations}
\end{widetext}
where $\sigma_1 = \sqrt{1 - \chi_1^2}$ and $\Omega_{\rm H}^1 = \dfrac{\chi_1}{2m_1 (1+\sigma_1)}$ is the horizon frequency
of the primary \ac{bh}. The physical interpretation of this form is clear when looking at $\dot{S}_1$, where
the superradiance prefactor ties the overall sign of the flux to whether the binary's orbital frequency ($\simeq p_\varphi/r^2$ to \ac{lo})
exceeds the \ac{bh}'s horizon frequency. In the test-mass limit, results for circular dynamics, known to much higher \ac{pn} order, show that the exact orbital frequency
of the smaller body appears in this prefactor. In the expressions specialized to quasicircular orbits of comparable-mass systems, a ``tidal frequency"
appears instead, representing the phasing of the tidal perturbation seen by the primary \ac{bh} in its own rest frame,
which incorporates time dilation and frame-dragging effects relative to the system's barycentric reference frame.
In the present case, the form of the generic-orbit expressions naturally leads us to write different prefactors for the energy and
angular momentum fluxes, predicting a role for the radial momentum in determining the sign of the former.

Horizon fluxes for non-spinning \acp{bh} start 1.5\ac{pn} beyond the leading general results we are considering here.
Analytical expressions for fluxes in that limit can nonetheless be derived that reach up to $O(c^{-3})$, which is 3\ac{pn} above our \ac{lo} results.
App.~C of Paper~I exploited this to attempt to compute higher-order, spin-independent contributions to the fluxes, parametrized as corrections to the superradiance prefactors.
Adopting the same notation, in general we can write factorized fluxes as:
\begin{equation}
    \label{eq:general_hfact}
    \dot{X}_1 = -\dfrac{16}{5} \nu^2 \biggl(\dfrac{m_1}{M}\biggr)^4 \dfrac{1+\sigma_1}{r^6} \biggl(\Omega_{\rm H}^1 - \dfrac{1}{c^3} \Omega_{\rm T}^{X}\biggr) \dot{X}_1^{\rm PN}\, ,
\end{equation}
where $X$ is either $M$ or $S$, and the $\dot{X}_1^{\rm PN}$ factor collects the remaining \ac{pn}-expanded terms.
The $\Omega_{\rm T}$ functions, so-called to harken back to the ``tidal frequency" of the quasi-circular limit, are either the leading terms that can be read off from Eqs.~\eqref{eq:pn_flux_hfact},
\begin{align}
    \Omega_{\rm T}^{S} = \dfrac{p_\varphi}{r^2} \, , \quad \Omega_{\rm T}^{M} = \dfrac{p_\varphi}{r^2} + 3 \dfrac{p_r^2}{p_\varphi} \, ,
    \label{eq:base_omgT}
\end{align}
or the versions completed by higher-order, nonspinning terms given in Eqs.~(B3) and~(B4) of Paper I, which we report here, transformed to \ac{eob} coordinates:
\begin{subequations}
\begin{align}
    \Omega_{\rm T}^{S, \rm{NS}} &= \dfrac{p_\varphi}{r^2} \biggl[1 - \dfrac{1}{c^2} \biggl(\dfrac{1}{r} - \dfrac{1-2m_1-3\nu}{2}p_r^2 \biggr. \biggr. \nonumber \\
    \biggl. \biggl. &+ \dfrac{1+\nu}{2} \dfrac{p_\varphi^2}{r^2} \biggr) - \dfrac{16m_1p_r}{3rc^3} \biggr] \, , \\
    \Omega_{\rm T}^{M, \rm{NS}} &= \Omega_{\rm T}^{S, \rm{NS}} + \dfrac{3p_r^2}{p_\varphi} \biggl[1 + \dfrac{1}{c^2} \biggl(\dfrac{47+6m_1+3\nu}{6}\dfrac{p_\varphi^2}{r^2} \biggr. \biggr. \nonumber \\
    \biggl. \biggl. &-\dfrac{1+\nu}{2}p_r^2 - \dfrac{9+11m_1+3\nu}{3r}\biggr) \biggr. \biggr. \nonumber \\
    \biggl. \biggl. &+ \left(p_\varphi^2-r\right)\dfrac{16m_1}{3r^3p_rc^3}\biggr] \, .
\end{align}
\end{subequations}

\subsection{Multipolar decomposition}
\label{subsec:multipolar_decomp}

Past works investigating analytical representations of fluxes at infinity and at the horizon have shown that decomposing
the total fluxes into multipolar contributions, to be treated individually, can lead to improved accuracy
when compared to numerical data~\citep{Damour:2008gu, Pan:2010hz, Messina:2018ghh, Nagar:2016ayt}.
Within the time-domain framework adopted here, this factorization is naturally connected to the multipolar
decomposition of the Teukolsky potential and, consequently, of the Newman--Penrose scalar $\psi_0$.
Previous works by Poisson and collaborators have shown that $\psi_0$ can be expressed as a sum over
spin-weighted spherical harmonics, with coefficients determined by the components of the tidal multipole
moments (and their time derivatives) with fixed azimuthal number $m$; see, for example, Eqs.~(28) and (29)
of~\citet{Chatziioannou:2012gq}.

Although we do not compute the fluxes directly from the expressions derived by Poisson,
this structural result can still be exploited to guide our factorization procedure. In practice, it suffices
to identify the terms in the tidal moments $\mathcal{E}, \mathcal{B}$, which form the basis of the derivation in Paper I, 
that are proportional to $\cos(m\phi)$ and $\sin(m\phi)$, track their propagation
through the flux calculation, and isolate their individual contributions to the final expressions.
At the perturbative order considered here, nonvanishing contributions arise from $\ell=2$ with $m=0,1,2$.
Although our calculation includes octupolar tidal fields, their effect enters indirectly through the induced
quadrupolar response; see~\citet{Saketh:2022xjb} for a detailed discussion.

To isolate the contribution associated with each azimuthal mode, we perform a Fourier decomposition of the tidal multipoles,
\begin{equation}
\mathcal{Q}_L(r, \dot{r}, \dot{\phi}, \phi) = \sum_m q_m \mathcal{Q}_L^m(r, \dot{r}, \dot{\phi}) e^{i m \phi} \, ,
\end{equation}
where $\mathcal{Q_L}$ denotes either the electric or magnetic tidal quadrupole or octupole, $q_m$ are bookkeeping parameters,
and $L$ is a multi-index. The individual mode coefficients are obtained via
\begin{equation}
\mathcal{Q}_L^m = \frac{1}{2\pi}\int_0^{2\pi} d \phi \ \mathcal{Q}_L(r, \dot{r}, \dot{\phi}, \phi) e^{-i m \phi} \, .
\end{equation}

Due to the parity properties of the electric and magnetic tidal multipoles, several modes vanish.
In particular:
\begin{equation}
\mathcal{E}_{ab}^1 = 0 \, , \quad \mathcal{B}_{ab}^0 = 0 \, , \quad \mathcal{B}_{ab}^2 = 0 \, ,
\end{equation}
and
\begin{equation}
\mathcal{E}_{abc}^{0} = 0 \, , \quad \mathcal{E}_{abc}^2 = 0 \, , \quad \mathcal{B}_{abc}^1 = 0 \, , \quad \mathcal{B}_{abc}^3 = 0 \, .
\end{equation}

With these decompositions in hand, the fluxes can be computed separately for each $m$-mode
by setting all $q_{m'} = 0$ for $m' \neq m$ in the tidal multipoles.
We obtain, in \ac{eob} coordinates and specified to the test-mass limit ($m_1 \rightarrow M$, $\chi_1 \rightarrow \abh$, $m_2 \rightarrow 0$), the following
expressions:
\begin{widetext}
\begin{subequations}
\label{eq:mode_by_mode_fluxes}
\begin{align}
    \dot{M}_{20} =& ~\frac{1}{c^3} \frac{12}{5} \nu^2 \dfrac{(1 + \sigma)}{r^8} (1-\abh^2)^2 p_r^2 \, , \\
    \dot{M}_{21} =& -\dfrac{1}{c^2}\frac{2}{5} \nu^2  \dfrac{\abh (4 - 3 \abh^2)}{r^{8}} \frac{p_{\varphi}^2}{r^2} \left(p_{\varphi} - \frac{4}{3 c} \abh\right) \, , \\
    \dot{M}_{22} =& -\frac{8}{5 r^8} \nu^2 \Bigg\{p_{\varphi}\abh(1+3\abh^2)\Bigg[1- \frac{1}{c^2}\red{\Bigg(\frac{7}{2}p_r^2 - \frac{3 p_\varphi^2}{2 r^2}  \Bigg)}\Bigg] + \frac{1}{c^3} \Bigg[ (1 + \sigma) (1 + 13 \abh^2 + 6 \abh^4) \Bigl(-\frac{9}{2} p_r^2 - \frac{2 p_\varphi^2}{r^2} \Bigr) \, \\
    & + \abh(1+3\abh^2) \Bigl(-\frac{64 p_r p_\varphi}{3 r}+ \frac{2}{r}\abh + 2 p_r^2 (5 \abh - 9 B_2(\abh)) - \red{\frac{8 p_\varphi^2 B_2(\abh)}{r^2}} \Bigr) \Bigg]\Bigg\}\notag \, , \\
    \dot{S}_{20} =& ~0 \, , \\
    \dot{S}_{21} =& -\dfrac{1}{c^2}\frac{2}{5}\nu^2 \dfrac{\abh (4-3\abh^2)}{r^8} p_{\varphi} \left(p_{\varphi} - \frac{4}{3 c}\abh\right) \, , \\
    \dot{S}_{22} =& -\frac{8}{5 r^6} \nu^2  \Bigg\{ \abh\left(1 + 3 \abh^2\right) \Bigg[ 1 - \frac{1}{c^2} \Bigl(3 p_r^2 - \frac{2 p_{\varphi}^2}{r^2} - \red{\frac{1}{r}}\Bigr)\Bigg] - \frac{1}{c^3}\Bigg[ 2\frac{p_\varphi}{r^2} (1 + \sigma) (1 + 13 \abh^2 + 6 \abh^4) \, \\
                               & + \abh (1 + 3 \abh^2) \Bigl(\frac{16 p_r}{r} + \frac{8 p_\varphi}{r^2} B_2(\abh)\Bigr)\Bigg]\Bigg\} \notag \, .
\end{align}
\end{subequations}
\end{widetext}
In the above expressions $B_2(\abh) = \Im\Bigl[\psi^0(3 + 2 i \abh/\sigma)\Bigr]$, where $\psi^0$ is the digamma function.
These expressions match the mode-split energy fluxes in~\citet{Fujita:2014eta} in the circular limit, and reveal that:
(i) the $m=0$ modes contribute only at \ac{nnlo}, and only to the energy flux as expected
from \eqref{eq:jdot}; (ii) considering the limit of $\abh\rightarrow 0$, $\dot{M}_{20}$ does not vanish, in agreement
with the prediction of~\citet{Poisson:2004cw}, which showed that in this limit the time-independent
Teukolsky equation for $s=2, \ell=2, m=0$ reduces to Teukolsky's radial equation in Schwarzschild spacetime; (iii)
for the $m=1$ mode the ``rigid rotation'' relation, $\dot{M} = \Omega \dot{S}$,
holds at both \ac{lo} and \ac{nlo}; (iv) this is not the case for the $m=2$ mode,
where this relation is satisfied only at \ac{lo}, with deviations appearing at \ac{nlo} and beyond.
We also see that the $(2,0)$ contribution to the energy flux is non-negative, in agreement with Eq.~\eqref{eq:mdot},
and it depends solely on the radial component of the velocity.
It thus always contributes a net increase in the \ac{bh} mass, irrespective of the processes that either add
or subtract rotational energy.

\subsection{Mode-by-mode factorization}
\label{subsec:mode_by_mode_superradiance}

As anticipated above, Eqs.~\eqref{eq:pn_flux_hfact} introduce a prefactor
that captures the sign change of the fluxes when the orbital frequency exceeds the \ac{bh}'s horizon frequency.
However, the multipolar decomposition discussed above reveals that the terms moved into this prefactor
come from the combined contributions of the $m=0$ and $m=2$ modes.
This clearly suggests that different $m$-modes may give contributions of differing sign to the total fluxes at any given time.
Past works that evaluated horizon fluxes on Kerr in the test-mass limit in the frequency domain, where the Teukolsky equation can be separated in all variables, gave hints to this.
In~\citet{OSullivan:2014ywd}, horizon fluxes in the case of a test particle on a bound orbit are decomposed into a Fourier series and into azimuthal, polar, and radial modes.
The sign of each of their contributions is determined by the relative size of the horizon frequency and a mode-dependent linear combination of the orbit's three fundamental frequencies ($\Omega_\varphi \equiv \Omega, \Omega_r, \Omega_\theta$), with different analytical details between energy and angular momentum.
Thus, summing over the polar and radial indices, they will combine differently in the individual $m$-modes, outside the case of circular orbits where only multiples of the orbital azimuthal frequency are relevant.
Whether this formalism can be exploited to guide sensible analytical forms for our mode-by-mode fluxes is an interesting question that we will explore in future work.
Here, we proceed for the multipolar fluxes similarly to how we identified the superradiance prefactor in Paper I for the total ones.
At the working perturbative order considered here, the only mode for which a similar structure can be meaningfully defined is the $(\ell, m)=(2,2)$.
We thus write:
\begin{subequations}
\label{eq:pn_flux_hfact_m22}
\begin{align}
\dot{M}_{22} \propto &~\Omega_{\rm H}- \frac{1}{c^3}\Bigl( \frac{p_{\varphi}}{r^2} + \frac{9}{4} \frac{p_r^2}{p_\varphi} \Bigr) \equiv \Omega_{\rm H} - \frac{1}{c^3} \Omega_{\rm T, 22}^{M} \, , \\
\dot{S}_{22} \propto &~\Omega_{\rm H}- \frac{1}{c^3}\frac{p_{\varphi}}{r^2} \equiv \Omega_{\rm H} - \frac{1}{c^3}\Omega_{\rm T, 22}^{S} \, .
\end{align}
\end{subequations}
where we define mode-specific ``tidal frequencies'' $\Omega_{\rm T, \ell m}^{X}$.
For the $(2, 1)$ mode, instead, motivated by the behavior on
circular orbits we impose:
\begin{equation}
\label{eq:pn_flux_hfact_s21}
\dot{S}_{21} \propto \Omega_{\rm H} - \frac{1}{c^3} \frac{p_{\varphi}}{r^2} \equiv \Omega_{\rm H} - \frac{1}{c^3}\Omega_{\rm T, 21}^{S}\, ,
\end{equation}
and the same for $\dot{M}_{21}$.

We can incorporate higher-order,
non-spinning terms into our mode-by-mode expressions by adopting the strategy
outlined above.
Denoting with $\Omega_{\rm T, \ell m}^{X, \rm{NS}}$ the
extensions of Eqs.~\eqref{eq:pn_flux_hfact_m22}--\eqref{eq:pn_flux_hfact_s21}
containing these additional factors, we compute them by requiring that they match multipolar
fluxes in the non-spinning limit.
This procedure is well defined only for the $(2,2)$ and $(2,1)$ modes, for which we find:
\begin{subequations}
\begin{align}
    \red{\Omega_{\rm T, 22}^{S, \rm{NS}}} &= \dfrac{p_\varphi}{r^2}\biggl[1 - \dfrac{1}{c^2} \biggl(
        \dfrac{1}{r} + \dfrac{1}{2}p_r^2 + \dfrac{1}{2} \dfrac{p_\varphi^2}{r^2} \biggr)
        - \dfrac{16 p_r}{3rc^3}
    \biggr] \, , \\
    \Omega_{\rm T, 21}^{S, \rm{NS}} &= \dfrac{p_\varphi}{r^2} \, ,\\
    \Omega_{\rm T, 22}^{M, \rm{NS}} &= \Omega_{\rm T, 22}^{S, \rm{NS}} + \dfrac{9p_r^2}{4p_\varphi} \biggl[1 + \dfrac{1}{c^2}\biggl(-\dfrac{20}{3r} + \dfrac{17}{6}\dfrac{p_\varphi^2}{r^2}  -\dfrac{1}{2}p_r^2\biggr) \biggr. \nonumber \\
    \biggl. &+ \dfrac{1}{c^3}\left(p_\varphi^2-r\right)\dfrac{16}{3r^3p_r}\biggr] \, ,\\
    \Omega_{\rm T, 21}^{M, \rm{NS}} &= \dfrac{p_\varphi}{r^2} + \dfrac{16p_r^2}{p_\varphi} \, .
\end{align}
\end{subequations}

\subsection{Circular-noncircular factorization}
\label{subsec:cnc_factorization}
In addition to the raw \ac{pn} expansions and the superradiance-factorized form, we consider in this work
additional analytical treatments meant to extend the domain of validity of our relatively low-order expressions toward
the strong-field regime.
First -- drawing from similar strategies
adopted in the literature for the fluxes at infinity on eccentric
orbits~\citep{Chiaramello:2020ehz, Khalil:2021txt, Albanesi:2021rby,Placidi:2021rkh, Albanesi:2022xge, Placidi:2023ofj, Faggioli:2024ugn, Gamboa:2024imd} --
we test a factorization that splits each flux into a circular term and a noncircular correction (``CNC'' factorization hereafter). 
We extract the circular factor by setting, in the \ac{pn}
terms of either form of either flux, $p_r = 0$ and $p_\varphi = j_c(r)$, where $j_c(r)$ is the angular momentum
of a circular orbit of radius $r$, given itself by a \ac{nnlo} \ac{pn} expression (see e.g.~\citet{Blanchet:2013haa}).
The noncircular correcting factor is then computed by dividing the complete expression by the circular version,
and reexpanding the result up to \ac{nnlo}:
\begin{subequations}
\label{eq:cnc_factorization}
\begin{align}
f^{\rm PN}_{\rm CNC} (r, p_\varphi, p_r) &= f_{\rm C}^{\rm PN} (r) f_{\rm NC}^{\rm PN} (r, p_\varphi, p_r) \, ,\\
f_{\rm C}^{\rm PN} (r) &= f^{\rm PN} (r, p_\varphi = j_c(r), p_r = 0) \, ,\\
f_{\rm NC}^{\rm PN} (r, p_\varphi, p_r) &= f^{\rm PN} (r,p_\varphi,p_r)/f^{\rm PN}_{\rm C} (r)\, ,
\end{align}
\end{subequations}
where $f^{\rm PN}(r, p_\varphi, p_r)$ is any of the expressions in curly brackets in Eqs.~\eqref{eq:pn_flux_raw},
~\eqref{eq:pn_flux_hfact} or ~\eqref{eq:mode_by_mode_fluxes}.
Applied, for instance, to the $(2,2)$ mode energy flux, to \ac{nlo} this procedure leads to:
\begin{equation}
    \label{eq:cnc_m22}
    \dot{M}_{22}^{\rm NC} = \frac{r^2}{p_\varphi^4} + \frac{1}{c^2} \Bigl( \frac{4}{p_\varphi^2} + 5 \frac{r}{p_\varphi^4} - \frac{p_r^2 r^2}{p_\varphi^4} - \red{\frac{3 r^{2/3}}{p_\varphi^{10/3}}} \Bigr) \, .
\end{equation}

\begin{table*}[t]
  \centering
  \begin{tabular}{lccccc}
    \multicolumn{5}{c}{Eccentric} \\
    \toprule
    & \multicolumn{2}{c}{$\dot{S}=0$ crossings} &
    \multicolumn{2}{c}{$\dot{M}=0$ crossings} \\
    \cmidrule(lr){2-3}\cmidrule(lr){4-5}
    Model
    &      $\dot{r}<0$ & $\dot{r}>0$ 
      & $\dot{r}<0$ & $\dot{r}>0$ \\
    \midrule
    \multicolumn{5}{l}{$m=0,1,2$} \\
    \midrule
    $\Omega_T$
      & 0/27~(0\%) & 0/27~(0\%)
      & 22/27~(81\%) & 18/31~(58\%) \\
    $\Omega_T^{\mathrm{dt}}$
      & 26/27~(96\%) & 19/27~(70\%)
      & 7/27~(26\%) & 14/31~(45\%) \\
    $\Omega_T^{\mathrm{NS}}$
      & 23/27~(85\%) & 9/27~(33\%)
      & 26/27~(96\%) & 16/31~(52\%) \\
    $\Omega_T^{\mathrm{NS,\,dt}}$
      & 25/27~(93\%) & 20/27~(74\%)
      & 26/27~(96\%) & 28/31~(90\%) \\
    \midrule
    \midrule
    \multicolumn{5}{l}{$m=1$} \\
    \midrule
    $\Omega_T$
      & 0/26~(0\%) & 0/27~(0\%)
      & 4/40~(10\%) & 3/37~(8\%) \\
    $\Omega_T^{\mathrm{dt}}$
      & 26/26~(100\%) & 19/27~(70\%)
      & 4/40~(10\%) & 3/37~(8\%) \\
    $\Omega_T^{\mathrm{NS}}$
      & 0/26~(0\%) & 0/27~(0\%)
      & 12/40~(30\%) & 4/37~(11\%) \\
    $\Omega_T^{\mathrm{NS,\,dt}}$
      & 26/26~(100\%) & 19/27~(70\%)
      & 38/40~(95\%) & 22/37~(59\%) \\
    \midrule
    \midrule
    \multicolumn{5}{l}{$m=2$} \\
    \midrule
    $\Omega_T$
      & 0/27~(0\%) & 0/28~(0\%)
      & 8/27~(30\%) & 1/28~(4\%) \\
    $\Omega_T^{\mathrm{dt}}$
      & 27/27~(100\%) & 19/28~(68\%)
      & 23/27~(85\%) & 25/28~(89\%) \\
    $\Omega_T^{\mathrm{NS}}$
      & 24/27~(89\%) & 10/28~(36\%)
      & 10/27~(37\%) & 17/28~(61\%) \\
    $\Omega_T^{\mathrm{NS,\,dt}}$
      & 26/27~(96\%) & 20/28~(71\%)
      & 27/27~(100\%) & 25/28~(89\%) \\
    \bottomrule
\end{tabular}

  \begin{tabular}{cccc}
    \multicolumn{4}{c}{Hyperbolic} \\
    \toprule
    \multicolumn{2}{c}{$\dot{S}=0$ crossings} &
    \multicolumn{2}{c}{$\dot{M}=0$ crossings} \\
    \cmidrule(lr){1-2}\cmidrule(lr){3-4}
      $\dot{r}<0$ & $\dot{r}>0$ 
      & $\dot{r}<0$ & $\dot{r}>0$ \\
    \midrule
    \multicolumn{4}{l}{$\hphantom{0,1,2}$} \\
    \midrule
    \vphantom{$\Omega_T$} 
        0/6~(0\%) & 0/8~(0\%)
      & 7/35~(20\%) & 8/29~(28\%) \\
    \vphantom{$\Omega_T^{\mathrm{dt}}$} 
        6/6~(100\%) & 1/8~(12\%)
      & 6/35~(17\%) & 6/29~(21\%) \\
    \vphantom{$\Omega_T^{\mathrm{NS}}$} 
        3/6~(50\%) & 0/8~(0\%)
      & 9/35~(26\%) & 4/29~(14\%) \\
    \vphantom{$\Omega_T^{\mathrm{NS,\,dt}}$} 
        3/6~(50\%) & 0/8~(0\%)
      & 7/35~(20\%) & 2/29~(7\%) \\
    \midrule
    \midrule
    \multicolumn{4}{l}{$\hphantom{1}$} \\
    \midrule
    \vphantom{$\Omega_T$} 
        0/8~(0\%) & 0/11~(0\%)
      & 1/40~(2\%) & 2/33~(6\%) \\
    \vphantom{$\Omega_T^{\mathrm{dt}}$} 
        6/8~(75\%) & 3/11~(27\%)
      & 0/40~(0\%) & 0/33~(0\%) \\
    \vphantom{$\Omega_T^{\mathrm{NS}}$} 
        0/8~(0\%) & 0/11~(0\%)
      & 17/40~(42\%) & 4/33~(12\%) \\
    \vphantom{$\Omega_T^{\mathrm{NS,\,dt}}$} 
        6/8~(75\%) & 3/11~(27\%)
      & 2/40~(5\%) & 1/33~(3\%) \\
    \midrule
    \midrule
    \multicolumn{4}{l}{$\hphantom{2}$} \\
    \midrule
    \vphantom{$\Omega_T$} 
        0/6~(0\%) & 0/7~(0\%)
      & 5/12~(42\%) & 8/16~(50\%) \\
    \vphantom{$\Omega_T^{\mathrm{dt}}$} 
        6/6~(100\%) & 1/7~(14\%)
      & 5/12~(42\%) & 3/16~(19\%) \\
    \vphantom{$\Omega_T^{\mathrm{NS}}$} 
        3/6~(50\%) & 0/7~(0\%)
      & 3/12~(25\%) & 4/16~(25\%) \\
    \vphantom{$\Omega_T^{\mathrm{NS,\,dt}}$} 
        3/6~(50\%) & 0/7~(0\%)
      & 8/12~(67\%) & 4/16~(25\%) \\
    \bottomrule
\end{tabular}

  \caption{\red{Fraction of points within the $\pm10\%$ band around
  $\Omega_T/\Omega_H = 1$ for the angular momentum ($\dot{S}=0$)
  and energy ($\dot{M}=0$) flux zero crossings, for the global
  fluxes and individual $m$-modes. We consider different models for the
  tidal frequency: ``dt'' indicated the use of exact time derivatives $\dot{r}$
  and $\dot{\varphi}$ in place of $p_r$ and $p_\varphi/r^2$; ``NS'' indicates the use of
  the nonspinning extensions of the tidal frequencies. We list figures for eccentric
  and hyperbolic orbits separately, and also split the ingoing and outgoing legs
  of the orbit to isolate the impact of post-periastron \ac{qnm} excitation.}
    \label{tab:superradiance}}
\end{table*}

By construction, the noncircular corrections computed this way reduce to unity for circular orbits, but only up to a given perturbative order.
It is then desirable to reparameterize them in a form that explicitly depends on $p_r$, $\dot{p}_r$ or any other
quantity that vanishes for circular orbits, so that the reduction to 1 is exact.
To do so, we define a new variable \red{$K \equiv \left[p_\varphi/j_c(r)\right]^2$}, 
substitute \red{$p_\varphi = j_c(r) \sqrt{K}$} into the noncircular flux factors and reexpand to the requisite order, treating $K$ as
a constant\footnote{This procedure is similar, but not identical, to
the treatment of $\hat{h}_{\ell m}^{\rm QK, nc}$ in~\citet{Placidi:2023ofj}, App.~C of~\citet{Placidi:2021rkh} and
~\citet{Khalil:2021txt}, where all terms are expanded in $r$, $p_r$ and $\dot{p}_r$.}.
In the test-mass limit we are considering here, we can then evaluate $K$ using the full, exact value of the angular momentum of a circular
geodesic in Kerr geometry, as given in~\citet{1983mtbh.book.....C}.
Applied for instance to Eq.~\eqref{eq:cnc_m22}, this reparameterization yields:
\begin{equation}
    \label{eq:cnc_r_m22}
    \dot{M}_{22}^{\rm NC, K} = \frac{1}{K^2} +  \frac{-1 - 3 K^{1/3} + 4 K - p_r^2 r}{c^2 r K^2}  \, .
\end{equation}
Equation~\eqref{eq:cnc_r_m22} manifestly displays the desired properties.
Full expressions for the noncircular multipolar fluxes are given in App.~\ref{sec:pn_expressions}.

This leaves us with fluxes of the form:
\begin{equation}
    \dot{X} = -\frac{8}{5} \nu^2 \Bigl( 1 -\frac{1}{c^3} \frac{\Omega_T^{X}}{\Omega_{\rm H}} \Bigr) \dot{X}^{\rm C} \dot{X}^{\rm NC, K} \, ,
\end{equation}
where \( X \in \{M, S, M_{\ell m}, S_{\ell m}\} \),  \( \Omega_T^{X} \) is the tidal frequency
entering the superradiance prefactor for each flux and \( \dot{X}^{\rm C} \) and \( \dot{X}^{\rm NC, K} \) are
the circular and noncircular factors defined above, the latter in its reparameterized form.

\subsection{Damour-Iyer-Nagar-like circular factorization}
\label{subsec:din_factorization}
Finally, we also consider factorized versions of the circular part of the \textit{multipolar} fluxes,
$\dot{X}_{\ell m}^{\rm C}$,
inspired by earlier works on their analytical representation~\citep{Damour:2007xr,Damour:2007yf, Damour:2008gu, Pan:2010hz}.
The idea is to replace the residual \ac{pn} series with a resummed form that accelerates 
convergence toward the exact result, following strategies that proved successful for the fluxes at infinity.
In particular, we use the 11\ac{pn} test-mass results of~\citet{Fujita:2014eta},
that were shown to provide accurate representations for circular orbits around a Kerr \ac{bh}
up to the \ac{lso} for arbitrarily large spins, and write:
\begin{equation}
    \dot{X}_{\ell m}^{\rm C} = x^6 \eta_{\ell m}^{X} (S_{\ell m}^{(\epsilon)})^2 f_{\ell m} \, .
\end{equation}
In the above, \( x = (M \Omega)^{2/3} \) is the standard PN parameter, \( S_{\ell m}^{(\epsilon)} \)
is the source term, equal to the specific energy or angular momentum of the particle for even and odd parity respectively, \( f_{\ell m} \)
are the residual amplitude corrections, 
and \( \eta_{\ell m}^{X} \) sets the leading order behavior of the flux for each
multipole, beyond the global leading factor of \( x^6 \).

\subsection{Resummation strategies}
\label{subsec:resummation_strategies}

The resummation of pure \ac{pn} expansions is a standard technique in the treatment of the relativistic two-body
problem, particularly within the \ac{eob} formalism, to extend their domain of validity toward the strong-field regime.
We thus also test resummation options for our generic flux expressions, following two avenues.

The first applies to the superradiance-factorized expressions only, and consists of replacing the \ac{lo}
orbital frequency and radial velocity that appear in the prefactors with their exact counterparts.
This replacement is well motivated: in the test-mass limit, the exact orbital frequency of the geodesic
governs the true superradiance threshold, so using $\Omega = \dot{\varphi}$
directly from the numerical trajectory rather than its \ac{pn} approximation $p_\varphi/r^2$ should improve the prediction
of the zero-crossings.
Explicitly, we replace:
\begin{subequations}
\label{eq:Hpref_resum}
\begin{align}
    \Omega_{\rm T}^{S} = \dfrac{p_\varphi}{r^2} \rightarrow& \, \Omega_{\rm T}^{S, \rm{dt}} = \dot{\varphi} \, , \\
    \Omega_{\rm T}^{M} = \dfrac{p_\varphi}{r^2} + 3 \dfrac{p_r^2}{p_\varphi} \rightarrow& \, \Omega_{\rm T}^{M, \rm{dt}} = \dot{\varphi} + 3 \dfrac{\dot{r}^2}{r^2 \dot{\varphi}} \, ,
\end{align}
\end{subequations}
and similarly for the $\Omega_{\rm T}^{X, \rm{NS}}$ and $\Omega_{\rm T, \ell m}^{X, \rm{NS}}$, which are mapped into $\Omega_{\rm T}^{X, \rm{NS}, \rm{dt}}, \Omega_{\rm T, \ell m}^{X, \rm{NS}, \rm{dt}}$,
\begin{subequations}
\begin{align}
    \Omega_{\rm T}^{S, \rm{NS, dt}} &= \dot{\varphi} \biggl(
        1 - \dfrac{16 \dot{r}}{3 r c^3}
    \biggr) \, , \\
    \Omega_{\rm T}^{M, \rm{NS, dt}} &= \Omega_{\rm T}^{S, \rm{NS,dt}} + \dfrac{3\dot{r}^2}{r^2 \dot{\varphi}} \biggl[
        1 + \dfrac{1}{c^2} \biggl(
            \dfrac{-5}{3 r} \nonumber \biggr. \biggr. \\
            \biggl. \biggl. &+ \dfrac{28}{3} r^2 \dot{\varphi}^2 
        \biggr) + \dfrac{16}{3 c^3} \dfrac{r^3 \dot{\varphi}^2 - 1}{r^2 \dot{r}}
    \biggr] \, , \\
    \Omega_{\rm T, 22}^{S, \rm{NS, dt}} &= \Omega_{\rm T}^{S, \rm{NS, dt}} \, , \\
    \Omega_{\rm T, 22}^{M, \rm{NS, dt}} &= \Omega_{\rm T, 22}^{S, \rm{NS,dt}} + \dfrac{9\dot{r}^2}{4r^2 \dot{\varphi}} \biggl[
        1 + \dfrac{1}{c^2} \biggl(
            \dfrac{-5}{3 r} \nonumber \biggr. \biggr. \\
            \biggl. \biggl. &+ \dfrac{10}{3} r^2 \dot{\varphi}^2 
        \biggr) + \dfrac{16}{3 c^3} \dfrac{r^3 \dot{\varphi}^2 - 1}{r^2 \dot{r}}
    \biggr] \, , \\
    \Omega_{\rm T, 21}^{S, \rm{NS, dt}} &= \dot{\varphi} \, , \\
    \Omega_{\rm T, 21}^{M, \rm{NS, dt}} &= \dot{\varphi} + \dfrac{16 \dot{r}^2}{r^2 \dot{\varphi}} \, .
\end{align}
\end{subequations}

The second resummation strategy deals with the residual \ac{pn} series appearing
in the circular factor of the fluxes, $f_{\ell m}$. Following earlier works~\citep{Nagar:2011aa, Fujita:2014eta},
we write $f_{\ell m} \equiv \rho_{\ell m}^{2\ell}$. This structure is motivated by analogy to the
factorization of the multipolar waveform amplitudes at infinity. Additional resummation strategies
for the $\rho_{\ell m}$ functions could be considered, such as Pad\'e approximants, but we leave
their investigation to future work.

The final expressions for the horizon fluxes that we consider in this work are then obtained combining
all of the above ingredients, which for the multipolar ones leads to:
\begin{equation}
    \label{eq:final_flux}
    \dot{X}_{\ell m} = -\frac{8}{5} \nu^2 x^6 \Bigl( 1 - \frac{1}{c^3}\frac{\Omega_T^{X}}{\Omega_{\rm H}} \Bigr) \dot{X}^{\rm NC, K} \eta_{\ell m}^{X} (S_{\ell m}^{(\epsilon)})^2 \rho_{\ell m}^{2\ell} \, .
\end{equation}

\begin{table*}
\begin{tabular}{lcccccc}
\toprule
Model & $\Omega_T$ model & CNC & non-circular, $\dot{r}\neq 0$ &  circular, $\dot{r} = 0$ & DIN & Eq. \\
\midrule
Model 1   & $\Omega = \dot{\varphi}$                          & circular only &  ---                            & 11PN  & $\checkmark$ & (36) of \citet{Fujita:2014eta} \\
Model 2   & $\Omega_{T, \ell m}^{\rm NS, dt}$ & $\checkmark$  & NNLO, with K-reparameterization & 11PN  & $\checkmark$ &\eqref{eq:final_flux} \\
Model 3   & $\Omega_{T, \ell m}^{\rm NS, dt}$ & $\checkmark$  & NNLO                            & 11PN  & $\checkmark$ &\eqref{eq:final_flux} \\
Model 4   & $\Omega_{T}^{\rm NS, dt}$         & ---           & NNLO                            & 1.5PN & --- &\eqref{eq:general_hfact} \\
Model 5   & $\Omega_{T}^{\rm NS, dt}$         & $\checkmark$  & NNLO                            & 1.5PN & --- &\eqref{eq:general_hfact},~\eqref{eq:cnc_factorization} \\
\bottomrule
\end{tabular}
\caption{Summary of the analytical models compared in Sec.~\ref{sec:resummation_assessment} to
numerical results.
Columns indicate which ingredients are included in each model: 
the superradiance-factorized tidal frequency ($\Omega_T$, Sec.~\ref{subsec:mode_by_mode_superradiance} and Sec.~\ref{subsec:resummation_strategies}), 
the circular-noncircular (CNC) factorization (Sec.~\ref{subsec:cnc_factorization}),
the presence of NNLO noncircular corrections ($\dot{r}\neq 0$),
the PN order of the circular part
and its Damour-Iyer-Nagar (DIN) resummation~(Sec.~\ref{subsec:din_factorization})
The last column indicates the equations that define the model.
A checkmark (\checkmark) indicates the ingredient is included; 
a dash (---) indicates it is absent.
\label{tab:models}}
\end{table*}

\section{Comparison of numerical and analytical results}
\label{sec:results}

\begin{figure}
    \includegraphics[width=0.49\textwidth]{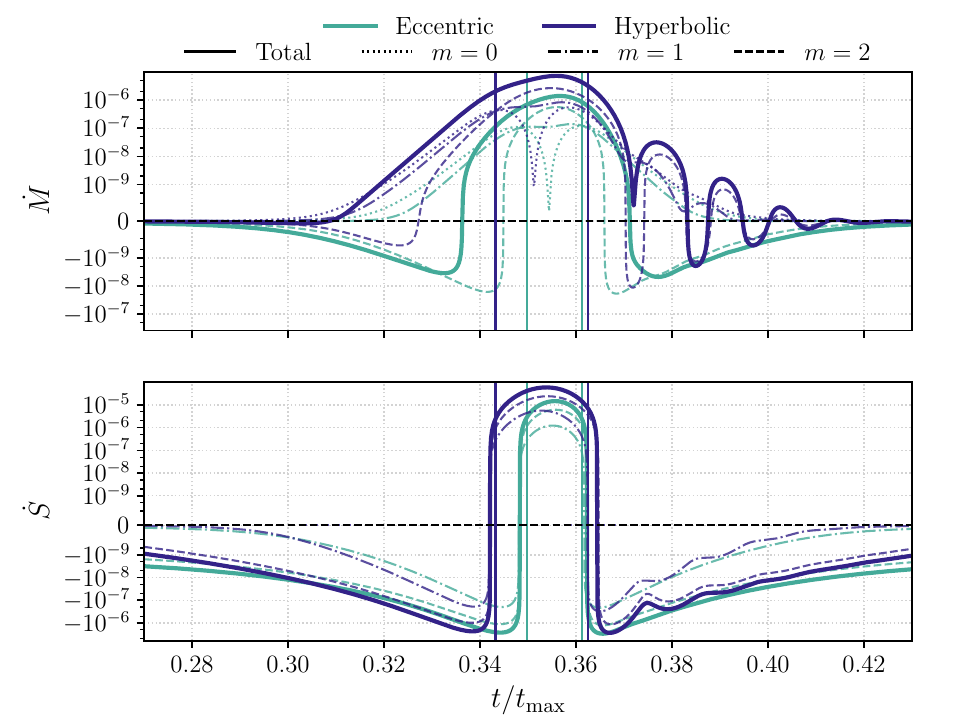}
    \caption{Examples of energy (top) and angular momentum (bottom) fluxes for an eccentric (green) and a hyperbolic (blue) orbit, showcasing 
    their sign changes around the time of periastron passage/closest approach. Vertical lines mark times when the orbital frequency 
    $\Omega = \Omega_{\rm H}$. Remarkably, $\dot{S}_{m=1}$ changes sign at the same time as $\dot{S}_{m=2}$.}
    \label{fig:fluxes_ex_sign}
\end{figure}

\begin{figure*}
    \includegraphics[width=0.32\textwidth]{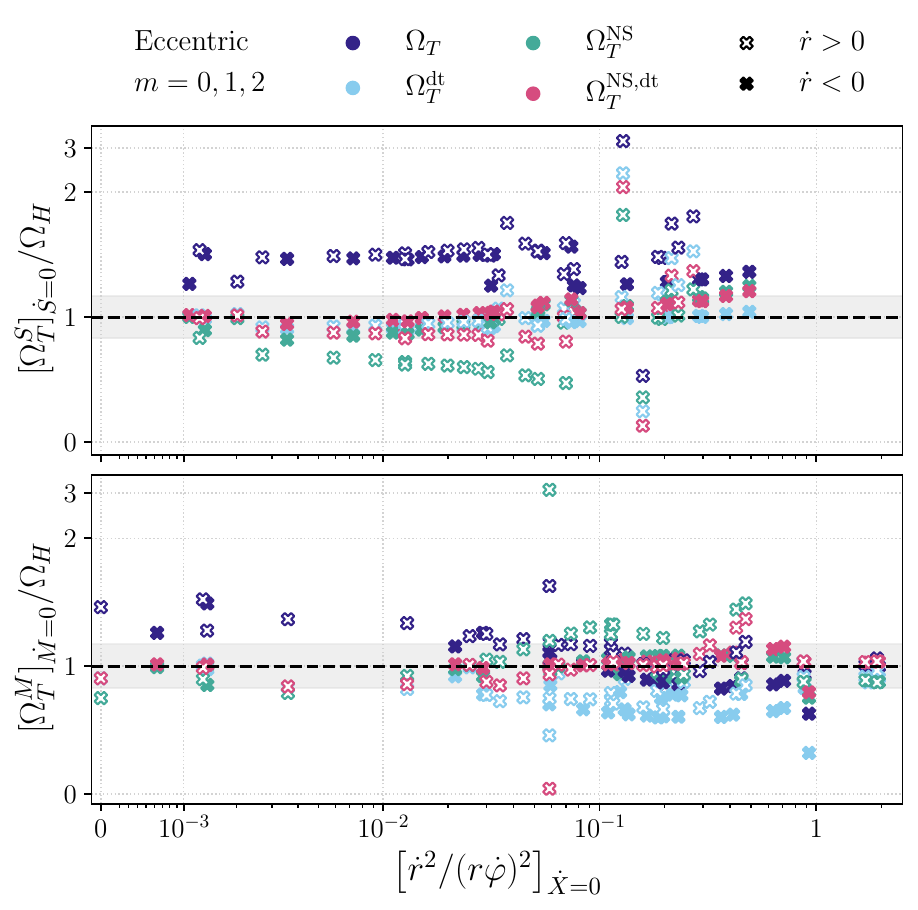}
    \includegraphics[width=0.32\textwidth]{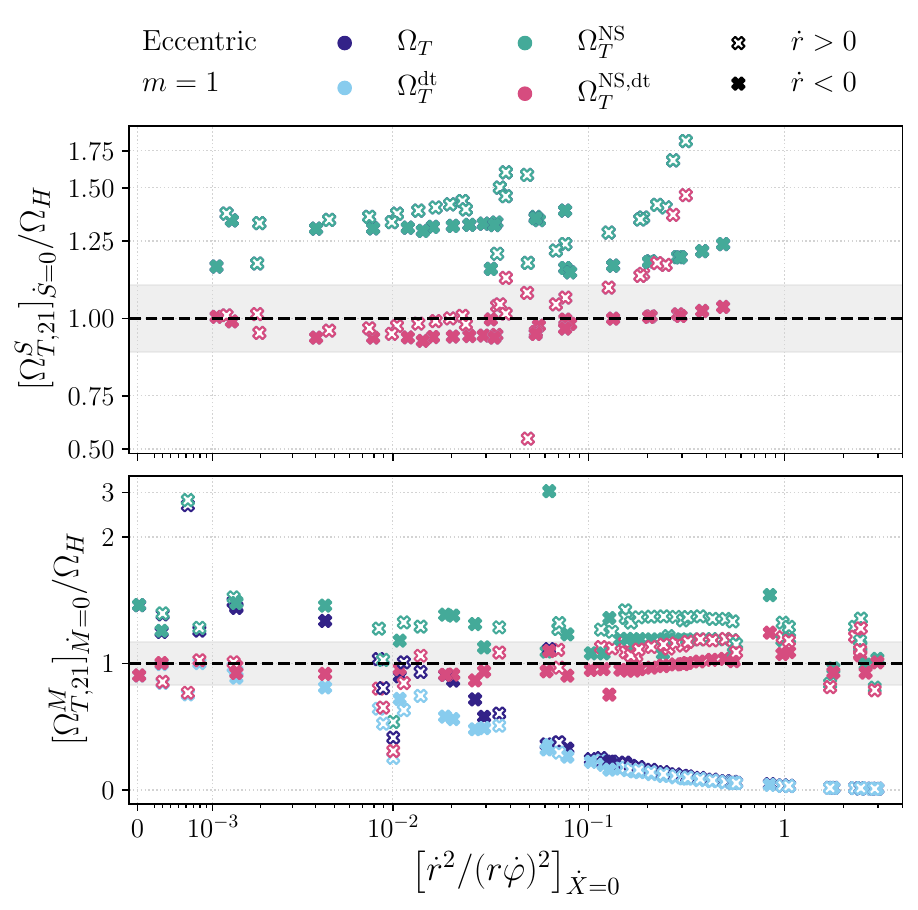}
    \includegraphics[width=0.32\textwidth]{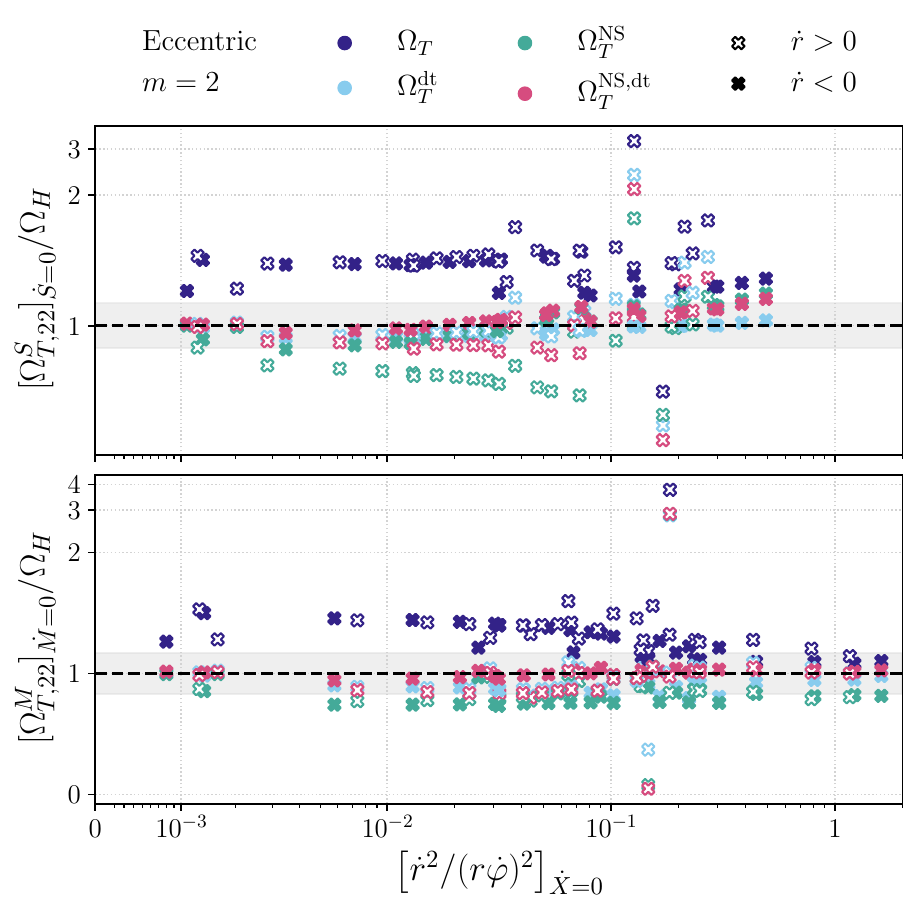}
    \includegraphics[width=0.32\textwidth]{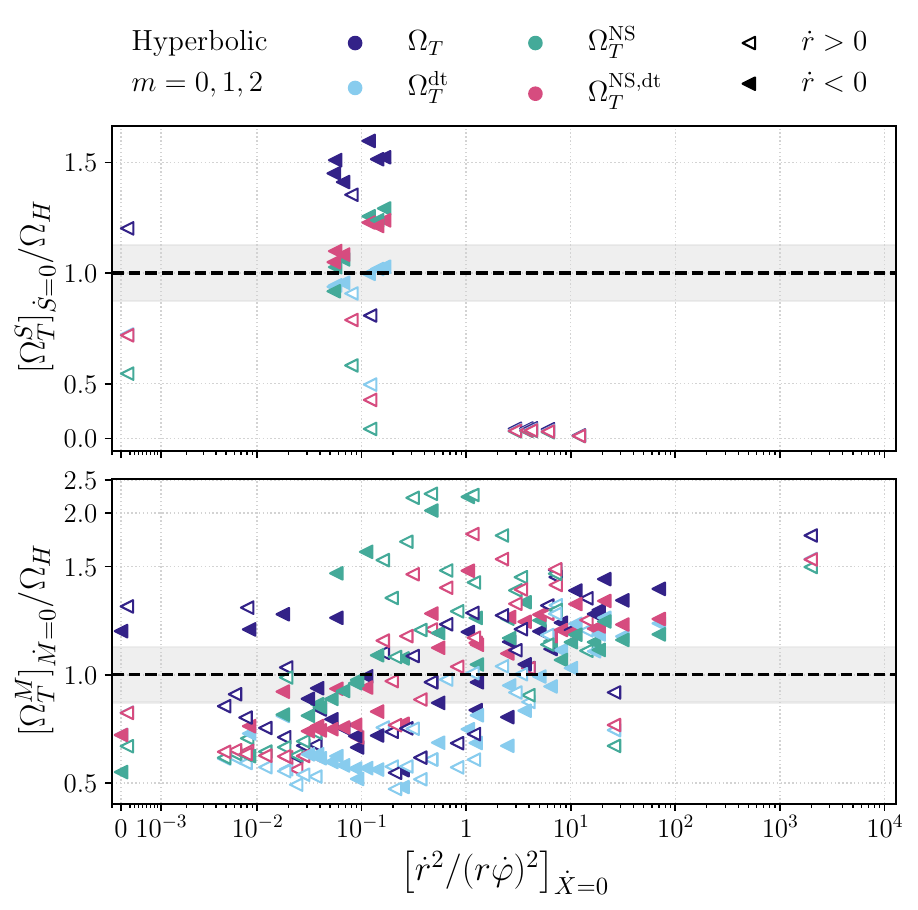}
    \includegraphics[width=0.32\textwidth]{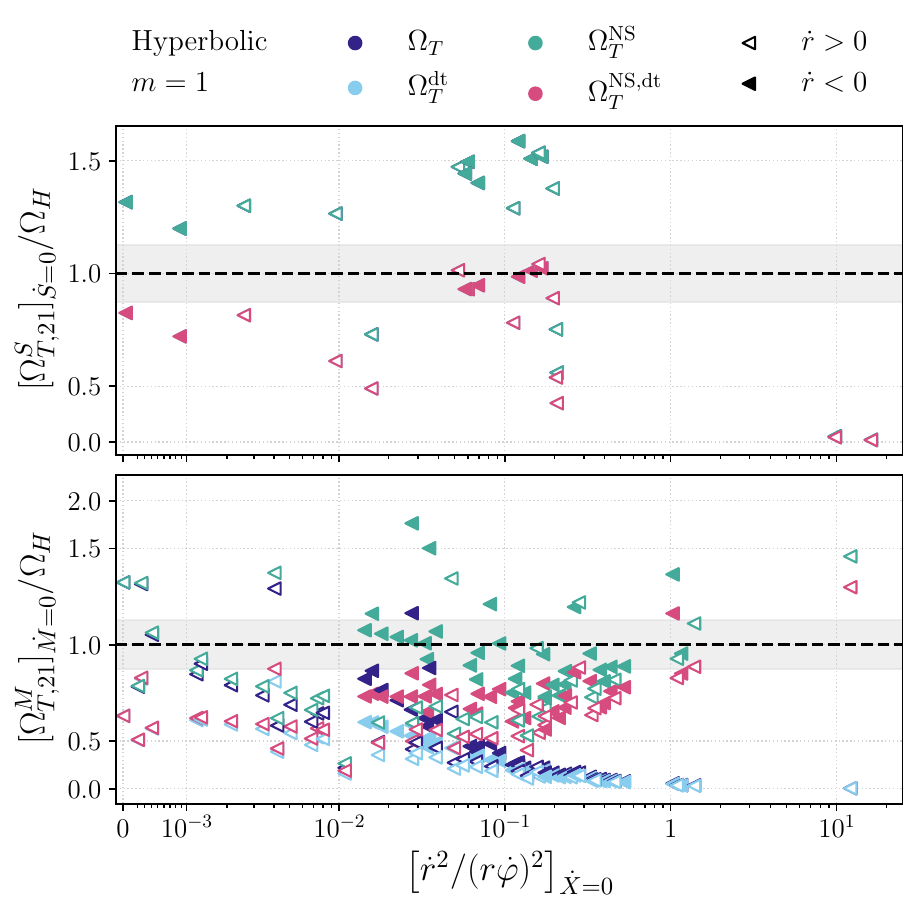}
    \includegraphics[width=0.32\textwidth]{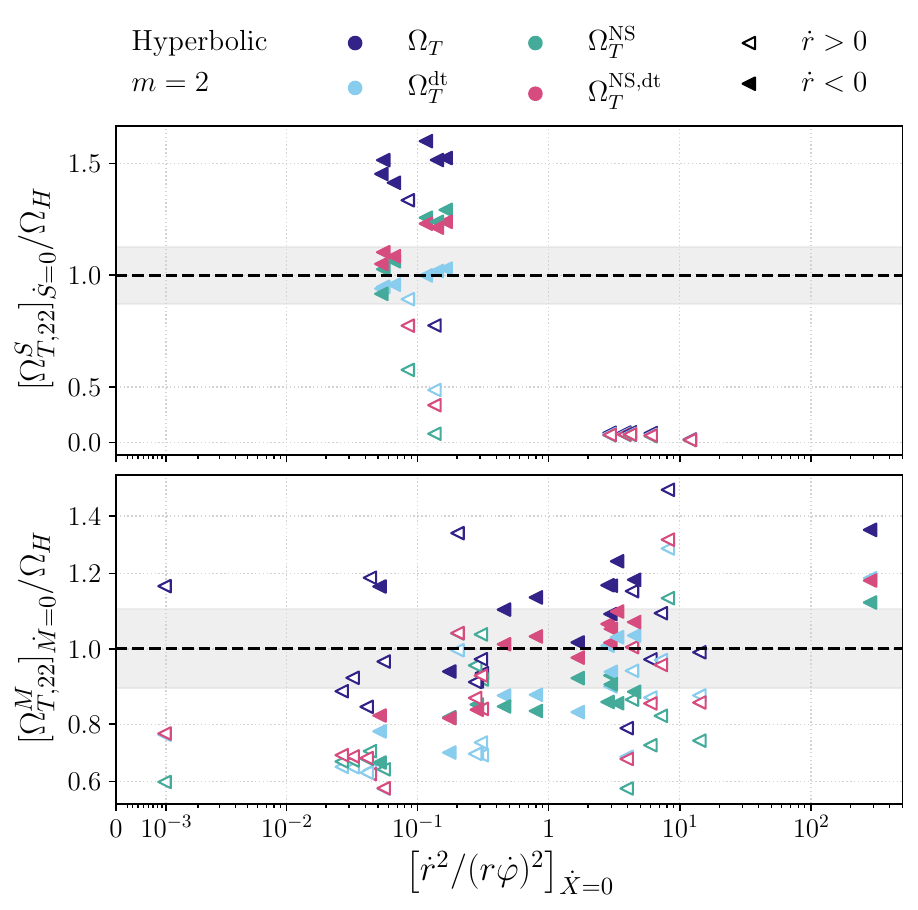}
    \caption{Comparing how well several models (see Sec.~\ref{subsec:mode_by_mode_superradiance}) for the tidal
    frequency entering the superradiance prefactors predict the sign change \red{in the fluxes for eccentric (first line)
    and hyperbolic (second line) orbits.}
    For each kind of orbit, we show the cases of $\dot{S}$ (top) and $\dot{M}$ (bottom), for the global fluxes (left), $m=1$ (center) and $m=2$ (right).
    \red{Filled} and unfilled markers differentiate between sign changes
    occurring before and after the \red{time of minimum separation}.
    \red{In each case, the ratio of the tidal frequency and $\Omega_{\rm H}$
    is plotted against the ratio of the radial and angular velocities;
    both are evaluated at times when the corresponding flux vanishes.}
    Globally, the $\Omega_{\rm T}^{\rm{NS, dt}}$ models (red) perform best for
    both $\dot{S}$ and $\dot{M}$. This is especially evident by looking at the $m=1$ modes: 
    \red{$\Omega_{\rm T, 21}^{\rm dt}$} models are necessary to capture the sign change in $\dot{S}_{21}$,
    while the nonspinning corrections are crucial for the energy flux.
    Note that for the former,  no $\dot{S}_{21}$ ``NS'' corrections are available,
    so the red and light blue models coincide, as do the dark blue and \red{teal} ones.}
    \label{fig:sr_prefactor}
\end{figure*}

\begin{figure}
    \includegraphics[width=0.47\textwidth]{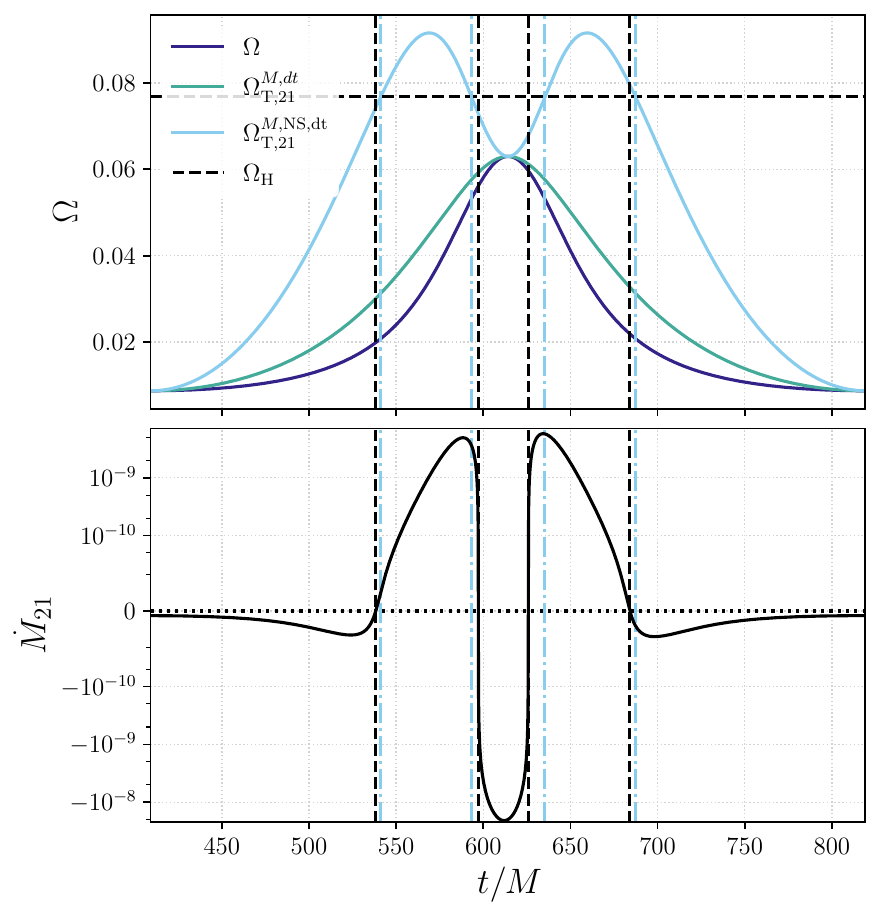}
    \caption{\red{Comparing analytical prediction and numerical results for the $(2, 1)$ energy flux for
    an eccentric orbit with $\hat{a} = 0.3$, $e = 0.5$, $p = 10$.
    Top panel: orbital frequency $\Omega$ and the $\Omega_{\rm T, 21}^{M, \rm dt}$ and $\Omega_{\rm T, 21}^{M, \rm NS, dt}$ tidal frequency models.
    Bottom panel: the $(2, 1)$ energy flux.
    Dashed black lines mark zeros in the flux; dashed colored lines times when the corresponding tidal frequency
    equals $\Omega_{\rm H}$.
    The peculiar pattern in $\Omega_{\rm T, 21}^{M, \rm NS, dt}$ caused by the $\dot{r}$ dependence of the
    non-spinning corrections is matched by the numerical flux.}}
    \label{fig:m1_E_flux}
\end{figure}

We now turn to the comparison of the numerical fluxes with the various analytical
models at our disposal, and attempt to determine the best combination of the 
different factorization and resummation strategies introduced in the previous section.
Given the large number of models that we can construct,
we proceed in a stepwise manner to reduce the model space efficiently.
We first focus on models for $\Omega_T$, and discard them based on their predictivity
(or lack thereof) for the sign and zeros of the fluxes.
Among the surviving models, we then rank them based on their performance against instantaneous fluxes
on circular orbits, where the analytical expressions are best controlled.
Finally, we extend the comparison to instantaneous fluxes for non-circular orbits, which provide the most
challenging test of the analytical expressions.
The best-performing model at this stage is then validated globally against the full set of systems
simulated in this work, using orbit-averaged fluxes for eccentric orbits and peak values for hyperbolic ones.

\subsection{The superradiance prefactor}
\label{sec:superradiance_prefactor}

\begin{figure*}
    \includegraphics[width=\textwidth]{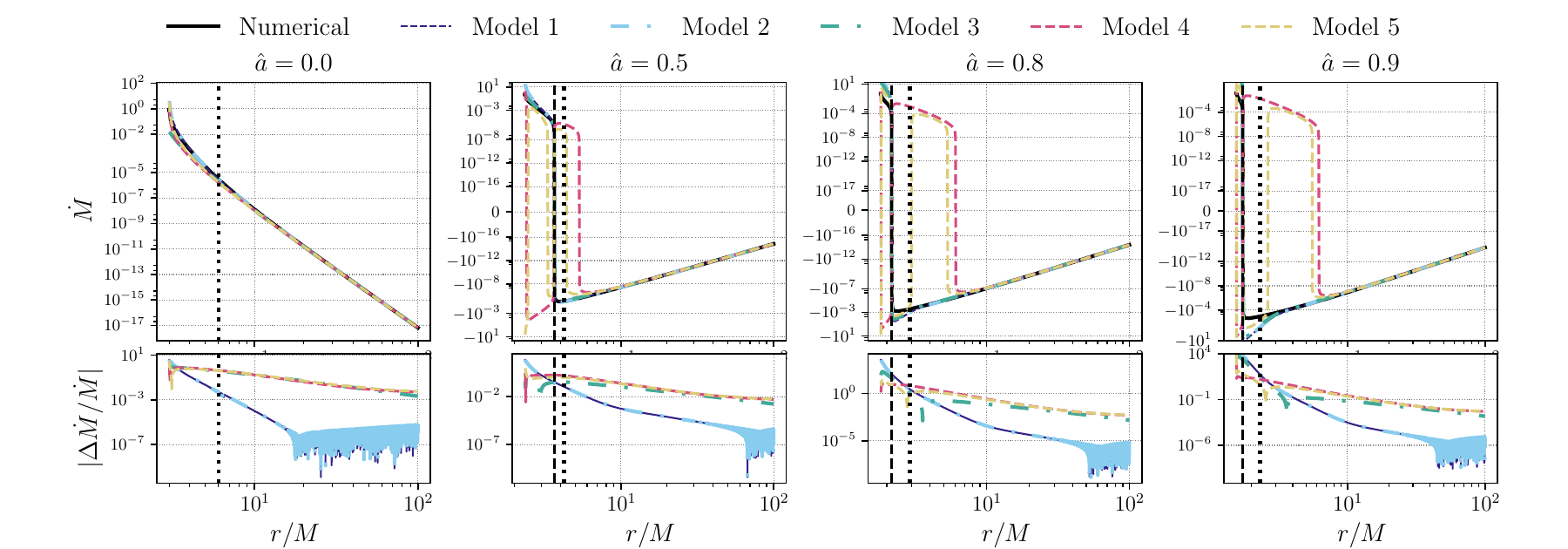}
    \caption{Relative differences between numerical and analytical fluxes for circular orbits for
    four different values of spin ($\abh=0$, $\abh=0.5$, $\abh=0.8$ and $\abh=0.9$ from left to right), as a function of the orbital separation
    and for different expressions of $\dot{M}$, as summarized in Tab.~\ref{tab:models}. The vertical dotted line marks the location
    of the \ac{lso} for each spin, while vertical dashed lines mark the location at which $\Omega = \Omega_{\rm H}$.
    Model 2 and model 1 are identical in this limit, thanks to the reparameterization of the CNC factor employed in the former.
    Models 3, 4 and 5, instead, perform significantly worse during the inspiral phase, with relative differences of order $\sim 1\%$.
    Moreover, models 4 and 5 display multiple sign changes in the strong-field regime, due to the low \ac{pn} order at which circular
    corrections are included.}
    \label{fig:circular_comparison}
\end{figure*}

\begin{figure*}
    \includegraphics[width=\textwidth]{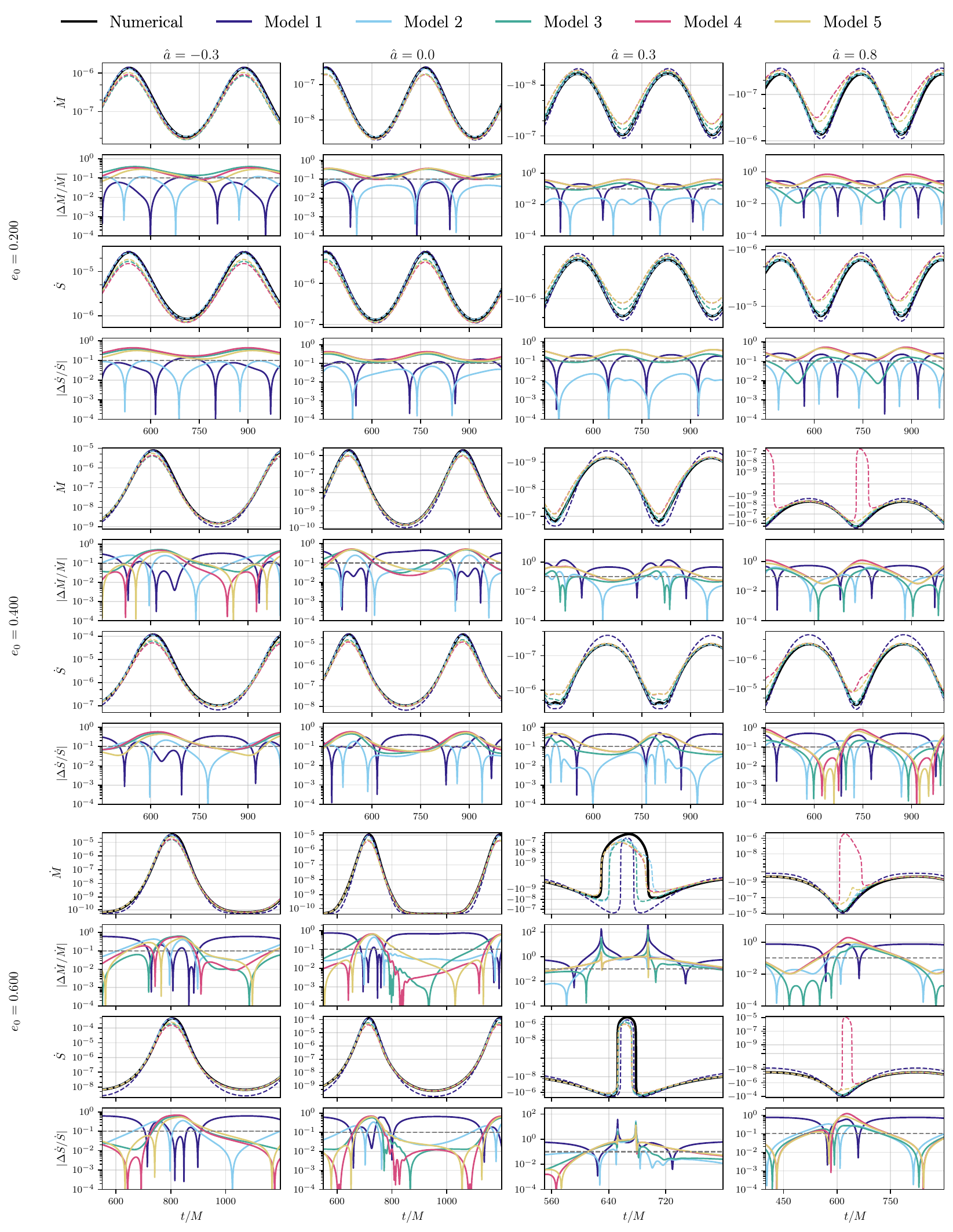}
    \caption{Energy and momentum fluxes and relative differences between numerical and analytical
    models for eccentric orbits with $p = 9$, $\abh=(-0.3,0.0,0.3,0.8)$ and varying eccentricity.
    We consider five different \ac{pn} expressions, as summarized in Tab.~\ref{tab:models}.
    At low eccentricities and for negative or zero spins, models 1 and 2 deliver the
    best performance. As eccentricity and spins increase, model 1 loses predictivity
    while model 2 remains overall the best-performing one, successfully interpolating
    between the low-$e_0$ and high-$e_0$ regimes.}
    \label{fig:eccentric_comparison}
\end{figure*}

\begin{figure*}
    \includegraphics[width=\textwidth]{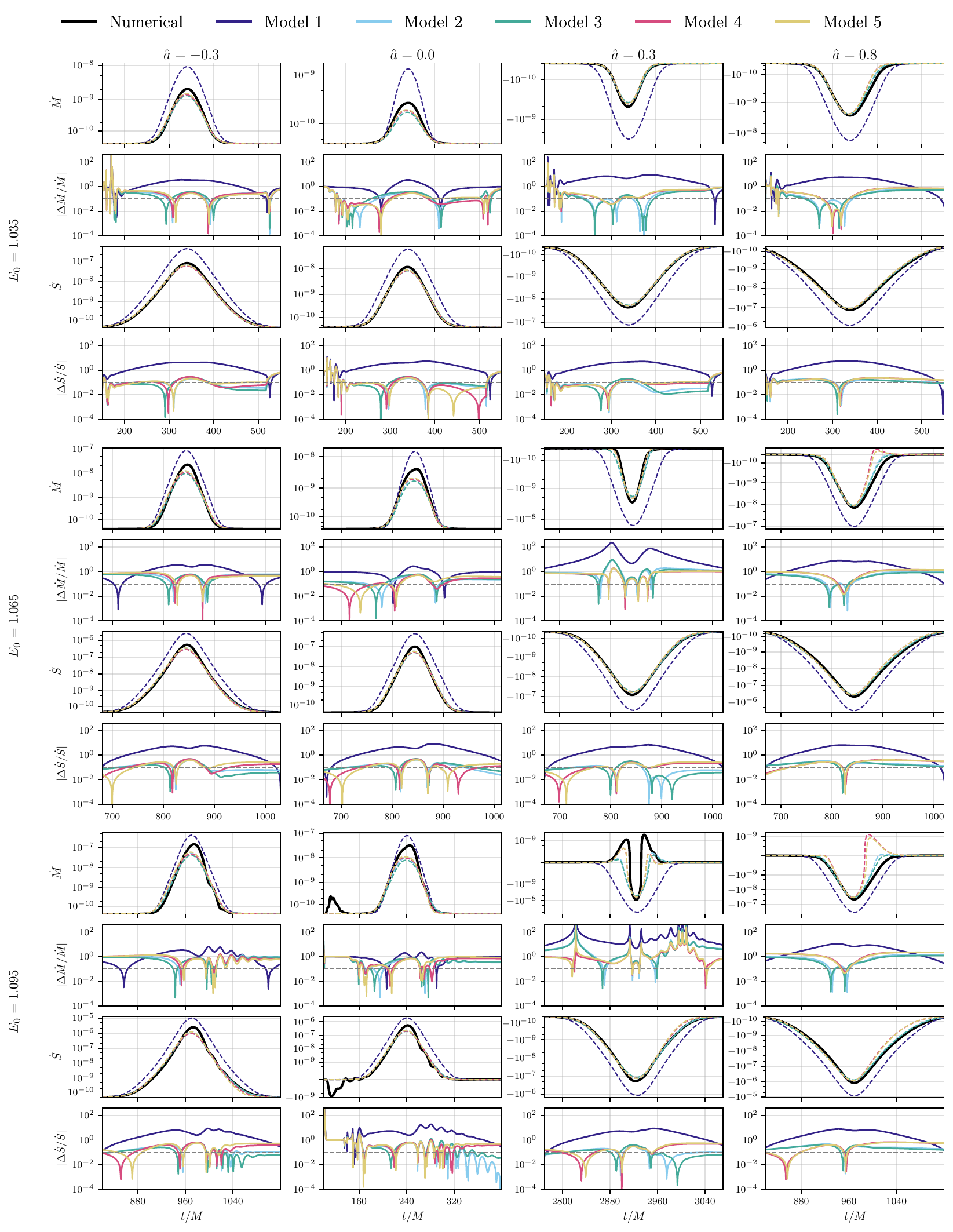}
    \caption{Energy and momentum flux and relative differences between numerical and analytical
    models for hyperbolic orbits with fixed $p_\varphi=8$, $\abh=(-0.3,0.0, 0.3,0.8)$ and varying initial energy.
    The models considered are summarized in Tab.~\ref{tab:models}. Model 2---5 are comparable in most
    cases, with models 4 and 5 approximating better the sign changes in $\dot{M}$ for 
    \red{$E_0 = 1.095$} and $\abh = 0.3$ with respect to 2 and 3. Model 1, instead, performs significantly worse than the others,
    particularly for $\dot{M}$. This is expected, as this expression does not include any noncircular correction.}
    \label{fig:hyperbolic_comparison}
\end{figure*}

\subsubsection{Zero Crossings of the Horizon Fluxes}

We begin our analysis of the superradiant behavior of the horizon fluxes
by once more considering two example cases with $\abh = 0.3$, one eccentric and one hyperbolic.
Figure~\ref{fig:fluxes_ex_sign} illustrates their key
features, common across all our simulations.
We find that:
(i) as expected, when looking at the total fluxes, $\dot{M} \neq \Omega \dot{S}$;
(ii) for $\dot{S}$, the $m=1$ and $m=2$ modes change sign at the same time, both before and after the minimum separation;
(iii) for $\dot{M}$, this is not generally the case; 
(iv) for all systems considered, $\dot{M}$ changes sign earlier than $\dot{S}$ during the approach, and later as the particle travels away from the center.

To quantify the performance of the different tidal frequency models
introduced in Sec.~\ref{sec:superradiance_prefactor},
we track the times $t_0^{\rm S}$ and $t_0^{\rm M}$ at which the numerical
$\dot{S}(t_0^{\rm S}) = 0$ and $\dot{M}(t_0^{\rm M}) = 0$, respectively.
We then evaluate the different $\Omega_{\rm T}$ models at these times, using the
underlying particle dynamics, and compare them with the horizon frequency
$\Omega_{\rm H}$, as displayed in Fig.~\ref{fig:sr_prefactor}.

\red{This apparently straightforward procedure is complicated by the presence of
numerical artifacts, especially during the initial and final stages of the simulation,
and of physical effects that our framework can not capture.
Indeed, since the simulations where sign changes occur are by necessity those
where the particle comes closest to the central \ac{bh}, the strong-field
perturbation at the turning point can excite its \acp{qnm}.
This has been observed already in the fluxes at infinity for similar
systems~\citep{1984PThPh..72..494K,Rifat:2019fkt,Thornburg:2019ukt,Albanesi:2021rby},
and is therefore expected to be a generic feature of close, highly
noncircular encounters.
The \ac{qnm} contribution can dominate the numerical horizon fluxes during the
outgoing leg, significantly altering their morphology and breaking the
clean correlation between the sign of the flux and the instantaneous orbital
dynamics: several zeros can appear in the \ac{qnm}-dominated phase,
as seen in a relatively mild example in the top panel of
Fig.~\ref{fig:fluxes_ex_sign}. This phenomenon occurs particularly often on hyperbolic orbits,
with stronger \ac{qnm} excitation at high energies and/or small
minimum separation; it is instead significant only in a handful of cases
on bound orbits, and only at the highest eccentricities we survey.
As such, to identify genuine sign changes due to the dynamics-driven component
in the presence of \ac{qnm} contamination, we visually inspect the fluxes and
retain only up to two zero crossings, informing our decision by
by the sign pattern during the first leg of the orbit and the predictions
of the various $\Omega_{\rm T}$ models. In general, we expect 
sign changes after the closest approach to be less reliable indicators
of the kind of superradiant effects that can help us identify the most
effective analytical model.}

Table~\ref{tab:superradiance} summarizes \red{our} results, reporting the fraction of zero crossings predicted
within $10\%$ of $\Omega_{\rm H}$ for each model, broken down by flux
channel ($\dot{S}=0$ and $\dot{M}=0$), radial direction ($\dot{r}<0$
and $\dot{r}>0$), and azimuthal mode.
\red{For the total spin flux, the \ac{lo} \ac{pn} tidal frequency $p_\varphi/r^2$ never falls within $10\%$ of $\Omega_{\rm H}$ at the zeros,
whereas each of the three corrected models does so for most ingoing-leg crossings on eccentric orbits:
$|\Omega_{\rm T}(t_0^{\rm S})/\Omega_{\rm H} - 1| \lesssim 0.1$ in $\srEccTotSDtInPct\%$ ($\Omega_{\rm T}^{\rm dt}$),
$\srEccTotSNsInPct\%$ ($\Omega_{\rm T}^{\rm NS}$) and $\srEccTotSNsdtInPct\%$ ($\Omega_{\rm T}^{\rm NS, dt}$) of cases with $\dot{r}<0$.
The two ingredients are not equivalent, however: the exact orbital frequency $\dot{\varphi}$ is what carries over to the outgoing leg
($\srEccTotSDtOutPct\%$--$\srEccTotSNsdtOutPct\%$ for the models employing it, against $\srEccTotSNsOutPct\%$ for $\Omega_{\rm T}^{\rm NS}$),
while the non-spinning corrections alone do not.
Only a handful of spin flux zeros occur in our hyperbolic sample; there $\Omega_{\rm T}^{\rm dt}$ captures $\srHypTotSDtInFrac$ of the ingoing-leg zeros,
and performance of all $\Omega_{\rm T}$ models drops starkly for times after closest approach.}

Turning to the energy flux, we observe that \red{the orbital frequency evaluated at the first flux zeros always satisfies $[\dot{\varphi}]_{\dot{M} = 0} < [\dot{\varphi}]_{\dot{S} = 0}$}, reflecting our point (iv) in the list above.
This hierarchy neatly aligns with our models for $\Omega_{\rm T}^{\rm M}$, which at leading order include a \emph{positive} correction to $\dot{\varphi}$
that depends on the radial velocity\red{, which pushes the tidal frequency above $\Omega_{\rm H}$ earlier.}
Out of the different $\Omega_{\rm T}^{\rm M}$ models, the best performance \red{on eccentric dynamics} is \red{that of} 
$\Omega_{\rm T}^{\rm{NS, dt}}$, with $\srEccTotMNsdtInPct\%$ ($\dot{r} < 0$)/$\srEccTotMNsdtOutPct\%$ ($\dot{r}>0$) of zero
crossings predicted within $10\%$.
\red{The roles of the two ingredients are reversed with respect to the spin flux.
The non-spinning corrections are essential: $\Omega_{\rm T}^{\rm NS}$ alone already reaches $\srEccTotMNsInPct\%$ on the ingoing leg,
while $\Omega_{\rm T}^{\rm dt}$ alone is the worst-performing model ($\srEccTotMDtInPct\%$), since replacing $p_\varphi/r^2$ with $\dot{\varphi}$
systematically lowers the predicted frequency.
The exact velocities matter mostly after periastron, where $\Omega_{\rm T}^{\rm NS, dt}$ reaches $\srEccTotMNsdtOutPct\%$
against $\srEccTotMNsOutPct\%$ for $\Omega_{\rm T}^{\rm NS}$.}
\red{No tidal frequency model realiably predicts the zeros of the energy flux on the hyperbolic orbits we consider.
Between $\srHypTotMOutOfBandMinPct\%$ and $\srHypTotMOutOfBandMaxPct\%$ of flux zeros occur when $|\Omega_{\rm T} / \Omega_{\rm H} - 1| > 10\%$;
only the base $\Omega_{\rm T}$ and $\Omega_{\rm T}^{\rm dt}$ appear to retain some level of accuracy across periastron passage.}

\red{We repeat the same analysis for the single-mode $\dot{M}$ and $\dot{S}$; results are shown in the middle and right panels of 
Fig.~\ref{fig:sr_prefactor}, and summarized in the lower two thirds of Table~\ref{tab:superradiance}.}
\red{Beginning with the eccentric simulations, for the $(2,2)$ mode, models employing the exact orbital frequency $\dot{\varphi}$ 
clearly outperform their PN counterparts for $\dot{S}$; for $\dot{M}$, the additional inclusion
of the non-spinning corrections delivers the best agreement with the numerical data, although here it is the exact orbital frequency
that does most of the work ($\srEccMtwoMDtInPct\%$ for $\Omega_{\rm T}^{\rm dt}$ alone against $\srEccMtwoMNsInPct\%$ for $\Omega_{\rm T}^{\rm NS}$ alone
on the ingoing leg), the mirror image of what we found for the total flux.}
\red{The sign of the $(2, 1)$ mode angular momentum flux is only captured by the $\Omega_{\rm T}^{\rm dt}, \Omega_{\rm T}^{\rm NS, dt}$ models; that of $\dot{M}$ is best reproduced by the latter, the non-spinning corrections proving especially important here.}
\red{Hyperbolic orbits again pose a much greater challenge. We find few zero crossings in the angular momentum fluxes; results for both modes closely match those for the total flux.
It is hard to draw clear conclusions from the $(2,2)$-mode energy flux data. The $\Omega_{\rm T}$ model performs best by our metric, though much of this is due to the zero crossings in the outgoing leg of the orbit, where we find significant contamination by \acp{qnm} excitation; $\Omega_{\rm T}^{\rm NS, dt}$ most often reproduces zeros before periastron.
The $(2,1)$-mode energy flux on hyperbolic orbits displays the most complex morphology,
and its zero crossings are clearly the toughest to predict accurately.
Data points in the lower middle panel of Fig.~\ref{fig:sr_prefactor} do appear to display the clearest patterns out of the hyperbolic orbit energy fluxes; all tidal frequency models are typically lower than $\Omega_{\rm H}$ at the flux zeros, and the tendency of $\Omega_{\rm T}^{\rm NS}$ to provide higher estimates than its competing models helps it nominally perform best.
This is the only instance where this happens, and we suspect is more likely an accidental result due to our analytical expressions being incapable of accurately describing the flux's properties on hyperbolic orbits rather than a genuine hint to that model's faithfulness over the others.}

\subsubsection{Sign patterns}
\red{The analysis presented above relies on heuristic criteria to identify
the zero crossings of the horizon fluxes. Here, we broaden our comparison
by examining the overall sign pattern in 
$\Omega_{\rm T} - \Omega_{\rm H}$, irrespective of how accurately the \emph{times}
of the fluxes' zero crossings are reproduced, and assess its consistency with the
numerical data.}

\red{We find mismatches in the predicted and observed total flux signs at the time of closest approach
in only a handful of cases for each mode combination (4 eccentric and 4 hyperbolic).
All of these we observe when using the base $\Omega_{\rm T}$ models of Eqs.~\eqref{eq:base_omgT}, which depend on the canonical
variables rather than the exact azimuthal and radial velocities.
Evaluating how accurately our tidal frequency models reproduce the observed flux sign
outside of the time of closest approach is harder,
with the post-peak \ac{qnm} contamination once more muddling the picture significantly.
Eccentric orbits are free of this additional contribution in all but 6 cases.
In this sub-sample, predicted and observed sign patterns in the total fluxes match
in almost all cases. Exceptions are simulations located near transition regions in
parameter space: between configurations where the fluxes change sign near
periastron and those where they do not.
The same holds for the $m = 2$ fluxes, and the $m=1$ angular momentum flux.
The $m=1$ energy flux is more interesting: in many cases, both eccentric and hyperbolic,
we see the flux change sign multiple times over the orbit, from negative at large
distances, to positive in a short time interval before periastron, to again
negative at closest approach, repeating quasi-symmetrically afterwards.
Remarkably, the tidal frequency functions that incorporate the nonspinning corrections,
because of their dependence on the radial velocity or momentum, match this pattern,
as seen in a \ac{qnm}-free example in Fig.~\ref{fig:m1_E_flux}.
The $\Omega_{T, 21}^{M, \rm NS, dt}$ model most faithfully predicts this more
complex behavior; exceptions include orbits where the apastron or asymptotic value of
$\Omega_{\rm 21T}^{M, \rm NS, dt}$ is close to $\Omega_{\rm H}$.}

Globally, then, the $\Omega^{\rm NS,\,dt}_{\rm T}$ models appear to best capture the link between the orbital
dynamics and the superradiant behavior of the fluxes, motivating their use as the default choice in the comparisons that follow.

\subsection{Instantaneous fluxes}
\label{sec:resummation_assessment}

\begin{figure*}
    \includegraphics[width=\textwidth]{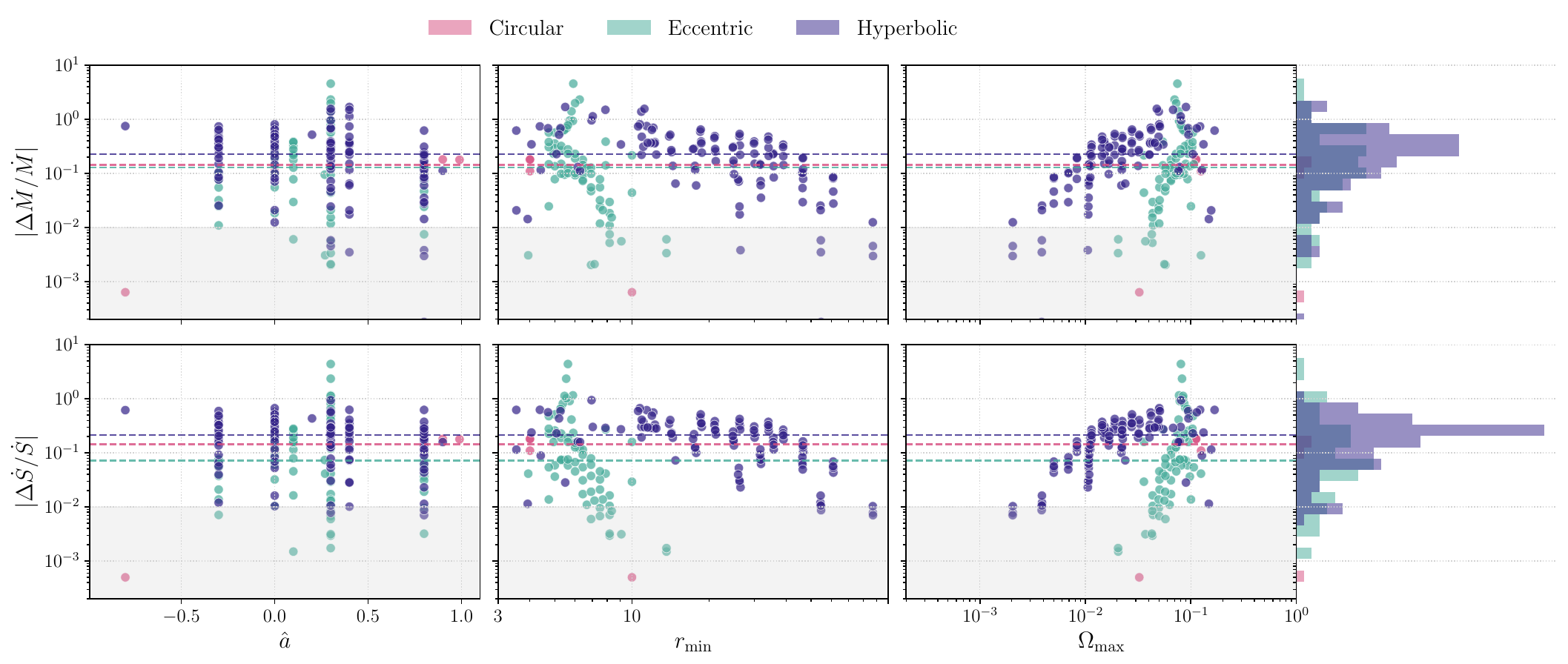}
    \caption{Relative differences between numerical and analytical fluxes for the full set of orbits considered in this work,
    for model 2 in Table~\ref{tab:models}. We show relative differences for the orbit-averaged fluxes in the eccentric and circular
    cases, and for the peak fluxes in the hyperbolic ones. Different colors indicate different orbital configurations, as indicated in the legend, while
    the gray shaded area marks the region where we estimate the numerical error to be. Horizontal, dashed lines mark the median relative difference for each configuration kind.}
    \label{fig:summary_comparison}
\end{figure*}

\subsubsection{Circular orbits}
\label{sec:circular_comparison}
Numerical results for circular orbits were obtained in~\citet{Taracchini:2013wfa} using a frequency-domain
Teukolsky solver. Analytical comparisons against these results were already presented in a number
of works~\citep{Nagar:2011aa, Bernuzzi:2012ku, Taracchini:2013wfa, Fujita:2014eta},
which showed that resummed expressions for the horizon fluxes, based on the factorization of the superradiance prefactor
and DIN-resummation of the remaining \ac{pn} series, provided a reasonable match (with relative differences of order unity or below)
to the numerical data up to the \ac{lso} for spins up to $\sim 0.99$ (see for instance Figs.~8 and~9 of~\citet{Taracchini:2013wfa}).
While these prior results were obtained using expressions specifically tuned to circular orbits,
a key question here is how well the generic-orbit expressions we have developed perform in this well-understood limit.

We consider a set of five models, summarized in Table~\ref{tab:models}, that include
different combinations of the ingredients
described in Sec.~\ref{sec:framework}.
This relatively minimal set of models allows us to isolate the impact of the CNC factorization,
of the high-order \ac{pn} information included in the circular
part, of the \ac{nnlo} noncircular corrections and its K-reparameterization.

Figure~\ref{fig:circular_comparison} shows the relative differences between the numerical fluxes
and these analytical models, summed over $m=1,2$\footnote{Recall that, for circular orbits,
m=0 fluxes are identically zero}, as a function of the orbital separation
and for four different spin values.
The central finding is that, thanks to the reparameterization of the NC corrections,
the performance of model 2 is identical to that of model 1, with relative differences with
respect to the exact fluxes at the level of $\sim \mathcal{O}(10^{-4})$ during the inspiral
and $\mathcal{O}(1)$ approaching the \ac{lso} for spins below $\sim 0.9$.
Importantly, it correctly captures the \emph{single} sign change in the fluxes close to the \ac{lso} crossing.
This feature is not correctly reproduced by models 4 and 5, which instead predict spurious sign changes at large separations,
and perform significantly worse than the first three.
This difference is a direct consequence of the high-order, resummed \ac{pn} information included in the circular part of models 1-3.
The poor performance of models 3,4,5 during the inspiral, instead, is a consequence of the low-order \ac{pn} noncircular corrections,
which introduce spurious terms beyond \ac{nnlo} that worsen the agreement with numerical data when not properly reparameterized.

\subsubsection{Noncircular orbits}
\label{sec:noncircular_comparison}
Moving on to the more interesting case of eccentric and hyperbolic orbits,
we first focus on the series of eccentric configurations with $p=9$ and varying
eccentricity and spin. This choice is motivated by the fact that these orbits
reach low minimum separations for large $e_0$, thus probing the strong-field
regime, and --- for positive $\abh$ values --- will also feature the pattern
of sign changes analyzed in Sec.~\ref{sec:superradiance_prefactor}.
We compare the \red{five} models with the numerical fluxes as a function of time,
examining both the fluxes themselves and their relative differences.
Figure~\ref{fig:eccentric_comparison} shows the results for both $\dot{M}$
(upper panels) and $\dot{S}$ (lower panels), with relative differences
$|\Delta\dot{M}/\dot{M}|$ and $|\Delta\dot{S}/\dot{S}|$ shown alongside.

For negative and zero spins at low eccentricities ($e_0 = 0.2$), model~2
matches the accuracy of model~1, both yielding relative differences of
$\mathcal{O}(10^{-1}\text{--}10^{-3})$, while models~3--5 perform
comparably or worse. This reflects the fact that at small $e_0$ the
orbit deviates only mildly from circularity, so the high-order circular
contributions dominate and the \ac{nnlo} noncircular terms shared by
models~2--5 do not yet provide a decisive advantage.
Once more, the reparameterization of the noncircular corrections -- present in model~2
but not in model 3 -- appears crucial. As eccentricity
and spin increase, model~2 improves upon the accuracy of
model~1 and performs at least as well as models~3, 4 and~5, making
it the most consistently reliable choice across the eccentric parameter
space. This advantage is particularly evident when the particle's radial velocity
is large and the deviation from circularity cannot be treated as a small perturbation.
All models struggle near pericenter for $\abh=0.3$ and $\abh=0.8$ at eccentricities
$e_0 \gtrsim 0.4$, the threshold at which sign changes begin to appear
in the numerical fluxes. At these points, relative differences can reach
$\mathcal{O}(1)$ or more, reflecting the breakdown of the low-order \ac{pn}
expressions in the strong field. For the most eccentric configurations
($e_0 = 0.9$, not shown), \ac{qnm} excitations in $\psi_0$ appear quite prominently
and further degrade the agreement by introducing oscillatory features in
the flux time series that are not captured by any of our analytical expressions.

Turning to hyperbolic orbits, results are shown in
Fig.~\ref{fig:hyperbolic_comparison} for both $\dot{M}$ and $\dot{S}$,
across three encounter energies ($E_0 = 1.035,\, 1.065,\, 1.095$) and
four spin values. Here the advantage of models~2-5 over model~1 is
systematic and present across the entire parameter space: hyperbolic
orbits deviate dramatically from circular motion at any separation, so
the quasi-circular model~1 loses its accuracy even for gentle encounters
($E_0 = 1.035$), while model~2 maintains the level of agreement
achieved by models~3, 4 and~5 throughout. The behavior of
$\dot{S}$ is qualitatively consistent with that of $\dot{M}$, though
with slightly better accuracy overall. Nonetheless, the
near-approach region remains the most challenging for all models,
owing to the same combination of strong-field \ac{pn} breakdown and
\ac{qnm} excitation discussed above.

\subsection{Averaged and peak fluxes}
\label{sec:global_comparison}
We now extend the comparison to the full set of
orbits, with a focus on non-circular trajectories, using orbit-averaged fluxes for
eccentric configurations and peak fluxes for hyperbolic ones.
Throughout this subsection we use model~2 with the $\Omega_T^{\rm NS,dt}$ prefactor
identified as optimal in Sec.~\ref{sec:superradiance_prefactor}.
Figure~\ref{fig:summary_comparison} summarizes the relative differences
$|\Delta\dot{X}/\dot{X}|$ as a function of spin $\abh$, minimum
separation $r_{\rm min}$, and peak orbital frequency
$\Omega_{\rm max}$, for both $\dot{M}$ and $\dot{S}$.

Globally, the analytical model performs comparably for both non-circular orbit
families, with better accuracy for the angular momentum flux than
for the energy flux in both cases. For hyperbolic orbits, the
median relative differences are \red{${\sim}\medianHyperbolicdE\%$ ($\dot{M}$) and
${\sim}\medianHyperbolicdJ\%$ ($\dot{S}$), with $\fractionHyperbolicdETen\%$ and $\fractionHyperbolicdJTen\%$} of configurations
respectively achieving agreement within $10\%$, and \red{$\fractionHyperbolicdEFifty\%$ and
$\fractionHyperbolicdJFifty\%$} within $50\%$. On eccentric orbits performance improves,
with median relative differences of \red{${\sim}\medianEccentricdE\%$ ($\dot{M}$)
and ${\sim}\medianEccentricdJ\%$ ($\dot{S}$), and $\fractionEccentricdETen\%$ and $\fractionEccentricdJTen\%$} of configurations
within $10\%$, and \red{$\fractionEccentricdEFifty\%$ and $\fractionEccentricdJFifty\%$} within $50\%$.
When restricting to the strong-field subset $r_{\rm min} < \EccentricrStrong$, the median errors
for hyperbolic orbits increase substantially to \red{${\sim}\medianHyperbolicdEStrong\%$
($\dot{M}$) and ${\sim}\medianHyperbolicdJStrong\%$ ($\dot{S}$)}, though this subset
contains only \red{$N=\HyperbolicNStrong$} configurations. By contrast, the eccentric
strong-field subset \red{($N=\EccentricNStrong$)} shows nearly unchanged median errors
of \red{${\sim}\medianEccentricdEStrong\%$ and ${\sim}\medianEccentricdJStrong\%$}, indicating that the bulk of the
eccentric sample already probes the strong-field regime and that
close approaches represent a more acute challenge for hyperbolic
trajectories.

The three panels of Fig.~\ref{fig:summary_comparison} reveal clear
correlations between the orbital parameters and the
relative errors. To quantify this, we compute the Spearman rank
correlation coefficient $\rho$ between each parameter and the
relative differences~\citep{spearman1904, zwillinger2000, kendall1973}; $|\rho|=1$ indicates a perfect monotonic
relationship while $\rho=0$ indicates no correlation.
The strongest single predictor is $\Omega_{\rm max}$, with \red{$\rho \approx \spearmanOmgmaxdE$
for $|\Delta\dot{M}/\dot{M}|$ and $\rho \approx \spearmanOmgmaxdJ$ for
$|\Delta\dot{S}/\dot{S}|$} across all orbits. The relatively modest
global values reflect the mixing of different orbital configurations rather than a weak
underlying trend: stratifying by family yields \red{$\rho \approx \spearmanOmgMaxHyperbolicdE$
and $\spearmanOmgMaxHyperbolicdJ$} in hyperbolic runs individually, and \red{$\rho \approx \spearmanOmgMaxEccentricdE$
and $\spearmanOmgMaxEccentricdJ$} for $|\Delta\dot{M}/\dot{M}|$ and
$|\Delta\dot{S}/\dot{S}|$ in eccentric ones, confirming that
$\Omega_{\rm max}$ is a reliable predictor within each population.
This is physically transparent: $\Omega_{\rm max}$ directly
measures how relativistic the orbit becomes at closest approach,
and thus how deeply the \ac{pn} expansion is pushed beyond its domain
of validity. The minimum separation $r_{\rm min}$ carries
comparable information \red{($\rho \approx \spearmanRmindE$ for $\dot{M}$ and
$\spearmanRmindJ$ for $\dot{S}$)}, as expected from its inverse relationship
with $\Omega_{\rm max}$. The spin dependence is \red{weaker
($\rho \approx \spearmanAdE$ and $\spearmanAdJ$ for $\dot{M}$ and $\dot{S}$ respectively)}:
no significant monotonic trend with $\hat{a}$ is identified across the
parameter space we surveyed.

\section{Conclusions}
\label{sec:conclusions}

In this work, we presented a detailed calculation of the fluxes of energy and angular momentum through the horizon of a Kerr \ac{bh} orbited by a test-mass along generic (circular, eccentric, hyperbolic) orbits.
Using a time-domain Teukolsky solver, we computed these for a large
set of configurations, varying the orbital parameters and the spin of the central \ac{bh}.
This constitutes the most extensive numerical survey of horizon fluxes for generic orbits in Kerr
spacetime to date, and provides a reference dataset for future modeling efforts.
In parallel, we decomposed existing generic-orbit \ac{pn} expressions for these fluxes
in multipolar contributions, and tested several factorizations and resummations
to improve their behavior in the strong-field regime.
Of particular conceptual importance was the realization that, unlike the case of circular orbits,
multipolar flux contributions may feature different superradiant behavior,
captured by different analytical factorizations.
Additionally, we developed a new reparameterization strategy that allows us to 
incorporate the noncircular corrections to the fluxes as multiplicative factors
that reduce to 1 exactly on circular dynamics,
and can be straightforwardly applied to any orbital configuration.

Comparing numerical and analytical results, we found that sign changes in the fluxes
can be quantitatively captured to within $\sim 10\%$ for \red{$\geq \srEccNsdtTotAllMinPct\%$ of the
bound, eccentric simulations} by our factorized expressions so long as
the $p_{\varphi}/r^2$ and $p_r$ terms in the prefactors are replaced by the exact orbital frequency $\Omega$
and the radial velocity $\dot{r}$, respectively,
and terms are added to reproduce non-spinning results.
\red{The same analytical treatment performs best on open, hyperbolic orbits,
although these proved intrinsically harder to accurately model, partly due
to significant \ac{qnm} excitation in close encounters dominating post-periastron
fluxes. Flux zeros were predicted within 10\% in between $\srHypNsdtAllMinPct\%$ and $\srHypNsdtAllMaxPct\%$ of cases,
depending on mode content and kind of flux.}
Circular-orbit comparisons, instead, confirmed earlier findings in the literature regarding
the performance of resummed analytical fluxes up to the \ac{lso} for
moderate-to-high spins.
Finally, comparisons of both instantaneous and orbit-averaged/peak fluxes
for eccentric and hyperbolic orbits revealed that our newly-introduced
noncircular correction factor is essential to describing the correct behavior
of the fluxes for large eccentricities or scatterings, but that further improvements are needed to achieve
quantitative agreement for very eccentric orbits $(e \geq 0.5)$ or close encounters.
The accuracy of the best-performing model (model~2) can be summarized as follows:
\begin{itemize}[nosep]
    \item \textit{Eccentric orbits:} median relative differences of \red{${\sim}\medianEccentricdE\%$ ($\dot{M}$) and ${\sim}\medianEccentricdJ\%$ ($\dot{S}$),
    with  $\fractionEccentricdETen\%$--$\fractionEccentricdJTen\%$} of configurations within $10\%$;
    \item \textit{Hyperbolic orbits:} median relative differences of \red{${\sim}\medianHyperbolicdE\%$ ($\dot{M}$) and ${\sim}\medianHyperbolicdJ\%$ ($\dot{S}$),
    with  $\fractionHyperbolicdETen\%$--$\fractionHyperbolicdJTen\%$} of configurations within $10\%$.
\end{itemize}
Among the orbital parameters, the peak orbital frequency $\Omega_{\rm max}$ emerges as 
the strongest predictor of the model error, with Spearman rank correlations 
\red{$\rho \approx \spearmanOmgmaxdE$ and $\spearmanOmgmaxdJ$} for $|\Delta\dot{M}/\dot{M}|$ and $|\Delta\dot{S}/\dot{S}|$ 
across all orbits, rising to \red{$\rho \approx \spearmanOmgMaxHyperbolicdE$ ($\spearmanOmgMaxHyperbolicdJ$)} for hyperbolic runs and 
\red{$\rho \approx \spearmanOmgMaxEccentricdE$ ($\spearmanOmgMaxEccentricdJ$)} for eccentric ones.

These results pave the way for improved analytical modeling of horizon fluxes
in both the test-mass and comparable-mass regimes. In particular, the analytical
expressions presented here can be straightforwardly implemented within
the \ac{eob} formalism for binaries on generic orbits in the large-mass-ratio
limit~\citep{Albertini:2022rfe,Albertini:2022dmc,vandeMeent:2023ols,Albertini:2023aol,Albertini:2024rrs,Albertini:2024agg,Leather:2025nhu, Albanesi:2026qtx, Faggioli:2026alx}
that is relevant for future space-based detectors such as the \ac{lisa}.
Future work will focus on extending these results to comparable-mass binaries,
attempting to combine the test-mass analytical expressions
and existing comparable-mass \ac{pn} results, as well as exploring
alternative resummation strategies~\citep{Cipriani:2026myb, Cipriani:2026xmx, Nishimura:2026nse},
calibration to numerical data, spin-induced precession~\citep{Nagni:2025cdw} and
comparison against orbit-averaged analytical 
results in the non-spinning test-mass limit~\citep{Forseth:2015oua, Munna:2020juq, Munna:2023vds}.
Moreover, we foresee that the new reparameterization strategy introduced here for the noncircular corrections
could lead to improved analytical models for the flux and waveform at infinity in the comparable-mass
limit as well.

Beyond the direct application to waveform modeling, some of the less-explored features
of the horizon fluxes unveiled in this work, such as quasi-normal mode excitations
during close encounters, also open up interesting avenues for future research.
Recently, it has been suggested that the dynamics of dynamical horizons during binary
mergers, encoded by the shear of the horizon, could be linked to features in the
waveform at infinity~\citep{Prasad:2020xgr, Prasad:2023bwa, Prasad:2024vsz}. Extending these studies to eccentric mergers,
and connecting them to the horizon fluxes studied here, could provide new insights into the nonlinear dynamics
of spacetime in these extreme scenarios. Additionally, recasting some of our findings
within the framework of dynamical horizons rather than event horizons could
provide a better understanding of the local physics at play during these interactions,
also in the case of plunging orbits where our computations are not directly applicable
due to the teleological nature of the event horizon~\citep{Ashtekar:2025wnu}.

\begin{figure*}
    \includegraphics[width=\textwidth]{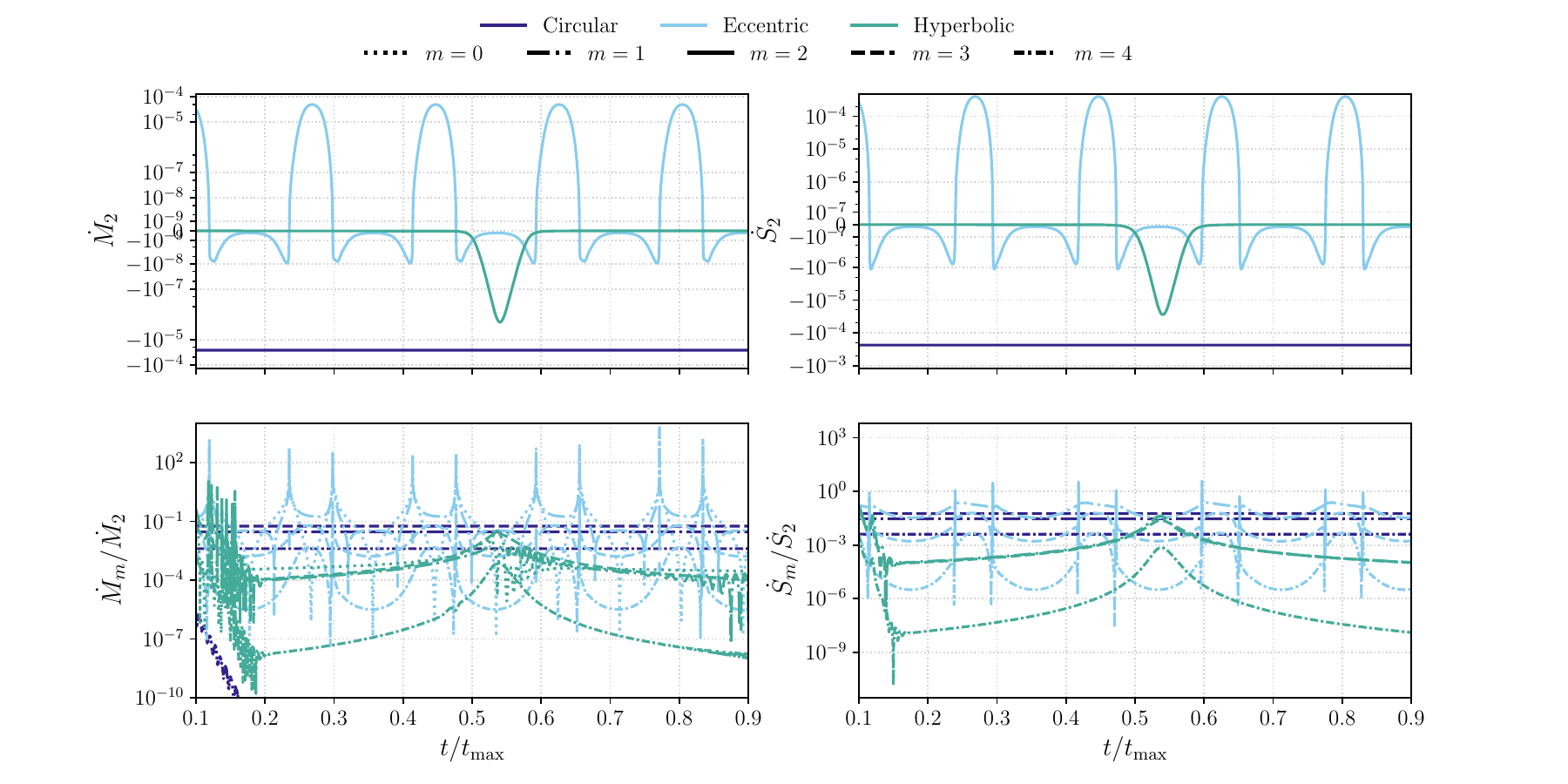}
    \caption{
        Dependence of the horizon fluxes on the number of $m$-modes included in the calculation.
        We consider three representative configurations: circular (pink), eccentric (green),
        and hyperbolic (blue) orbits. For each case, we show the energy (left panels) and angular momentum (right panels)
        fluxes as function of normalized time for $m=2$ (top panels), and
        their ratios with respect to the $m=2$ contribution for $m=1,3,4$ (bottom panels).  
        For eccentric (hyperbolic) orbits, higher modes can contribute up to $\sim 10\%$  ($\sim 1\%$)  of the $m=2$ flux
        at periastron passage/closest approach.
    }
    \label{fig:m_hierarchy}
\end{figure*}

\begin{figure}
    \includegraphics[width=0.49\textwidth]{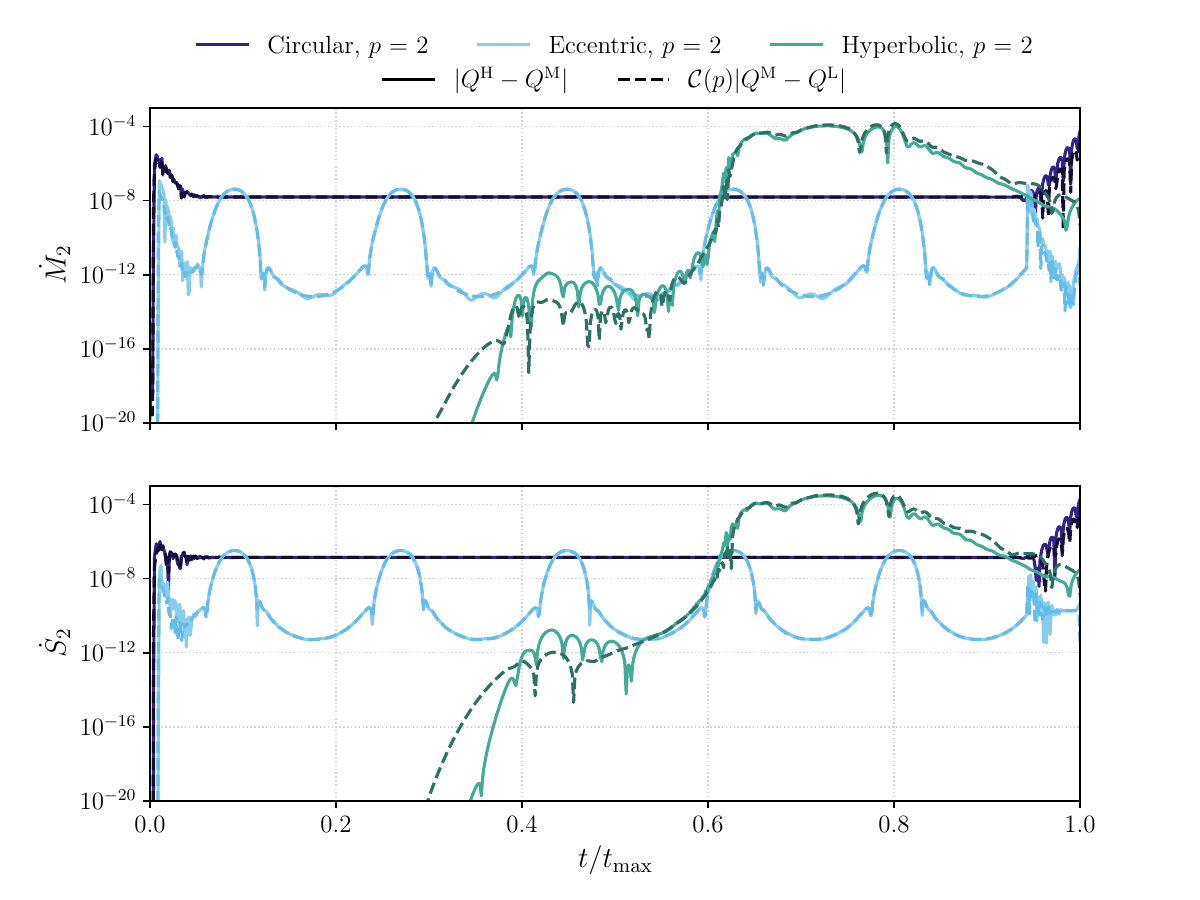}
    \caption{
        Convergence test for three representative configurations (circular, eccentric, hyperbolic orbits)
        evolved at three different grid resolutions: \( n_r \times n_{\theta} = 1801 \times 161,\; 3601 \times 161,\; 7201 \times 161 \).
        Top and bottom panels show the horizon fluxes of energy and angular momentum, respectively, for each configuration; absolute
        differences between the \red{highest} and medium resolutions are represented via straight lines, while the dashed lines
        show the rescaled middle-to-low resolution differences assuming a convergence order \(p\) (reported in the legend).
    }
    \label{fig:convergence_tests}
\end{figure}

\acknowledgments
The authors would like to thank T.~Damour for insightful discussions and suggestions regarding
factorization and resummation strategies. Additionally, the authors would like to warmly
thank A.~Nagar and S.~Bernuzzi for sharing \texttt{Teukode} with them and offering support during
the development of the entire work. Finally, R.~G would like to thank V.~Prasad for fruitful discussions
on dynamical horizons and numerical relativity, and S.~Hound, M.~Carducci for inspiration throughout the project.
R.~G. acknowledges support from NSF Grant PHY-2020275
(Network for Neutrinos, Nuclear Astrophysics, and Symmetries (N3AS)). 
D.~C. acknowledges support from the Italian Ministry of University and Research (MUR) via the PRIN 2022ZHYFA2,
{\it GRavitational wavEform models for coalescing compAct binaries
with eccenTricity} (GREAT).
E.~S. acknowledges support from NASA Grant 80NSSC21K1720 and the Charles E. Kaufman Foundation of The Pittsburgh Foundation.
S.~A. acknowledges support from the Deutsche Forschungsgemeinschaft (DFG) project ``GROOVHY'' 
(BE 6301/5-1 Projektnummer: 523180871) and from the Alexander von Humboldt Foundation through a Humboldt Research Fellowship
for postdoctoral researchers. 
The authors recognize the Penn State Institute for Computational and Data Sciences (ICDS) for providing support
through the ICDS Roar Core Facility (RRID:SCR\_026424).

\paragraph*{Data availability}
Data and scripts underlying this work are available at
\url{https://github.com/teobresums/sbc-II} and on zenodo with DOI
\url{10.5281/zenodo.19800342}

\paragraph*{Software} Analyses in this work made use of \texttt{NumPy}~\citep{Harris:2020xlr}, \texttt{SciPy}~\citep{Virtanen:2019joe},
\texttt{SymPy}~\citep{Meurer:2017yhf} and \texttt{mpmath}~\citep{mpmath}.
Figures in this work were produced using \texttt{matplotlib}~\citep{Hunter:2007}
and \texttt{seaborn}~\citep{Waskom:2021}.

\appendix

\section{Code tests}
\label{sec:code_tests}

\subsection{Flux dependence on $m$}
\label{sec:m_dependence}

It is well known that for quasi-circular orbits the leading contribution to the horizon fluxes
comes from the $\ell = m=2$ mode, with higher multipoles providing progressively smaller corrections~\citep{Breuer:1973kt, Chrzanowski:1974nr}.
In the case of eccentric or hyperbolic orbits, however, such hierarchy is not guaranteed to hold a priori.
Therefore, to assess the dependence of our results on the number of $m$-modes included in the flux calculation,
we consider 3 representative configurations (circular, eccentric, hyperbolic) and compare
each fixed $m\neq 2$ contribution to the energy and momentum fluxes to the $m=2$ one.
Figure~\ref{fig:m_hierarchy} shows the results of this analysis. For eccentric orbits, we find that
the $m=2$ mode dominates overall, with $m=1$ providing up to $\sim 10\%$ of $\dot{M}_2$ and $\dot{S}_2$ at periastron passage,
and higher modes being progressively suppressed. For hyperbolic orbits, $m=1$ and $m=3$ modes are comparable,
but overall of the order of $10^{-2}$ of the $m=2$ contribution at closest approach. Notably, for the eccentric
configuration considered here, at times when $\dot{M}_2$ vanishes, the $m=1$ and $m=0$ modes do not,
and become the dominant contributions.

Given these results, we simulate all configurations including only the \( m = 0, 1, 2 \) modes
in the flux calculation. 
We estimate the error associated with this choice to be of the order of
$1\%$ for eccentric orbits and $0.1\%$ for hyperbolic orbits.

\subsection{Self-convergence}
\label{sec:convergence}

We now move to estimating the self-convergence order of our numerical results, and the associated numerical error due to
grid discretization.
To this end, we consider once more three representative configurations (one for each orbital type)
and evolve them at five different grid resolutions, namely
\( n_r \times n_{\theta} = 1801 \times 161,\; 3601 \times 161,\; 7201 \times 161 ,\; 3061 \times 81 ,\; 3061 \times 321 \),
where \( n_r \) and \( n_{\theta} \) are the number of grid points in the radial and polar directions, respectively.

Focusing on the three resolutions with the same angular (radial) grid spacing and varying radial (angular) resolution,
we compute the horizon fluxes of energy and angular momentum for each resolution
and estimate the convergence order \(p\) by solving:
\begin{align}
    \biggl(\dfrac{Q^{\rm high} - Q^{\rm med}}{Q^{\rm med} - Q^{\rm low}}\biggr) = \frac{(\delta^{\rm high})^p - (\delta^{\rm med})^p}{(\delta^{\rm med})^p - (\delta^{\rm low})^p} \, ,
\end{align}
where \(Q^{\rm low}\), \(Q^{\rm med}\), and \(Q^{\rm high}\) are the fluxes computed at low, medium, and high resolution,
respectively, and \(\delta^{\rm low}\), \(\delta^{\rm med}\), and \(\delta^{\rm high}\) are the corresponding grid spacings.
Assuming a convergence order \(p\), we can then estimate the leading-order numerical error on \(Q\), $\Delta_Q$, via Richardson extrapolation:
\begin{equation}
    \Delta_Q = Q^{\rm exact} - Q^{\rm med} \sim \frac{|Q^{\rm high} - Q^{\rm med}|}{|(\delta^{\rm high}/\delta^{\rm med})^p - 1|} \, .
\end{equation}

The results are reported in Fig.~\ref{fig:convergence_tests}. We find that
our simulations converge at order \( p \sim 2 \) both for radial and angular resolution studies, irrespective of the orbital
configuration considered. When varying the radial resolution, the associated $\Delta_{M}$
and $\Delta_{S}$ are respectively of the order of 
$2~\times 10^{-8}$ and $2~\times 10^{-7}$ for circular orbits,
$7~\times 10^{-9}$ and $6~\times 10^{-8}$ for eccentric orbits,
and $2 \times 10^{-5}$ and $5 \times 10^{-5}$ for hyperbolic orbits.
In terms of relative errors, this translates to \(\Delta_{M}/\dot{M} \) and
and \(\Delta_{S}/\dot{S} \) of the order of $\sim 10^{-4}-10^{-2}$.

Similar results are obtained when varying the angular resolution,
with relative errors of the order of $\sim 10^{-4}-10^{-3}$ for eccentric and hyperbolic orbits
as well. The angular grid resolution therefore is subdominant with respect to the radial one, by
roughly an order of magnitude in terms of relative errors.

Given these results, we run all remaining simulations at the medium resolution
\( n_r \times n_{\theta} = 3601 \times 161 \),
and estimate the relative numerical error on the horizon fluxes to be of the order of
$10^{-4}-10^{-2}$, where the upper limit is chosen to be conservative.

\subsection{Comparison with earlier results}
\label{sec:lit_comparisons}
As a further check of our numerical implementation, we compare our results for circular orbits
with earlier data available in the literature~\citep{Sundararajan:2007jg, Taracchini:2013wfa} (see Tab.~\ref{tab:circular_comparison}).
These were obtained using frequency-domain solvers. 
To ease comparison, we orbit-average our fluxes over a sufficiently long time window after
the initial transient junk radiation has died out. The overall agreement between
the different results is reported in Tab.~\ref{tab:circular_comparison}. We find relative differences of the order of 
$10^{-3} - 10^{-4}$ for both energy and angular momentum fluxes, confirming the reliability of our implementation
and settings.

\section{PN expressions for non-circular corrections}
\label{sec:pn_expressions}
In this appendix we collect the \ac{pn} expressions for the non-circular corrections to the fluxes that were omitted from the main text for brevity.
We also provide their explicit K-reparameterized versions, as used in model~2.
\begin{widetext}
\begin{align}
    \dot{M}_{22}^{\rm NC} &= \frac{r^2}{p_\varphi^4} + \frac{8 p_\varphi^2 + 10r - 2p_r^2 r^2 - 6\left(p_\varphi\, r\right)^{2/3}}{2\, p_\varphi^4\, c^2} \nonumber \\
                            & -\frac{1}{6\,p_\varphi^5\,c^{3}} \left[
                                128\,p_\varphi p_r r
                                + 24\hat{a}\!\left(p_\varphi^2+2r\right)
                                + \frac{3p_r^2 r^2
                                    \!\left(3\hat{a}^3(-2+18\sigma)
                                            \red{+\hat{a}(70+90\sigma)}\right)}
                                        {1+3\hat{a}^2}
                                + 108\,p_r^2 r^2\,B_2(\hat{a})
                                \right] \, ,\\
    \dot{M}_{21}^{\rm NC} &= \frac{r^{4/3}}{p_\varphi^{8/3}} + \frac{4\hat{a}\!\left(p_\varphi^{4/3}-r^{2/3}\right)r^{2/3}}{3\,p_\varphi^{11/3}\,c} \, ,\\
    \dot{S}_{22}^{\rm NC} &= \frac{r^2}{p_\varphi^4} + \frac{8p_\varphi^2 + 10r - 2p_r^2 r^2 - 6\left(p_\varphi\, r\right)^{2/3}}{2\,p_\varphi^4\,c^2} -\frac{\left(16\,p_\varphi p_r r + 4\hat{a}\!\left(p_\varphi^2+2r\right)\right)}{p_\varphi^5\,c^{3}} \, ,\\
    \dot{S}_{21}^{\rm NC} &= \frac{r^{4/3}}{p_\varphi^{8/3}} + \frac{4\hat{a}\!\left(p_\varphi^{4/3}-r^{2/3}\right)r^{2/3}}{3\,p_\varphi^{11/3}\,c} \, .
\end{align}

\begin{align}
    \dot{M}_{22}^{\rm NC,K} &= \frac{1}{K^2} - \frac{1 + 3K^{1/3} - 4K + p_r^2 r}{K^2\,r\,c^2} + \frac{1}{3K^{5/2}r^{3/2}(1+3\hat{a}^2)\,c^{3}}
                                \Big[{-64\,p_r\sqrt{Kr}\!\left(1+3\hat{a}^2\right)}- 12\!\left(2-3\sqrt{K}+K\right)\hat{a}\!\left(1+3\hat{a}^2\right) \notag\\
                            &- 3p_r^2 r\,\hat{a}\!\left(35 - 3\hat{a}^2 + 9\sigma\!\left(5+3\hat{a}^2\right)\right) - 54\,p_r^2 r\!\left(1+3\hat{a}^2\right)B_2(\hat{a})\Big]\, ,\\
    \dot{M}_{21}^{\rm NC,K} &= \frac{1}{K^{4/3}} + \frac{4\!\left(K^{2/3}-1\right)\hat{a}}{3\,K^{11/6}\sqrt{r} c} \, ,\\
    \dot{S}_{22}^{\rm NC,K} &= \frac{1}{K^2}- \frac{1 + 3K^{1/3} - 4K + p_r^2 r}{K^2\,r\,c^2}- \frac{4\!\left(4p_r\sqrt{Kr} +\left(2-3\sqrt{K}+K\right)\hat{a}\right)}{K^{5/2}r^{3/2}\,c^{3}} \, ,\\
    \dot{S}_{21}^{\rm NC,K} &= \frac{1}{K^{4/3}} + \frac{4\!\left(K^{2/3}-1\right)\hat{a}}{3\,K^{11/6}\sqrt{r} c} \, .
\end{align}

\end{widetext}

\begin{table*}[t]
	\caption{
        Comparison of energy and angular momentum fluxes from~\cite{Sundararajan:2007jg, Taracchini:2013wfa} with data computed with Teukode.
        The latter fluxes are found by dividing $\dot{E}_m$ by the orbital frequency $\Omega$, and
        compared with our directly computed ones. Our data is the sum of all modes with the given value of $m$,
        while the reference values only include up to $\ell = 8$. We average our fluxes over the orbit, starting from a time when
        the initial burst of junk radiation has sufficiently died out up to the retarded time corresponding
        to the end of the dynamics.
        \label{tab:circular_comparison}}
	\begin{center}
    \begin{tabular}{c c c  c c c  c c c}
        \toprule
        \bmath{$a$}  & \bmath{$r_0$} & \bmath{$m$} & \bmath{$\dot{E}_m^{\rm ref}$} & \bmath{$\langle \dot{E}_m \rangle$} & \bmath{$\Delta \dot{E}_{m}/\dot{E}_m^{\rm ref}$} & \bmath{$\dot{J}_m^{\rm ref}$} & \bmath{$\langle \dot{J}_m \rangle$} & \bmath{$\Delta \dot{J}_{m}/\dot{J}_m^{\rm ref}$} \\
        \midrule
        0.0 & 4.0 & 1 & $\minus7.77607\times10^{-5}$ & $\minus7.78025\times10^{-5}$ & $5.37\times10^{-4}$ & $\minus6.22086\times10^{-4}$ & $\minus6.22421\times10^{-4}$ & $5.39\times10^{-4}$\\
        0.0 & 4.0 & 2 & $\minus5.65064\times10^{-4}$ & $\minus5.64595\times10^{-4}$ & $8.29\times10^{-4}$ & $\minus4.52051\times10^{-3}$ & $\minus4.51678\times10^{-3}$ & $8.25\times10^{-4}$\\
        0.0 & 6.0 & 2 & $\minus2.62826\times10^{-6}$ & $\minus2.62249\times10^{-6}$ & $2.19\times10^{-3}$ & $\minus3.86274\times10^{-5}$ & $\minus3.85427\times10^{-5}$ & $2.19\times10^{-3}$\\[0.5em]
        0.9 & 4.0 & 1 & $-1.52489\times10^{-6}$ & $-1.52639\times10^{-6}$ & $9.84\times10^{-4}$ & $-1.35715\times10^{-5}$ & $-1.35846\times10^{-5}$ & $9.71\times10^{-4}$\\
        0.9 & 4.0 & 2 & $-5.28398\times10^{-5}$ & $-5.28063\times10^{-5}$ & $6.34\times10^{-4}$ & $-4.70274\times10^{-4}$ & $-4.69970\times10^{-4}$ & $6.47\times10^{-4}$\\
        0.9 & 4.0 & 3 & $-3.00690\times10^{-6}$ & $-3.00981\times10^{-6}$ & $9.67\times10^{-4}$ & $-2.67614\times10^{-5}$ & $-2.67864\times10^{-5}$ & $9.34\times10^{-4}$\\
        0.9 & 10.0 & 2 & $-1.19691\times10^{-7}$ & $-1.19591\times10^{-7}$ & $8.41\times10^{-4}$ & $-3.89269\times10^{-6}$ & $-3.88859\times10^{-6}$ & $1.05\times10^{-3}$\\[0.5em]
        0.99 & 4.0 & 1 & $-7.81063\times10^{-7}$ & $-7.81874\times10^{-7}$ & $1.04\times10^{-3}$ & $-7.02176\times10^{-6}$ & $-7.02891\times10^{-6}$ & $1.02\times10^{-3}$\\
        0.99 & 4.0 & 2 & $-6.66107\times10^{-5}$ & $-6.65596\times10^{-5}$ & $7.67\times10^{-4}$ & $-5.98830\times10^{-4}$ & $-5.98346\times10^{-4}$ & $8.09\times10^{-4}$\\
        0.99 & 10.0 & 2 & $-1.50988\times10^{-7}$ & $-1.50890\times10^{-7}$ & $6.52\times10^{-4}$ & $-4.92414\times10^{-6}$ & $-4.91894\times10^{-6}$ & $1.06\times10^{-3}$\\
        \bottomrule
    \end{tabular}
    \end{center}
\end{table*}


\begin{thebibliography}{111}%
\makeatletter
\providecommand \@ifxundefined [1]{%
 \@ifx{#1\undefined}
}%
\providecommand \@ifnum [1]{%
 \ifnum #1\expandafter \@firstoftwo
 \else \expandafter \@secondoftwo
 \fi
}%
\providecommand \@ifx [1]{%
 \ifx #1\expandafter \@firstoftwo
 \else \expandafter \@secondoftwo
 \fi
}%
\providecommand \natexlab [1]{#1}%
\providecommand \enquote  [1]{``#1''}%
\providecommand \bibnamefont  [1]{#1}%
\providecommand \bibfnamefont [1]{#1}%
\providecommand \citenamefont [1]{#1}%
\providecommand \href@noop [0]{\@secondoftwo}%
\providecommand \href [0]{\begingroup \@sanitize@url \@href}%
\providecommand \@href[1]{\@@startlink{#1}\@@href}%
\providecommand \@@href[1]{\endgroup#1\@@endlink}%
\providecommand \@sanitize@url [0]{\catcode `\\12\catcode `\$12\catcode
  `\&12\catcode `\#12\catcode `\^12\catcode `\_12\catcode `\%12\relax}%
\providecommand \@@startlink[1]{}%
\providecommand \@@endlink[0]{}%
\providecommand \url  [0]{\begingroup\@sanitize@url \@url }%
\providecommand \@url [1]{\endgroup\@href {#1}{\urlprefix }}%
\providecommand \urlprefix  [0]{URL }%
\providecommand \Eprint [0]{\href }%
\providecommand \doibase [0]{http://dx.doi.org/}%
\providecommand \selectlanguage [0]{\@gobble}%
\providecommand \bibinfo  [0]{\@secondoftwo}%
\providecommand \bibfield  [0]{\@secondoftwo}%
\providecommand \translation [1]{[#1]}%
\providecommand \BibitemOpen [0]{}%
\providecommand \bibitemStop [0]{}%
\providecommand \bibitemNoStop [0]{.\EOS\space}%
\providecommand \EOS [0]{\spacefactor3000\relax}%
\providecommand \BibitemShut  [1]{\csname bibitem#1\endcsname}%
\let\auto@bib@innerbib\@empty
\bibitem [{\citenamefont {Hayward}(1994)}]{Hayward:1994yy}%
  \BibitemOpen
  \bibfield  {author} {\bibinfo {author} {\bibfnamefont {Sean~A.}\ \bibnamefont
  {Hayward}},\ }\bibfield  {title} {\enquote {\bibinfo {title} {{Spin
  coefficient form of the new laws of black hole dynamics}},}\ }\href {\doibase
  10.1088/0264-9381/11/12/016} {\bibfield  {journal} {\bibinfo  {journal}
  {Class. Quant. Grav.}\ }\textbf {\bibinfo {volume} {11}},\ \bibinfo {pages}
  {3025--3036} (\bibinfo {year} {1994})},\ \Eprint
  {http://arxiv.org/abs/gr-qc/9406033} {arXiv:gr-qc/9406033} \BibitemShut
  {NoStop}%
\bibitem [{\citenamefont {Thornburg}(1996)}]{Thornburg:1995cp}%
  \BibitemOpen
  \bibfield  {author} {\bibinfo {author} {\bibfnamefont {Jonathan}\
  \bibnamefont {Thornburg}},\ }\bibfield  {title} {\enquote {\bibinfo {title}
  {{Finding apparent horizons in numerical relativity}},}\ }\href {\doibase
  10.1103/PhysRevD.54.4899} {\bibfield  {journal} {\bibinfo  {journal} {Phys.
  Rev. D}\ }\textbf {\bibinfo {volume} {54}},\ \bibinfo {pages} {4899--4918}
  (\bibinfo {year} {1996})},\ \Eprint {http://arxiv.org/abs/gr-qc/9508014}
  {arXiv:gr-qc/9508014} \BibitemShut {NoStop}%
\bibitem [{\citenamefont {Ashtekar}\ and\ \citenamefont
  {Krishnan}(2003)}]{Ashtekar:2003hk}%
  \BibitemOpen
  \bibfield  {author} {\bibinfo {author} {\bibfnamefont {Abhay}\ \bibnamefont
  {Ashtekar}}\ and\ \bibinfo {author} {\bibfnamefont {Badri}\ \bibnamefont
  {Krishnan}},\ }\bibfield  {title} {\enquote {\bibinfo {title} {{Dynamical
  horizons and their properties}},}\ }\href {\doibase
  10.1103/PhysRevD.68.104030} {\bibfield  {journal} {\bibinfo  {journal} {Phys.
  Rev. D}\ }\textbf {\bibinfo {volume} {68}},\ \bibinfo {pages} {104030}
  (\bibinfo {year} {2003})},\ \Eprint {http://arxiv.org/abs/gr-qc/0308033}
  {arXiv:gr-qc/0308033} \BibitemShut {NoStop}%
\bibitem [{\citenamefont {Ashtekar}\ and\ \citenamefont
  {Krishnan}(2025)}]{Ashtekar:2025wnu}%
  \BibitemOpen
  \bibfield  {author} {\bibinfo {author} {\bibfnamefont {Abhay}\ \bibnamefont
  {Ashtekar}}\ and\ \bibinfo {author} {\bibfnamefont {Badri}\ \bibnamefont
  {Krishnan}},\ }\bibfield  {title} {\enquote {\bibinfo {title} {{Quasi-local
  black hole horizons: recentadvances}},}\ }\href {\doibase
  10.1007/s41114-025-00061-4} {\bibfield  {journal} {\bibinfo  {journal}
  {Living Rev. Rel.}\ }\textbf {\bibinfo {volume} {28}},\ \bibinfo {pages} {8}
  (\bibinfo {year} {2025})},\ \Eprint {http://arxiv.org/abs/2502.11825}
  {arXiv:2502.11825 [gr-qc]} \BibitemShut {NoStop}%
\bibitem [{\citenamefont {Christodoulou}(1970)}]{Christodoulou:1970wf}%
  \BibitemOpen
  \bibfield  {author} {\bibinfo {author} {\bibfnamefont {D.}~\bibnamefont
  {Christodoulou}},\ }\bibfield  {title} {\enquote {\bibinfo {title}
  {{Reversible and irreversible transformations in black hole physics}},}\
  }\href {\doibase 10.1103/PhysRevLett.25.1596} {\bibfield  {journal} {\bibinfo
   {journal} {Phys. Rev. Lett.}\ }\textbf {\bibinfo {volume} {25}},\ \bibinfo
  {pages} {1596--1597} (\bibinfo {year} {1970})}\BibitemShut {NoStop}%
\bibitem [{\citenamefont {Hawking}(1971)}]{Hawking:1971tu}%
  \BibitemOpen
  \bibfield  {author} {\bibinfo {author} {\bibfnamefont {S.~W.}\ \bibnamefont
  {Hawking}},\ }\bibfield  {title} {\enquote {\bibinfo {title} {{Gravitational
  radiation from colliding black holes}},}\ }\href {\doibase
  10.1103/PhysRevLett.26.1344} {\bibfield  {journal} {\bibinfo  {journal}
  {Phys. Rev. Lett.}\ }\textbf {\bibinfo {volume} {26}},\ \bibinfo {pages}
  {1344--1346} (\bibinfo {year} {1971})}\BibitemShut {NoStop}%
\bibitem [{\citenamefont {Bardeen}\ \emph {et~al.}(1973)\citenamefont
  {Bardeen}, \citenamefont {Carter},\ and\ \citenamefont
  {Hawking}}]{Bardeen:1973gs}%
  \BibitemOpen
  \bibfield  {author} {\bibinfo {author} {\bibfnamefont {James~M.}\
  \bibnamefont {Bardeen}}, \bibinfo {author} {\bibfnamefont {B.}~\bibnamefont
  {Carter}}, \ and\ \bibinfo {author} {\bibfnamefont {S.~W.}\ \bibnamefont
  {Hawking}},\ }\bibfield  {title} {\enquote {\bibinfo {title} {{The Four laws
  of black hole mechanics}},}\ }\href {\doibase 10.1007/BF01645742} {\bibfield
  {journal} {\bibinfo  {journal} {Commun. Math. Phys.}\ }\textbf {\bibinfo
  {volume} {31}},\ \bibinfo {pages} {161--170} (\bibinfo {year}
  {1973})}\BibitemShut {NoStop}%
\bibitem [{\citenamefont {Del~Pozzo}\ and\ \citenamefont
  {Nagar}(2017)}]{DelPozzo:2016kmd}%
  \BibitemOpen
  \bibfield  {author} {\bibinfo {author} {\bibfnamefont {Walter}\ \bibnamefont
  {Del~Pozzo}}\ and\ \bibinfo {author} {\bibfnamefont {Alessandro}\
  \bibnamefont {Nagar}},\ }\bibfield  {title} {\enquote {\bibinfo {title}
  {{Analytic family of post-merger template waveforms}},}\ }\href {\doibase
  10.1103/PhysRevD.95.124034} {\bibfield  {journal} {\bibinfo  {journal} {Phys.
  Rev. D}\ }\textbf {\bibinfo {volume} {95}},\ \bibinfo {pages} {124034}
  (\bibinfo {year} {2017})},\ \Eprint {http://arxiv.org/abs/1606.03952}
  {arXiv:1606.03952 [gr-qc]} \BibitemShut {NoStop}%
\bibitem [{\citenamefont {Isi}\ \emph {et~al.}(2021)\citenamefont {Isi},
  \citenamefont {Farr}, \citenamefont {Giesler}, \citenamefont {Scheel},\ and\
  \citenamefont {Teukolsky}}]{Isi:2020tac}%
  \BibitemOpen
  \bibfield  {author} {\bibinfo {author} {\bibfnamefont {Maximiliano}\
  \bibnamefont {Isi}}, \bibinfo {author} {\bibfnamefont {Will~M.}\ \bibnamefont
  {Farr}}, \bibinfo {author} {\bibfnamefont {Matthew}\ \bibnamefont {Giesler}},
  \bibinfo {author} {\bibfnamefont {Mark~A.}\ \bibnamefont {Scheel}}, \ and\
  \bibinfo {author} {\bibfnamefont {Saul~A.}\ \bibnamefont {Teukolsky}},\
  }\bibfield  {title} {\enquote {\bibinfo {title} {{Testing the Black-Hole Area
  Law with GW150914}},}\ }\href {\doibase 10.1103/PhysRevLett.127.011103}
  {\bibfield  {journal} {\bibinfo  {journal} {Phys. Rev. Lett.}\ }\textbf
  {\bibinfo {volume} {127}},\ \bibinfo {pages} {011103} (\bibinfo {year}
  {2021})},\ \Eprint {http://arxiv.org/abs/2012.04486} {arXiv:2012.04486
  [gr-qc]} \BibitemShut {NoStop}%
\bibitem [{\citenamefont {Abbott}\ \emph
  {et~al.}(2016{\natexlab{a}})\citenamefont {Abbott} \emph
  {et~al.}}]{KAGRA:2013rdx}%
  \BibitemOpen
  \bibfield  {author} {\bibinfo {author} {\bibfnamefont {B.~P.}\ \bibnamefont
  {Abbott}} \emph {et~al.} (\bibinfo {collaboration} {KAGRA, LIGO Scientific,
  Virgo}),\ }\bibfield  {title} {\enquote {\bibinfo {title} {{Prospects for
  observing and localizing gravitational-wave transients with Advanced LIGO,
  Advanced Virgo and KAGRA}},}\ }\href {\doibase 10.1007/s41114-020-00026-9}
  {\bibfield  {journal} {\bibinfo  {journal} {Living Rev. Rel.}\ }\textbf
  {\bibinfo {volume} {19}},\ \bibinfo {pages} {1} (\bibinfo {year}
  {2016}{\natexlab{a}})},\ \Eprint {http://arxiv.org/abs/1304.0670}
  {arXiv:1304.0670 [gr-qc]} \BibitemShut {NoStop}%
\bibitem [{\citenamefont {Aasi}\ \emph {et~al.}(2015)\citenamefont {Aasi} \emph
  {et~al.}}]{LIGOScientific:2014pky}%
  \BibitemOpen
  \bibfield  {author} {\bibinfo {author} {\bibfnamefont {J.}~\bibnamefont
  {Aasi}} \emph {et~al.} (\bibinfo {collaboration} {LIGO Scientific}),\
  }\bibfield  {title} {\enquote {\bibinfo {title} {{Advanced LIGO}},}\ }\href
  {\doibase 10.1088/0264-9381/32/7/074001} {\bibfield  {journal} {\bibinfo
  {journal} {Class. Quant. Grav.}\ }\textbf {\bibinfo {volume} {32}},\ \bibinfo
  {pages} {074001} (\bibinfo {year} {2015})},\ \Eprint
  {http://arxiv.org/abs/1411.4547} {arXiv:1411.4547 [gr-qc]} \BibitemShut
  {NoStop}%
\bibitem [{\citenamefont {Acernese}\ \emph {et~al.}(2015)\citenamefont
  {Acernese} \emph {et~al.}}]{VIRGO:2014yos}%
  \BibitemOpen
  \bibfield  {author} {\bibinfo {author} {\bibfnamefont {F.}~\bibnamefont
  {Acernese}} \emph {et~al.} (\bibinfo {collaboration} {VIRGO}),\ }\bibfield
  {title} {\enquote {\bibinfo {title} {{Advanced Virgo: a second-generation
  interferometric gravitational wave detector}},}\ }\href {\doibase
  10.1088/0264-9381/32/2/024001} {\bibfield  {journal} {\bibinfo  {journal}
  {Class. Quant. Grav.}\ }\textbf {\bibinfo {volume} {32}},\ \bibinfo {pages}
  {024001} (\bibinfo {year} {2015})},\ \Eprint {http://arxiv.org/abs/1408.3978}
  {arXiv:1408.3978 [gr-qc]} \BibitemShut {NoStop}%
\bibitem [{\citenamefont {Abbott}\ \emph
  {et~al.}(2016{\natexlab{b}})\citenamefont {Abbott} \emph
  {et~al.}}]{LIGOScientific:2016vlm}%
  \BibitemOpen
  \bibfield  {author} {\bibinfo {author} {\bibfnamefont {B.~P.}\ \bibnamefont
  {Abbott}} \emph {et~al.} (\bibinfo {collaboration} {LIGO Scientific,
  Virgo}),\ }\bibfield  {title} {\enquote {\bibinfo {title} {{Properties of the
  Binary Black Hole Merger GW150914}},}\ }\href {\doibase
  10.1103/PhysRevLett.116.241102} {\bibfield  {journal} {\bibinfo  {journal}
  {Phys. Rev. Lett.}\ }\textbf {\bibinfo {volume} {116}},\ \bibinfo {pages}
  {241102} (\bibinfo {year} {2016}{\natexlab{b}})},\ \Eprint
  {http://arxiv.org/abs/1602.03840} {arXiv:1602.03840 [gr-qc]} \BibitemShut
  {NoStop}%
\bibitem [{\citenamefont {Abbott}\ \emph
  {et~al.}(2016{\natexlab{c}})\citenamefont {Abbott} \emph
  {et~al.}}]{LIGOScientific:2016wkq}%
  \BibitemOpen
  \bibfield  {author} {\bibinfo {author} {\bibfnamefont {Thomas~D.}\
  \bibnamefont {Abbott}} \emph {et~al.} (\bibinfo {collaboration} {LIGO
  Scientific, Virgo}),\ }\bibfield  {title} {\enquote {\bibinfo {title}
  {{Improved analysis of GW150914 using a fully spin-precessing waveform
  Model}},}\ }\href {\doibase 10.1103/PhysRevX.6.041014} {\bibfield  {journal}
  {\bibinfo  {journal} {Phys. Rev. X}\ }\textbf {\bibinfo {volume} {6}},\
  \bibinfo {pages} {041014} (\bibinfo {year} {2016}{\natexlab{c}})},\ \Eprint
  {http://arxiv.org/abs/1606.01210} {arXiv:1606.01210 [gr-qc]} \BibitemShut
  {NoStop}%
\bibitem [{\citenamefont {Abac}\ \emph {et~al.}(2025)\citenamefont {Abac} \emph
  {et~al.}}]{LIGOScientific:2025rid}%
  \BibitemOpen
  \bibfield  {author} {\bibinfo {author} {\bibfnamefont {A.~G.}\ \bibnamefont
  {Abac}} \emph {et~al.} (\bibinfo {collaboration} {LIGO Scientific, Virgo,
  KAGRA}),\ }\bibfield  {title} {\enquote {\bibinfo {title} {{GW250114: Testing
  Hawking{\textquoteright}s Area Law and the Kerr Nature of Black Holes}},}\
  }\href {\doibase 10.1103/kw5g-d732} {\bibfield  {journal} {\bibinfo
  {journal} {Phys. Rev. Lett.}\ }\textbf {\bibinfo {volume} {135}},\ \bibinfo
  {pages} {111403} (\bibinfo {year} {2025})},\ \Eprint
  {http://arxiv.org/abs/2509.08054} {arXiv:2509.08054 [gr-qc]} \BibitemShut
  {NoStop}%
\bibitem [{LIG(2025)}]{LIGOScientific:2025obp}%
  \BibitemOpen
  \bibfield  {title} {\enquote {\bibinfo {title} {{Black Hole Spectroscopy and
  Tests of General Relativity with GW250114}},}\ }\href@noop {} {\  (\bibinfo
  {year} {2025})},\ \Eprint {http://arxiv.org/abs/2509.08099} {arXiv:2509.08099
  [gr-qc]} \BibitemShut {NoStop}%
\bibitem [{\citenamefont {Prasad}(2026)}]{Prasad:2026imj}%
  \BibitemOpen
  \bibfield  {author} {\bibinfo {author} {\bibfnamefont {Vaishak}\ \bibnamefont
  {Prasad}},\ }\bibfield  {title} {\enquote {\bibinfo {title} {{Multi-Segment
  Consistency Tests of General Relativity}},}\ }\href@noop {} {\  (\bibinfo
  {year} {2026})},\ \Eprint {http://arxiv.org/abs/2603.05835} {arXiv:2603.05835
  [gr-qc]} \BibitemShut {NoStop}%
\bibitem [{\citenamefont {Le~Tiec}\ \emph {et~al.}(2012)\citenamefont
  {Le~Tiec}, \citenamefont {Blanchet},\ and\ \citenamefont
  {Whiting}}]{LeTiec:2011ab}%
  \BibitemOpen
  \bibfield  {author} {\bibinfo {author} {\bibfnamefont {Alexandre}\
  \bibnamefont {Le~Tiec}}, \bibinfo {author} {\bibfnamefont {Luc}\ \bibnamefont
  {Blanchet}}, \ and\ \bibinfo {author} {\bibfnamefont {Bernard~F.}\
  \bibnamefont {Whiting}},\ }\bibfield  {title} {\enquote {\bibinfo {title}
  {{The First Law of Binary Black Hole Mechanics in General Relativity and
  Post-Newtonian Theory}},}\ }\href {\doibase 10.1103/PhysRevD.85.064039}
  {\bibfield  {journal} {\bibinfo  {journal} {Phys. Rev. D}\ }\textbf {\bibinfo
  {volume} {85}},\ \bibinfo {pages} {064039} (\bibinfo {year} {2012})},\
  \Eprint {http://arxiv.org/abs/1111.5378} {arXiv:1111.5378 [gr-qc]}
  \BibitemShut {NoStop}%
\bibitem [{\citenamefont {Blanchet}\ \emph {et~al.}(2013)\citenamefont
  {Blanchet}, \citenamefont {Buonanno},\ and\ \citenamefont
  {Le~Tiec}}]{Blanchet:2012at}%
  \BibitemOpen
  \bibfield  {author} {\bibinfo {author} {\bibfnamefont {Luc}\ \bibnamefont
  {Blanchet}}, \bibinfo {author} {\bibfnamefont {Alessandra}\ \bibnamefont
  {Buonanno}}, \ and\ \bibinfo {author} {\bibfnamefont {Alexandre}\
  \bibnamefont {Le~Tiec}},\ }\bibfield  {title} {\enquote {\bibinfo {title}
  {{First law of mechanics for black hole binaries with spins}},}\ }\href
  {\doibase 10.1103/PhysRevD.87.024030} {\bibfield  {journal} {\bibinfo
  {journal} {Phys. Rev. D}\ }\textbf {\bibinfo {volume} {87}},\ \bibinfo
  {pages} {024030} (\bibinfo {year} {2013})},\ \Eprint
  {http://arxiv.org/abs/1211.1060} {arXiv:1211.1060 [gr-qc]} \BibitemShut
  {NoStop}%
\bibitem [{\citenamefont {Friedman}\ \emph {et~al.}(2002)\citenamefont
  {Friedman}, \citenamefont {Uryu},\ and\ \citenamefont
  {Shibata}}]{Friedman:2001pf}%
  \BibitemOpen
  \bibfield  {author} {\bibinfo {author} {\bibfnamefont {John~L.}\ \bibnamefont
  {Friedman}}, \bibinfo {author} {\bibfnamefont {Koji}\ \bibnamefont {Uryu}}, \
  and\ \bibinfo {author} {\bibfnamefont {Masaru}\ \bibnamefont {Shibata}},\
  }\bibfield  {title} {\enquote {\bibinfo {title} {{Thermodynamics of binary
  black holes and neutron stars}},}\ }\href {\doibase
  10.1103/PhysRevD.70.129904} {\bibfield  {journal} {\bibinfo  {journal} {Phys.
  Rev. D}\ }\textbf {\bibinfo {volume} {65}},\ \bibinfo {pages} {064035}
  (\bibinfo {year} {2002})},\ \bibinfo {note} {[Erratum: Phys.Rev.D 70, 129904
  (2004)]},\ \Eprint {http://arxiv.org/abs/gr-qc/0108070} {arXiv:gr-qc/0108070}
  \BibitemShut {NoStop}%
\bibitem [{\citenamefont {Uryu}\ \emph {et~al.}(2010)\citenamefont {Uryu},
  \citenamefont {Gourgoulhon},\ and\ \citenamefont {Markakis}}]{Uryu:2010su}%
  \BibitemOpen
  \bibfield  {author} {\bibinfo {author} {\bibfnamefont {Koji}\ \bibnamefont
  {Uryu}}, \bibinfo {author} {\bibfnamefont {Eric}\ \bibnamefont
  {Gourgoulhon}}, \ and\ \bibinfo {author} {\bibfnamefont {Charalampos}\
  \bibnamefont {Markakis}},\ }\bibfield  {title} {\enquote {\bibinfo {title}
  {{Thermodynamics of magnetized binary compact objects}},}\ }\href {\doibase
  10.1103/PhysRevD.82.104054} {\bibfield  {journal} {\bibinfo  {journal} {Phys.
  Rev. D}\ }\textbf {\bibinfo {volume} {82}},\ \bibinfo {pages} {104054}
  (\bibinfo {year} {2010})},\ \Eprint {http://arxiv.org/abs/1010.4409}
  {arXiv:1010.4409 [gr-qc]} \BibitemShut {NoStop}%
\bibitem [{\citenamefont {Gonzo}\ \emph {et~al.}(2025)\citenamefont {Gonzo},
  \citenamefont {Lewis},\ and\ \citenamefont {Pound}}]{Gonzo:2024xjk}%
  \BibitemOpen
  \bibfield  {author} {\bibinfo {author} {\bibfnamefont {Riccardo}\
  \bibnamefont {Gonzo}}, \bibinfo {author} {\bibfnamefont {Jack}\ \bibnamefont
  {Lewis}}, \ and\ \bibinfo {author} {\bibfnamefont {Adam}\ \bibnamefont
  {Pound}},\ }\bibfield  {title} {\enquote {\bibinfo {title} {{First Law of
  Binary Black Hole Scattering}},}\ }\href {\doibase 10.1103/s85p-gh7b}
  {\bibfield  {journal} {\bibinfo  {journal} {Phys. Rev. Lett.}\ }\textbf
  {\bibinfo {volume} {135}},\ \bibinfo {pages} {131401} (\bibinfo {year}
  {2025})},\ \Eprint {http://arxiv.org/abs/2409.03437} {arXiv:2409.03437
  [gr-qc]} \BibitemShut {NoStop}%
\bibitem [{\citenamefont {Wang}\ \emph {et~al.}(2024)\citenamefont {Wang},
  \citenamefont {Zhu}, \citenamefont {Huang},\ and\ \citenamefont
  {Shu}}]{Wang:2023jah}%
  \BibitemOpen
  \bibfield  {author} {\bibinfo {author} {\bibfnamefont {Chao-Wan-Zhen}\
  \bibnamefont {Wang}}, \bibinfo {author} {\bibfnamefont {Jin-Bao}\
  \bibnamefont {Zhu}}, \bibinfo {author} {\bibfnamefont {Guo-Qing}\
  \bibnamefont {Huang}}, \ and\ \bibinfo {author} {\bibfnamefont {Fu-Wen}\
  \bibnamefont {Shu}},\ }\bibfield  {title} {\enquote {\bibinfo {title}
  {{Testing the first law of black hole mechanics with gravitational waves}},}\
  }\href {\doibase 10.1007/s11433-024-2442-3} {\bibfield  {journal} {\bibinfo
  {journal} {Sci. China Phys. Mech. Astron.}\ }\textbf {\bibinfo {volume}
  {67}},\ \bibinfo {pages} {100413} (\bibinfo {year} {2024})},\ \Eprint
  {http://arxiv.org/abs/2304.10117} {arXiv:2304.10117 [gr-qc]} \BibitemShut
  {NoStop}%
\bibitem [{\citenamefont {Hartle}(1973)}]{Hartle:1973zz}%
  \BibitemOpen
  \bibfield  {author} {\bibinfo {author} {\bibfnamefont {James~B.}\
  \bibnamefont {Hartle}},\ }\bibfield  {title} {\enquote {\bibinfo {title}
  {{Tidal Friction in Slowly Rotating Black Holes}},}\ }\href {\doibase
  10.1103/PhysRevD.8.1010} {\bibfield  {journal} {\bibinfo  {journal} {Phys.
  Rev. D}\ }\textbf {\bibinfo {volume} {8}},\ \bibinfo {pages} {1010--1024}
  (\bibinfo {year} {1973})}\BibitemShut {NoStop}%
\bibitem [{\citenamefont {Alvi}(2001)}]{Alvi:2001mx}%
  \BibitemOpen
  \bibfield  {author} {\bibinfo {author} {\bibfnamefont {Kashif}\ \bibnamefont
  {Alvi}},\ }\bibfield  {title} {\enquote {\bibinfo {title} {{Energy and
  angular momentum flow into a black hole in a binary}},}\ }\href {\doibase
  10.1103/PhysRevD.64.104020} {\bibfield  {journal} {\bibinfo  {journal} {Phys.
  Rev. D}\ }\textbf {\bibinfo {volume} {64}},\ \bibinfo {pages} {104020}
  (\bibinfo {year} {2001})},\ \Eprint {http://arxiv.org/abs/gr-qc/0107080}
  {arXiv:gr-qc/0107080} \BibitemShut {NoStop}%
\bibitem [{\citenamefont {Poisson}\ and\ \citenamefont
  {Sasaki}(1995)}]{Poisson:1994yf}%
  \BibitemOpen
  \bibfield  {author} {\bibinfo {author} {\bibfnamefont {Eric}\ \bibnamefont
  {Poisson}}\ and\ \bibinfo {author} {\bibfnamefont {Misao}\ \bibnamefont
  {Sasaki}},\ }\bibfield  {title} {\enquote {\bibinfo {title} {{Gravitational
  radiation from a particle in circular orbit around a black hole. 5: Black
  hole absorption and tail corrections}},}\ }\href {\doibase
  10.1103/PhysRevD.51.5753} {\bibfield  {journal} {\bibinfo  {journal} {Phys.
  Rev. D}\ }\textbf {\bibinfo {volume} {51}},\ \bibinfo {pages} {5753--5767}
  (\bibinfo {year} {1995})},\ \Eprint {http://arxiv.org/abs/gr-qc/9412027}
  {arXiv:gr-qc/9412027} \BibitemShut {NoStop}%
\bibitem [{\citenamefont {Tagoshi}\ \emph {et~al.}(1997)\citenamefont
  {Tagoshi}, \citenamefont {Mano},\ and\ \citenamefont
  {Takasugi}}]{Tagoshi:1997jy}%
  \BibitemOpen
  \bibfield  {author} {\bibinfo {author} {\bibfnamefont {Hideyuki}\
  \bibnamefont {Tagoshi}}, \bibinfo {author} {\bibfnamefont {Shuhei}\
  \bibnamefont {Mano}}, \ and\ \bibinfo {author} {\bibfnamefont {Eiichi}\
  \bibnamefont {Takasugi}},\ }\bibfield  {title} {\enquote {\bibinfo {title}
  {{PostNewtonian expansion of gravitational waves from a particle in circular
  orbits around a rotating black hole: Effects of black hole absorption}},}\
  }\href {\doibase 10.1143/PTP.98.829} {\bibfield  {journal} {\bibinfo
  {journal} {Prog. Theor. Phys.}\ }\textbf {\bibinfo {volume} {98}},\ \bibinfo
  {pages} {829--850} (\bibinfo {year} {1997})},\ \Eprint
  {http://arxiv.org/abs/gr-qc/9711072} {arXiv:gr-qc/9711072} \BibitemShut
  {NoStop}%
\bibitem [{\citenamefont {Mino}\ \emph {et~al.}(1997)\citenamefont {Mino},
  \citenamefont {Sasaki}, \citenamefont {Shibata}, \citenamefont {Tagoshi},\
  and\ \citenamefont {Tanaka}}]{Mino:1997bx}%
  \BibitemOpen
  \bibfield  {author} {\bibinfo {author} {\bibfnamefont {Yasushi}\ \bibnamefont
  {Mino}}, \bibinfo {author} {\bibfnamefont {Misao}\ \bibnamefont {Sasaki}},
  \bibinfo {author} {\bibfnamefont {Masaru}\ \bibnamefont {Shibata}}, \bibinfo
  {author} {\bibfnamefont {Hideyuki}\ \bibnamefont {Tagoshi}}, \ and\ \bibinfo
  {author} {\bibfnamefont {Takahiro}\ \bibnamefont {Tanaka}},\ }\bibfield
  {title} {\enquote {\bibinfo {title} {{Black hole perturbation: Chapter 1}},}\
  }\href {\doibase 10.1143/PTPS.128.1} {\bibfield  {journal} {\bibinfo
  {journal} {Prog. Theor. Phys. Suppl.}\ }\textbf {\bibinfo {volume} {128}},\
  \bibinfo {pages} {1--121} (\bibinfo {year} {1997})},\ \Eprint
  {http://arxiv.org/abs/gr-qc/9712057} {arXiv:gr-qc/9712057} \BibitemShut
  {NoStop}%
\bibitem [{\citenamefont {Teukolsky}(1972)}]{Teukolsky:1972my}%
  \BibitemOpen
  \bibfield  {author} {\bibinfo {author} {\bibfnamefont {S.~A.}\ \bibnamefont
  {Teukolsky}},\ }\bibfield  {title} {\enquote {\bibinfo {title} {{Rotating
  black holes - separable wave equations for gravitational and electromagnetic
  perturbations}},}\ }\href {\doibase 10.1103/PhysRevLett.29.1114} {\bibfield
  {journal} {\bibinfo  {journal} {Phys. Rev. Lett.}\ }\textbf {\bibinfo
  {volume} {29}},\ \bibinfo {pages} {1114--1118} (\bibinfo {year}
  {1972})}\BibitemShut {NoStop}%
\bibitem [{\citenamefont {Teukolsky}(1973)}]{Teukolsky:1973ha}%
  \BibitemOpen
  \bibfield  {author} {\bibinfo {author} {\bibfnamefont {Saul~A.}\ \bibnamefont
  {Teukolsky}},\ }\bibfield  {title} {\enquote {\bibinfo {title}
  {{Perturbations of a rotating black hole. 1. Fundamental equations for
  gravitational electromagnetic and neutrino field perturbations}},}\ }\href
  {\doibase 10.1086/152444} {\bibfield  {journal} {\bibinfo  {journal}
  {Astrophys. J.}\ }\textbf {\bibinfo {volume} {185}},\ \bibinfo {pages}
  {635--647} (\bibinfo {year} {1973})}\BibitemShut {NoStop}%
\bibitem [{\citenamefont {Bernuzzi}\ \emph {et~al.}(2012)\citenamefont
  {Bernuzzi}, \citenamefont {Nagar},\ and\ \citenamefont
  {Zenginoglu}}]{Bernuzzi:2012ku}%
  \BibitemOpen
  \bibfield  {author} {\bibinfo {author} {\bibfnamefont {Sebastiano}\
  \bibnamefont {Bernuzzi}}, \bibinfo {author} {\bibfnamefont {Alessandro}\
  \bibnamefont {Nagar}}, \ and\ \bibinfo {author} {\bibfnamefont {Anil}\
  \bibnamefont {Zenginoglu}},\ }\bibfield  {title} {\enquote {\bibinfo {title}
  {{Horizon-absorption effects in coalescing black-hole binaries: An
  effective-one-body study of the non-spinning case}},}\ }\href {\doibase
  10.1103/PhysRevD.86.104038} {\bibfield  {journal} {\bibinfo  {journal} {Phys.
  Rev. D}\ }\textbf {\bibinfo {volume} {86}},\ \bibinfo {pages} {104038}
  (\bibinfo {year} {2012})},\ \Eprint {http://arxiv.org/abs/1207.0769}
  {arXiv:1207.0769 [gr-qc]} \BibitemShut {NoStop}%
\bibitem [{\citenamefont {Taracchini}\ \emph {et~al.}(2013)\citenamefont
  {Taracchini}, \citenamefont {Buonanno}, \citenamefont {Hughes},\ and\
  \citenamefont {Khanna}}]{Taracchini:2013wfa}%
  \BibitemOpen
  \bibfield  {author} {\bibinfo {author} {\bibfnamefont {Andrea}\ \bibnamefont
  {Taracchini}}, \bibinfo {author} {\bibfnamefont {Alessandra}\ \bibnamefont
  {Buonanno}}, \bibinfo {author} {\bibfnamefont {Scott~A.}\ \bibnamefont
  {Hughes}}, \ and\ \bibinfo {author} {\bibfnamefont {Gaurav}\ \bibnamefont
  {Khanna}},\ }\bibfield  {title} {\enquote {\bibinfo {title} {{Modeling the
  horizon-absorbed gravitational flux for equatorial-circular orbits in Kerr
  spacetime}},}\ }\href {\doibase 10.1103/PhysRevD.88.044001} {\bibfield
  {journal} {\bibinfo  {journal} {Phys. Rev. D}\ }\textbf {\bibinfo {volume}
  {88}},\ \bibinfo {pages} {044001} (\bibinfo {year} {2013})},\ \bibinfo {note}
  {[Erratum: Phys.Rev.D 88, 109903 (2013)]},\ \Eprint
  {http://arxiv.org/abs/1305.2184} {arXiv:1305.2184 [gr-qc]} \BibitemShut
  {NoStop}%
\bibitem [{\citenamefont {Fujita}(2015)}]{Fujita:2014eta}%
  \BibitemOpen
  \bibfield  {author} {\bibinfo {author} {\bibfnamefont {Ryuichi}\ \bibnamefont
  {Fujita}},\ }\bibfield  {title} {\enquote {\bibinfo {title} {{Gravitational
  Waves from a Particle in Circular Orbits around a Rotating Black Hole to the
  11th Post-Newtonian Order}},}\ }\href {\doibase 10.1093/ptep/ptv012}
  {\bibfield  {journal} {\bibinfo  {journal} {PTEP}\ }\textbf {\bibinfo
  {volume} {2015}},\ \bibinfo {pages} {033E01} (\bibinfo {year} {2015})},\
  \Eprint {http://arxiv.org/abs/1412.5689} {arXiv:1412.5689 [gr-qc]}
  \BibitemShut {NoStop}%
\bibitem [{\citenamefont {Shah}(2014)}]{Shah:2014tka}%
  \BibitemOpen
  \bibfield  {author} {\bibinfo {author} {\bibfnamefont {Abhay~G.}\
  \bibnamefont {Shah}},\ }\bibfield  {title} {\enquote {\bibinfo {title}
  {{Gravitational-wave flux for a particle orbiting a Kerr black hole to 20th
  post-Newtonian order: a numerical approach}},}\ }\href {\doibase
  10.1103/PhysRevD.90.044025} {\bibfield  {journal} {\bibinfo  {journal} {Phys.
  Rev. D}\ }\textbf {\bibinfo {volume} {90}},\ \bibinfo {pages} {044025}
  (\bibinfo {year} {2014})},\ \Eprint {http://arxiv.org/abs/1403.2697}
  {arXiv:1403.2697 [gr-qc]} \BibitemShut {NoStop}%
\bibitem [{\citenamefont {O'Sullivan}\ and\ \citenamefont
  {Hughes}(2016)}]{OSullivan:2015lni}%
  \BibitemOpen
  \bibfield  {author} {\bibinfo {author} {\bibfnamefont {Stephen}\ \bibnamefont
  {O'Sullivan}}\ and\ \bibinfo {author} {\bibfnamefont {Scott~A.}\ \bibnamefont
  {Hughes}},\ }\bibfield  {title} {\enquote {\bibinfo {title} {{Strong-field
  tidal distortions of rotating black holes: II. Horizon dynamics from
  eccentric and inclined orbits}},}\ }\href {\doibase
  10.1103/PhysRevD.94.044057} {\bibfield  {journal} {\bibinfo  {journal} {Phys.
  Rev. D}\ }\textbf {\bibinfo {volume} {94}},\ \bibinfo {pages} {044057}
  (\bibinfo {year} {2016})},\ \Eprint {http://arxiv.org/abs/1505.03809}
  {arXiv:1505.03809 [gr-qc]} \BibitemShut {NoStop}%
\bibitem [{\citenamefont {Poisson}(2004)}]{Poisson:2004cw}%
  \BibitemOpen
  \bibfield  {author} {\bibinfo {author} {\bibfnamefont {Eric}\ \bibnamefont
  {Poisson}},\ }\bibfield  {title} {\enquote {\bibinfo {title} {{Absorption of
  mass and angular momentum by a black hole: Time-domain formalisms for
  gravitational perturbations, and the small-hole / slow-motion
  approximation}},}\ }\href {\doibase 10.1103/PhysRevD.70.084044} {\bibfield
  {journal} {\bibinfo  {journal} {Phys. Rev. D}\ }\textbf {\bibinfo {volume}
  {70}},\ \bibinfo {pages} {084044} (\bibinfo {year} {2004})},\ \Eprint
  {http://arxiv.org/abs/gr-qc/0407050} {arXiv:gr-qc/0407050} \BibitemShut
  {NoStop}%
\bibitem [{\citenamefont {Taylor}\ and\ \citenamefont
  {Poisson}(2008)}]{Taylor:2008xy}%
  \BibitemOpen
  \bibfield  {author} {\bibinfo {author} {\bibfnamefont {Stephanne}\
  \bibnamefont {Taylor}}\ and\ \bibinfo {author} {\bibfnamefont {Eric}\
  \bibnamefont {Poisson}},\ }\bibfield  {title} {\enquote {\bibinfo {title}
  {{Nonrotating black hole in a post-Newtonian tidal environment}},}\ }\href
  {\doibase 10.1103/PhysRevD.78.084016} {\bibfield  {journal} {\bibinfo
  {journal} {Phys. Rev. D}\ }\textbf {\bibinfo {volume} {78}},\ \bibinfo
  {pages} {084016} (\bibinfo {year} {2008})},\ \Eprint
  {http://arxiv.org/abs/0806.3052} {arXiv:0806.3052 [gr-qc]} \BibitemShut
  {NoStop}%
\bibitem [{\citenamefont {Comeau}\ and\ \citenamefont
  {Poisson}(2009)}]{Comeau:2009bz}%
  \BibitemOpen
  \bibfield  {author} {\bibinfo {author} {\bibfnamefont {Simon}\ \bibnamefont
  {Comeau}}\ and\ \bibinfo {author} {\bibfnamefont {Eric}\ \bibnamefont
  {Poisson}},\ }\bibfield  {title} {\enquote {\bibinfo {title} {{Tidal
  interaction of a small black hole in the field of a large Kerr black
  hole}},}\ }\href {\doibase 10.1103/PhysRevD.80.087501} {\bibfield  {journal}
  {\bibinfo  {journal} {Phys. Rev. D}\ }\textbf {\bibinfo {volume} {80}},\
  \bibinfo {pages} {087501} (\bibinfo {year} {2009})},\ \Eprint
  {http://arxiv.org/abs/0908.4518} {arXiv:0908.4518 [gr-qc]} \BibitemShut
  {NoStop}%
\bibitem [{\citenamefont {Poisson}\ and\ \citenamefont
  {Vlasov}(2010)}]{Poisson:2009qj}%
  \BibitemOpen
  \bibfield  {author} {\bibinfo {author} {\bibfnamefont {Eric}\ \bibnamefont
  {Poisson}}\ and\ \bibinfo {author} {\bibfnamefont {Igor}\ \bibnamefont
  {Vlasov}},\ }\bibfield  {title} {\enquote {\bibinfo {title} {{Geometry and
  dynamics of a tidally deformed black hole}},}\ }\href {\doibase
  10.1103/PhysRevD.81.024029} {\bibfield  {journal} {\bibinfo  {journal} {Phys.
  Rev. D}\ }\textbf {\bibinfo {volume} {81}},\ \bibinfo {pages} {024029}
  (\bibinfo {year} {2010})},\ \Eprint {http://arxiv.org/abs/0910.4311}
  {arXiv:0910.4311 [gr-qc]} \BibitemShut {NoStop}%
\bibitem [{\citenamefont {Poisson}(2015)}]{Poisson:2014gka}%
  \BibitemOpen
  \bibfield  {author} {\bibinfo {author} {\bibfnamefont {Eric}\ \bibnamefont
  {Poisson}},\ }\bibfield  {title} {\enquote {\bibinfo {title} {{Tidal
  deformation of a slowly rotating black hole}},}\ }\href {\doibase
  10.1103/PhysRevD.91.044004} {\bibfield  {journal} {\bibinfo  {journal} {Phys.
  Rev. D}\ }\textbf {\bibinfo {volume} {91}},\ \bibinfo {pages} {044004}
  (\bibinfo {year} {2015})},\ \Eprint {http://arxiv.org/abs/1411.4711}
  {arXiv:1411.4711 [gr-qc]} \BibitemShut {NoStop}%
\bibitem [{\citenamefont {Poisson}\ and\ \citenamefont
  {Corrigan}(2018)}]{Poisson:2018qqd}%
  \BibitemOpen
  \bibfield  {author} {\bibinfo {author} {\bibfnamefont {Eric}\ \bibnamefont
  {Poisson}}\ and\ \bibinfo {author} {\bibfnamefont {Eamonn}\ \bibnamefont
  {Corrigan}},\ }\bibfield  {title} {\enquote {\bibinfo {title} {{Nonrotating
  black hole in a post-Newtonian tidal environment II}},}\ }\href {\doibase
  10.1103/PhysRevD.97.124048} {\bibfield  {journal} {\bibinfo  {journal} {Phys.
  Rev. D}\ }\textbf {\bibinfo {volume} {97}},\ \bibinfo {pages} {124048}
  (\bibinfo {year} {2018})},\ \Eprint {http://arxiv.org/abs/1804.01848}
  {arXiv:1804.01848 [gr-qc]} \BibitemShut {NoStop}%
\bibitem [{\citenamefont {Chatziioannou}\ \emph {et~al.}(2013)\citenamefont
  {Chatziioannou}, \citenamefont {Poisson},\ and\ \citenamefont
  {Yunes}}]{Chatziioannou:2012gq}%
  \BibitemOpen
  \bibfield  {author} {\bibinfo {author} {\bibfnamefont {Katerina}\
  \bibnamefont {Chatziioannou}}, \bibinfo {author} {\bibfnamefont {Eric}\
  \bibnamefont {Poisson}}, \ and\ \bibinfo {author} {\bibfnamefont {Nicolas}\
  \bibnamefont {Yunes}},\ }\bibfield  {title} {\enquote {\bibinfo {title}
  {{Tidal heating and torquing of a Kerr black hole to next-to-leading order in
  the tidal coupling}},}\ }\href {\doibase 10.1103/PhysRevD.87.044022}
  {\bibfield  {journal} {\bibinfo  {journal} {Phys. Rev. D}\ }\textbf {\bibinfo
  {volume} {87}},\ \bibinfo {pages} {044022} (\bibinfo {year} {2013})},\
  \Eprint {http://arxiv.org/abs/1211.1686} {arXiv:1211.1686 [gr-qc]}
  \BibitemShut {NoStop}%
\bibitem [{\citenamefont {Chatziioannou}\ \emph {et~al.}(2016)\citenamefont
  {Chatziioannou}, \citenamefont {Poisson},\ and\ \citenamefont
  {Yunes}}]{Chatziioannou:2016kem}%
  \BibitemOpen
  \bibfield  {author} {\bibinfo {author} {\bibfnamefont {Katerina}\
  \bibnamefont {Chatziioannou}}, \bibinfo {author} {\bibfnamefont {Eric}\
  \bibnamefont {Poisson}}, \ and\ \bibinfo {author} {\bibfnamefont {Nicolas}\
  \bibnamefont {Yunes}},\ }\bibfield  {title} {\enquote {\bibinfo {title}
  {{Improved next-to-leading order tidal heating and torquing of a Kerr black
  hole}},}\ }\href {\doibase 10.1103/PhysRevD.94.084043} {\bibfield  {journal}
  {\bibinfo  {journal} {Phys. Rev. D}\ }\textbf {\bibinfo {volume} {94}},\
  \bibinfo {pages} {084043} (\bibinfo {year} {2016})},\ \Eprint
  {http://arxiv.org/abs/1608.02899} {arXiv:1608.02899 [gr-qc]} \BibitemShut
  {NoStop}%
\bibitem [{\citenamefont {Saketh}\ \emph {et~al.}(2023)\citenamefont {Saketh},
  \citenamefont {Steinhoff}, \citenamefont {Vines},\ and\ \citenamefont
  {Buonanno}}]{Saketh:2022xjb}%
  \BibitemOpen
  \bibfield  {author} {\bibinfo {author} {\bibfnamefont {M.~V.~S.}\
  \bibnamefont {Saketh}}, \bibinfo {author} {\bibfnamefont {Jan}\ \bibnamefont
  {Steinhoff}}, \bibinfo {author} {\bibfnamefont {Justin}\ \bibnamefont
  {Vines}}, \ and\ \bibinfo {author} {\bibfnamefont {Alessandra}\ \bibnamefont
  {Buonanno}},\ }\bibfield  {title} {\enquote {\bibinfo {title} {{Modeling
  horizon absorption in spinning binary black holes using effective worldline
  theory}},}\ }\href {\doibase 10.1103/PhysRevD.107.084006} {\bibfield
  {journal} {\bibinfo  {journal} {Phys. Rev. D}\ }\textbf {\bibinfo {volume}
  {107}},\ \bibinfo {pages} {084006} (\bibinfo {year} {2023})},\ \Eprint
  {http://arxiv.org/abs/2212.13095} {arXiv:2212.13095 [gr-qc]} \BibitemShut
  {NoStop}%
\bibitem [{\citenamefont {Datta}(2024)}]{Datta:2023wsn}%
  \BibitemOpen
  \bibfield  {author} {\bibinfo {author} {\bibfnamefont {Sayak}\ \bibnamefont
  {Datta}},\ }\bibfield  {title} {\enquote {\bibinfo {title} {{Horizon fluxes
  of binary black holes in eccentric orbits}},}\ }\href {\doibase
  10.1140/epjc/s10052-024-13371-8} {\bibfield  {journal} {\bibinfo  {journal}
  {Eur. Phys. J. C}\ }\textbf {\bibinfo {volume} {84}},\ \bibinfo {pages}
  {1077} (\bibinfo {year} {2024})},\ \Eprint {http://arxiv.org/abs/2305.03771}
  {arXiv:2305.03771 [gr-qc]} \BibitemShut {NoStop}%
\bibitem [{\citenamefont {Chiaramello}\ and\ \citenamefont
  {Gamba}(2025)}]{Chiaramello:2024unv}%
  \BibitemOpen
  \bibfield  {author} {\bibinfo {author} {\bibfnamefont {Danilo}\ \bibnamefont
  {Chiaramello}}\ and\ \bibinfo {author} {\bibfnamefont {Rossella}\
  \bibnamefont {Gamba}},\ }\bibfield  {title} {\enquote {\bibinfo {title}
  {{Horizon absorption on noncircular, planar binary black hole dynamics}},}\
  }\href {\doibase 10.1103/PhysRevD.111.024024} {\bibfield  {journal} {\bibinfo
   {journal} {Phys. Rev. D}\ }\textbf {\bibinfo {volume} {111}},\ \bibinfo
  {pages} {024024} (\bibinfo {year} {2025})},\ \Eprint
  {http://arxiv.org/abs/2408.15322} {arXiv:2408.15322 [gr-qc]} \BibitemShut
  {NoStop}%
\bibitem [{\citenamefont {Scheel}\ \emph {et~al.}(2015)\citenamefont {Scheel},
  \citenamefont {Giesler}, \citenamefont {Hemberger}, \citenamefont {Lovelace},
  \citenamefont {Kuper}, \citenamefont {Boyle}, \citenamefont {Szil{\'a}gyi},\
  and\ \citenamefont {Kidder}}]{Scheel:2014ina}%
  \BibitemOpen
  \bibfield  {author} {\bibinfo {author} {\bibfnamefont {Mark~A.}\ \bibnamefont
  {Scheel}}, \bibinfo {author} {\bibfnamefont {Matthew}\ \bibnamefont
  {Giesler}}, \bibinfo {author} {\bibfnamefont {Daniel~A.}\ \bibnamefont
  {Hemberger}}, \bibinfo {author} {\bibfnamefont {Geoffrey}\ \bibnamefont
  {Lovelace}}, \bibinfo {author} {\bibfnamefont {Kevin}\ \bibnamefont {Kuper}},
  \bibinfo {author} {\bibfnamefont {Michael}\ \bibnamefont {Boyle}}, \bibinfo
  {author} {\bibfnamefont {B.}~\bibnamefont {Szil{\'a}gyi}}, \ and\ \bibinfo
  {author} {\bibfnamefont {Lawrence~E.}\ \bibnamefont {Kidder}},\ }\bibfield
  {title} {\enquote {\bibinfo {title} {{Improved methods for simulating nearly
  extremal binary black holes}},}\ }\href {\doibase
  10.1088/0264-9381/32/10/105009} {\bibfield  {journal} {\bibinfo  {journal}
  {Class. Quant. Grav.}\ }\textbf {\bibinfo {volume} {32}},\ \bibinfo {pages}
  {105009} (\bibinfo {year} {2015})},\ \Eprint {http://arxiv.org/abs/1412.1803}
  {arXiv:1412.1803 [gr-qc]} \BibitemShut {NoStop}%
\bibitem [{\citenamefont {Nelson}\ \emph {et~al.}(2019)\citenamefont {Nelson},
  \citenamefont {Etienne}, \citenamefont {McWilliams},\ and\ \citenamefont
  {Nguyen}}]{Nelson:2019czq}%
  \BibitemOpen
  \bibfield  {author} {\bibinfo {author} {\bibfnamefont {Patrick~E.}\
  \bibnamefont {Nelson}}, \bibinfo {author} {\bibfnamefont {Zachariah~B.}\
  \bibnamefont {Etienne}}, \bibinfo {author} {\bibfnamefont {Sean~T.}\
  \bibnamefont {McWilliams}}, \ and\ \bibinfo {author} {\bibfnamefont
  {Viviana}\ \bibnamefont {Nguyen}},\ }\bibfield  {title} {\enquote {\bibinfo
  {title} {{Induced Spins from Scattering Experiments of Initially Nonspinning
  Black Holes}},}\ }\href {\doibase 10.1103/PhysRevD.100.124045} {\bibfield
  {journal} {\bibinfo  {journal} {Phys. Rev. D}\ }\textbf {\bibinfo {volume}
  {100}},\ \bibinfo {pages} {124045} (\bibinfo {year} {2019})},\ \Eprint
  {http://arxiv.org/abs/1909.08621} {arXiv:1909.08621 [gr-qc]} \BibitemShut
  {NoStop}%
\bibitem [{\citenamefont {Jaraba}\ and\ \citenamefont
  {Garcia-Bellido}(2021)}]{Jaraba:2021ces}%
  \BibitemOpen
  \bibfield  {author} {\bibinfo {author} {\bibfnamefont {Santiago}\
  \bibnamefont {Jaraba}}\ and\ \bibinfo {author} {\bibfnamefont {Juan}\
  \bibnamefont {Garcia-Bellido}},\ }\bibfield  {title} {\enquote {\bibinfo
  {title} {{Black hole induced spins from hyperbolic encounters in dense
  clusters}},}\ }\href {\doibase 10.1016/j.dark.2021.100882} {\bibfield
  {journal} {\bibinfo  {journal} {Phys. Dark Univ.}\ }\textbf {\bibinfo
  {volume} {34}},\ \bibinfo {pages} {100882} (\bibinfo {year} {2021})},\
  \Eprint {http://arxiv.org/abs/2106.01436} {arXiv:2106.01436 [gr-qc]}
  \BibitemShut {NoStop}%
\bibitem [{\citenamefont {Rodr{\'\i}guez-Monteverde}\ \emph
  {et~al.}(2025)\citenamefont {Rodr{\'\i}guez-Monteverde}, \citenamefont
  {Jaraba},\ and\ \citenamefont
  {Garc{\'\i}a-Bellido}}]{Rodriguez-Monteverde:2024tnt}%
  \BibitemOpen
  \bibfield  {author} {\bibinfo {author} {\bibfnamefont {Jorge~L.}\
  \bibnamefont {Rodr{\'\i}guez-Monteverde}}, \bibinfo {author} {\bibfnamefont
  {Santiago}\ \bibnamefont {Jaraba}}, \ and\ \bibinfo {author} {\bibfnamefont
  {Juan}\ \bibnamefont {Garc{\'\i}a-Bellido}},\ }\bibfield  {title} {\enquote
  {\bibinfo {title} {{Spin induction from scattering of two spinning black
  holes in dense clusters}},}\ }\href {\doibase 10.1016/j.dark.2024.101776}
  {\bibfield  {journal} {\bibinfo  {journal} {Phys. Dark Univ.}\ }\textbf
  {\bibinfo {volume} {47}},\ \bibinfo {pages} {101776} (\bibinfo {year}
  {2025})},\ \Eprint {http://arxiv.org/abs/2410.11634} {arXiv:2410.11634
  [gr-qc]} \BibitemShut {NoStop}%
\bibitem [{\citenamefont {Kogan}\ \emph {et~al.}(2025)\citenamefont {Kogan},
  \citenamefont {Pardoe},\ and\ \citenamefont {Witek}}]{Kogan:2025vml}%
  \BibitemOpen
  \bibfield  {author} {\bibinfo {author} {\bibfnamefont {Healey}\ \bibnamefont
  {Kogan}}, \bibinfo {author} {\bibfnamefont {Frederick C.~L.}\ \bibnamefont
  {Pardoe}}, \ and\ \bibinfo {author} {\bibfnamefont {Helvi}\ \bibnamefont
  {Witek}},\ }\bibfield  {title} {\enquote {\bibinfo {title} {{Spin-up and
  mass-gain in hyperbolic encounters of spinning black holes}},}\ }\href@noop
  {} {\  (\bibinfo {year} {2025})},\ \Eprint {http://arxiv.org/abs/2511.00307}
  {arXiv:2511.00307 [gr-qc]} \BibitemShut {NoStop}%
\bibitem [{\citenamefont {Datta}\ \emph {et~al.}(2024)\citenamefont {Datta},
  \citenamefont {Brito}, \citenamefont {Hughes}, \citenamefont {Klinger},\ and\
  \citenamefont {Pani}}]{Datta:2024vll}%
  \BibitemOpen
  \bibfield  {author} {\bibinfo {author} {\bibfnamefont {Sayak}\ \bibnamefont
  {Datta}}, \bibinfo {author} {\bibfnamefont {Richard}\ \bibnamefont {Brito}},
  \bibinfo {author} {\bibfnamefont {Scott~A.}\ \bibnamefont {Hughes}}, \bibinfo
  {author} {\bibfnamefont {Talya}\ \bibnamefont {Klinger}}, \ and\ \bibinfo
  {author} {\bibfnamefont {Paolo}\ \bibnamefont {Pani}},\ }\bibfield  {title}
  {\enquote {\bibinfo {title} {{Tidal heating as a discriminator for horizons
  in equatorial eccentric extreme mass ratio inspirals}},}\ }\href {\doibase
  10.1103/PhysRevD.110.024048} {\bibfield  {journal} {\bibinfo  {journal}
  {Phys. Rev. D}\ }\textbf {\bibinfo {volume} {110}},\ \bibinfo {pages}
  {024048} (\bibinfo {year} {2024})},\ \Eprint
  {http://arxiv.org/abs/2404.04013} {arXiv:2404.04013 [gr-qc]} \BibitemShut
  {NoStop}%
\bibitem [{\citenamefont {Harms}\ \emph {et~al.}(2013)\citenamefont {Harms},
  \citenamefont {Bernuzzi},\ and\ \citenamefont {Br{\"u}gmann}}]{Harms:2013ib}%
  \BibitemOpen
  \bibfield  {author} {\bibinfo {author} {\bibfnamefont {Enno}\ \bibnamefont
  {Harms}}, \bibinfo {author} {\bibfnamefont {Sebastiano}\ \bibnamefont
  {Bernuzzi}}, \ and\ \bibinfo {author} {\bibfnamefont {Bernd}\ \bibnamefont
  {Br{\"u}gmann}},\ }\bibfield  {title} {\enquote {\bibinfo {title} {{Numerical
  solution of the 2+1 Teukolsky equation on a hyperboloidal and horizon
  penetrating foliation of Kerr and application to late-time decays}},}\ }\href
  {\doibase 10.1088/0264-9381/30/11/115013} {\bibfield  {journal} {\bibinfo
  {journal} {Class. Quant. Grav.}\ }\textbf {\bibinfo {volume} {30}},\ \bibinfo
  {pages} {115013} (\bibinfo {year} {2013})},\ \Eprint
  {http://arxiv.org/abs/1301.1591} {arXiv:1301.1591 [gr-qc]} \BibitemShut
  {NoStop}%
\bibitem [{\citenamefont {Harms}\ \emph {et~al.}(2014)\citenamefont {Harms},
  \citenamefont {Bernuzzi}, \citenamefont {Nagar},\ and\ \citenamefont
  {Zenginoglu}}]{Harms:2014dqa}%
  \BibitemOpen
  \bibfield  {author} {\bibinfo {author} {\bibfnamefont {Enno}\ \bibnamefont
  {Harms}}, \bibinfo {author} {\bibfnamefont {Sebastiano}\ \bibnamefont
  {Bernuzzi}}, \bibinfo {author} {\bibfnamefont {Alessandro}\ \bibnamefont
  {Nagar}}, \ and\ \bibinfo {author} {\bibfnamefont {An}~\bibnamefont
  {Zenginoglu}},\ }\bibfield  {title} {\enquote {\bibinfo {title} {{A new
  gravitational wave generation algorithm for particle perturbations of the
  Kerr spacetime}},}\ }\href {\doibase 10.1088/0264-9381/31/24/245004}
  {\bibfield  {journal} {\bibinfo  {journal} {Class. Quant. Grav.}\ }\textbf
  {\bibinfo {volume} {31}},\ \bibinfo {pages} {245004} (\bibinfo {year}
  {2014})},\ \Eprint {http://arxiv.org/abs/1406.5983} {arXiv:1406.5983 [gr-qc]}
  \BibitemShut {NoStop}%
\bibitem [{\citenamefont {Damour}\ and\ \citenamefont
  {Nagar}(2014)}]{Damour:2014sva}%
  \BibitemOpen
  \bibfield  {author} {\bibinfo {author} {\bibfnamefont {Thibault}\
  \bibnamefont {Damour}}\ and\ \bibinfo {author} {\bibfnamefont {Alessandro}\
  \bibnamefont {Nagar}},\ }\bibfield  {title} {\enquote {\bibinfo {title} {{New
  effective-one-body description of coalescing nonprecessing spinning
  black-hole binaries}},}\ }\href {\doibase 10.1103/PhysRevD.90.044018}
  {\bibfield  {journal} {\bibinfo  {journal} {Phys. Rev. D}\ }\textbf {\bibinfo
  {volume} {90}},\ \bibinfo {pages} {044018} (\bibinfo {year} {2014})},\
  \Eprint {http://arxiv.org/abs/1406.6913} {arXiv:1406.6913 [gr-qc]}
  \BibitemShut {NoStop}%
\bibitem [{\citenamefont {Zenginoglu}(2008)}]{Zenginoglu:2008wc}%
  \BibitemOpen
  \bibfield  {author} {\bibinfo {author} {\bibfnamefont {Anil}\ \bibnamefont
  {Zenginoglu}},\ }\bibfield  {title} {\enquote {\bibinfo {title} {{A
  Hyperboloidal study of tail decay rates for scalar and Yang-Mills fields}},}\
  }\href {\doibase 10.1088/0264-9381/25/17/175013} {\bibfield  {journal}
  {\bibinfo  {journal} {Class. Quant. Grav.}\ }\textbf {\bibinfo {volume}
  {25}},\ \bibinfo {pages} {175013} (\bibinfo {year} {2008})},\ \Eprint
  {http://arxiv.org/abs/0803.2018} {arXiv:0803.2018 [gr-qc]} \BibitemShut
  {NoStop}%
\bibitem [{\citenamefont {Bernuzzi}\ \emph {et~al.}(2011)\citenamefont
  {Bernuzzi}, \citenamefont {Nagar},\ and\ \citenamefont
  {Zenginoglu}}]{Bernuzzi:2011aj}%
  \BibitemOpen
  \bibfield  {author} {\bibinfo {author} {\bibfnamefont {Sebastiano}\
  \bibnamefont {Bernuzzi}}, \bibinfo {author} {\bibfnamefont {Alessandro}\
  \bibnamefont {Nagar}}, \ and\ \bibinfo {author} {\bibfnamefont {Anil}\
  \bibnamefont {Zenginoglu}},\ }\bibfield  {title} {\enquote {\bibinfo {title}
  {{Binary black hole coalescence in the large-mass-ratio limit: the
  hyperboloidal layer method and waveforms at null infinity}},}\ }\href
  {\doibase 10.1103/PhysRevD.84.084026} {\bibfield  {journal} {\bibinfo
  {journal} {Phys. Rev. D}\ }\textbf {\bibinfo {volume} {84}},\ \bibinfo
  {pages} {084026} (\bibinfo {year} {2011})},\ \Eprint
  {http://arxiv.org/abs/1107.5402} {arXiv:1107.5402 [gr-qc]} \BibitemShut
  {NoStop}%
\bibitem [{\citenamefont {Fontbut{\'e}}\ \emph {et~al.}(2025)\citenamefont
  {Fontbut{\'e}}, \citenamefont {Bernuzzi}, \citenamefont {Albanesi},
  \citenamefont {Radice}, \citenamefont {Rashti}, \citenamefont {Cook},
  \citenamefont {Daszuta},\ and\ \citenamefont {Nagar}}]{Fontbute:2025ixd}%
  \BibitemOpen
  \bibfield  {author} {\bibinfo {author} {\bibfnamefont {Joan}\ \bibnamefont
  {Fontbut{\'e}}}, \bibinfo {author} {\bibfnamefont {Sebastiano}\ \bibnamefont
  {Bernuzzi}}, \bibinfo {author} {\bibfnamefont {Simone}\ \bibnamefont
  {Albanesi}}, \bibinfo {author} {\bibfnamefont {David}\ \bibnamefont
  {Radice}}, \bibinfo {author} {\bibfnamefont {Alireza}\ \bibnamefont
  {Rashti}}, \bibinfo {author} {\bibfnamefont {William}\ \bibnamefont {Cook}},
  \bibinfo {author} {\bibfnamefont {Boris}\ \bibnamefont {Daszuta}}, \ and\
  \bibinfo {author} {\bibfnamefont {Alessandro}\ \bibnamefont {Nagar}},\
  }\bibfield  {title} {\enquote {\bibinfo {title} {{Covariant and
  Gauge-invariant Metric-based Gravitational-waves Extraction in Numerical
  Relativity}},}\ }\href@noop {} {\  (\bibinfo {year} {2025})},\ \Eprint
  {http://arxiv.org/abs/2508.03799} {arXiv:2508.03799 [gr-qc]} \BibitemShut
  {NoStop}%
\bibitem [{\citenamefont {Bernuzzi}\ \emph {et~al.}(2025)\citenamefont
  {Bernuzzi}, \citenamefont {Fontbut{\'e}}, \citenamefont {Albanesi},\ and\
  \citenamefont {Zengino{\u{g}}lu}}]{Bernuzzi:2025hhu}%
  \BibitemOpen
  \bibfield  {author} {\bibinfo {author} {\bibfnamefont {Sebastiano}\
  \bibnamefont {Bernuzzi}}, \bibinfo {author} {\bibfnamefont {Joan}\
  \bibnamefont {Fontbut{\'e}}}, \bibinfo {author} {\bibfnamefont {Simone}\
  \bibnamefont {Albanesi}}, \ and\ \bibinfo {author} {\bibfnamefont {Anil}\
  \bibnamefont {Zengino{\u{g}}lu}},\ }\bibfield  {title} {\enquote {\bibinfo
  {title} {{Perturbative hyperboloidal extraction of gravitational waves in 3+1
  numerical relativity}},}\ }\href {\doibase 10.1103/g5cp-lzgw} {\bibfield
  {journal} {\bibinfo  {journal} {Phys. Rev. D}\ }\textbf {\bibinfo {volume}
  {112}},\ \bibinfo {pages} {084036} (\bibinfo {year} {2025})},\ \Eprint
  {http://arxiv.org/abs/2508.05743} {arXiv:2508.05743 [gr-qc]} \BibitemShut
  {NoStop}%
\bibitem [{\citenamefont {Damour}\ \emph {et~al.}(2009)\citenamefont {Damour},
  \citenamefont {Iyer},\ and\ \citenamefont {Nagar}}]{Damour:2008gu}%
  \BibitemOpen
  \bibfield  {author} {\bibinfo {author} {\bibfnamefont {Thibault}\
  \bibnamefont {Damour}}, \bibinfo {author} {\bibfnamefont {Bala~R.}\
  \bibnamefont {Iyer}}, \ and\ \bibinfo {author} {\bibfnamefont {Alessandro}\
  \bibnamefont {Nagar}},\ }\bibfield  {title} {\enquote {\bibinfo {title}
  {{Improved resummation of post-Newtonian multipolar waveforms from
  circularized compact binaries}},}\ }\href {\doibase
  10.1103/PhysRevD.79.064004} {\bibfield  {journal} {\bibinfo  {journal} {Phys.
  Rev. D}\ }\textbf {\bibinfo {volume} {79}},\ \bibinfo {pages} {064004}
  (\bibinfo {year} {2009})},\ \Eprint {http://arxiv.org/abs/0811.2069}
  {arXiv:0811.2069 [gr-qc]} \BibitemShut {NoStop}%
\bibitem [{\citenamefont {Pan}\ \emph {et~al.}(2011)\citenamefont {Pan},
  \citenamefont {Buonanno}, \citenamefont {Fujita}, \citenamefont {Racine},\
  and\ \citenamefont {Tagoshi}}]{Pan:2010hz}%
  \BibitemOpen
  \bibfield  {author} {\bibinfo {author} {\bibfnamefont {Yi}~\bibnamefont
  {Pan}}, \bibinfo {author} {\bibfnamefont {Alessandra}\ \bibnamefont
  {Buonanno}}, \bibinfo {author} {\bibfnamefont {Ryuichi}\ \bibnamefont
  {Fujita}}, \bibinfo {author} {\bibfnamefont {Etienne}\ \bibnamefont
  {Racine}}, \ and\ \bibinfo {author} {\bibfnamefont {Hideyuki}\ \bibnamefont
  {Tagoshi}},\ }\bibfield  {title} {\enquote {\bibinfo {title} {{Post-Newtonian
  factorized multipolar waveforms for spinning, non-precessing black-hole
  binaries}},}\ }\href {\doibase 10.1103/PhysRevD.83.064003} {\bibfield
  {journal} {\bibinfo  {journal} {Phys. Rev. D}\ }\textbf {\bibinfo {volume}
  {83}},\ \bibinfo {pages} {064003} (\bibinfo {year} {2011})},\ \bibinfo {note}
  {[Erratum: Phys.Rev.D 87, 109901 (2013)]},\ \Eprint
  {http://arxiv.org/abs/1006.0431} {arXiv:1006.0431 [gr-qc]} \BibitemShut
  {NoStop}%
\bibitem [{\citenamefont {Messina}\ \emph {et~al.}(2018)\citenamefont
  {Messina}, \citenamefont {Maldarella},\ and\ \citenamefont
  {Nagar}}]{Messina:2018ghh}%
  \BibitemOpen
  \bibfield  {author} {\bibinfo {author} {\bibfnamefont {Francesco}\
  \bibnamefont {Messina}}, \bibinfo {author} {\bibfnamefont {Alberto}\
  \bibnamefont {Maldarella}}, \ and\ \bibinfo {author} {\bibfnamefont
  {Alessandro}\ \bibnamefont {Nagar}},\ }\bibfield  {title} {\enquote {\bibinfo
  {title} {{Factorization and resummation: A new paradigm to improve
  gravitational wave amplitudes. II: the higher multipolar modes}},}\ }\href
  {\doibase 10.1103/PhysRevD.97.084016} {\bibfield  {journal} {\bibinfo
  {journal} {Phys. Rev. D}\ }\textbf {\bibinfo {volume} {97}},\ \bibinfo
  {pages} {084016} (\bibinfo {year} {2018})},\ \Eprint
  {http://arxiv.org/abs/1801.02366} {arXiv:1801.02366 [gr-qc]} \BibitemShut
  {NoStop}%
\bibitem [{\citenamefont {Nagar}\ and\ \citenamefont
  {Shah}(2016)}]{Nagar:2016ayt}%
  \BibitemOpen
  \bibfield  {author} {\bibinfo {author} {\bibfnamefont {Alessandro}\
  \bibnamefont {Nagar}}\ and\ \bibinfo {author} {\bibfnamefont {Abhay}\
  \bibnamefont {Shah}},\ }\bibfield  {title} {\enquote {\bibinfo {title}
  {{Factorization and resummation: A new paradigm to improve gravitational wave
  amplitudes}},}\ }\href {\doibase 10.1103/PhysRevD.94.104017} {\bibfield
  {journal} {\bibinfo  {journal} {Phys. Rev. D}\ }\textbf {\bibinfo {volume}
  {94}},\ \bibinfo {pages} {104017} (\bibinfo {year} {2016})},\ \Eprint
  {http://arxiv.org/abs/1606.00207} {arXiv:1606.00207 [gr-qc]} \BibitemShut
  {NoStop}%
\bibitem [{\citenamefont {O'Sullivan}\ and\ \citenamefont
  {Hughes}(2014)}]{OSullivan:2014ywd}%
  \BibitemOpen
  \bibfield  {author} {\bibinfo {author} {\bibfnamefont {Stephen}\ \bibnamefont
  {O'Sullivan}}\ and\ \bibinfo {author} {\bibfnamefont {Scott~A.}\ \bibnamefont
  {Hughes}},\ }\bibfield  {title} {\enquote {\bibinfo {title} {{Strong-field
  tidal distortions of rotating black holes: Formalism and results for
  circular, equatorial orbits}},}\ }\href {\doibase 10.1103/PhysRevD.91.109901}
  {\bibfield  {journal} {\bibinfo  {journal} {Phys. Rev. D}\ }\textbf {\bibinfo
  {volume} {90}},\ \bibinfo {pages} {124039} (\bibinfo {year} {2014})},\
  \bibinfo {note} {[Erratum: Phys.Rev.D 91, 109901 (2015)]},\ \Eprint
  {http://arxiv.org/abs/1407.6983} {arXiv:1407.6983 [gr-qc]} \BibitemShut
  {NoStop}%
\bibitem [{\citenamefont {Chiaramello}\ and\ \citenamefont
  {Nagar}(2020)}]{Chiaramello:2020ehz}%
  \BibitemOpen
  \bibfield  {author} {\bibinfo {author} {\bibfnamefont {Danilo}\ \bibnamefont
  {Chiaramello}}\ and\ \bibinfo {author} {\bibfnamefont {Alessandro}\
  \bibnamefont {Nagar}},\ }\bibfield  {title} {\enquote {\bibinfo {title}
  {{Faithful analytical effective-one-body waveform model for spin-aligned,
  moderately eccentric, coalescing black hole binaries}},}\ }\href {\doibase
  10.1103/PhysRevD.101.101501} {\bibfield  {journal} {\bibinfo  {journal}
  {Phys. Rev. D}\ }\textbf {\bibinfo {volume} {101}},\ \bibinfo {pages}
  {101501} (\bibinfo {year} {2020})},\ \Eprint
  {http://arxiv.org/abs/2001.11736} {arXiv:2001.11736 [gr-qc]} \BibitemShut
  {NoStop}%
\bibitem [{\citenamefont {Khalil}\ \emph {et~al.}(2021)\citenamefont {Khalil},
  \citenamefont {Buonanno}, \citenamefont {Steinhoff},\ and\ \citenamefont
  {Vines}}]{Khalil:2021txt}%
  \BibitemOpen
  \bibfield  {author} {\bibinfo {author} {\bibfnamefont {Mohammed}\
  \bibnamefont {Khalil}}, \bibinfo {author} {\bibfnamefont {Alessandra}\
  \bibnamefont {Buonanno}}, \bibinfo {author} {\bibfnamefont {Jan}\
  \bibnamefont {Steinhoff}}, \ and\ \bibinfo {author} {\bibfnamefont {Justin}\
  \bibnamefont {Vines}},\ }\bibfield  {title} {\enquote {\bibinfo {title}
  {{Radiation-reaction force and multipolar waveforms for eccentric,
  spin-aligned binaries in the effective-one-body formalism}},}\ }\href
  {\doibase 10.1103/PhysRevD.104.024046} {\bibfield  {journal} {\bibinfo
  {journal} {Phys. Rev. D}\ }\textbf {\bibinfo {volume} {104}},\ \bibinfo
  {pages} {024046} (\bibinfo {year} {2021})},\ \Eprint
  {http://arxiv.org/abs/2104.11705} {arXiv:2104.11705 [gr-qc]} \BibitemShut
  {NoStop}%
\bibitem [{\citenamefont {Albanesi}\ \emph {et~al.}(2021)\citenamefont
  {Albanesi}, \citenamefont {Nagar},\ and\ \citenamefont
  {Bernuzzi}}]{Albanesi:2021rby}%
  \BibitemOpen
  \bibfield  {author} {\bibinfo {author} {\bibfnamefont {Simone}\ \bibnamefont
  {Albanesi}}, \bibinfo {author} {\bibfnamefont {Alessandro}\ \bibnamefont
  {Nagar}}, \ and\ \bibinfo {author} {\bibfnamefont {Sebastiano}\ \bibnamefont
  {Bernuzzi}},\ }\bibfield  {title} {\enquote {\bibinfo {title} {{Effective
  one-body model for extreme-mass-ratio spinning binaries on eccentric
  equatorial orbits: Testing radiation reaction and waveform}},}\ }\href
  {\doibase 10.1103/PhysRevD.104.024067} {\bibfield  {journal} {\bibinfo
  {journal} {Phys. Rev. D}\ }\textbf {\bibinfo {volume} {104}},\ \bibinfo
  {pages} {024067} (\bibinfo {year} {2021})},\ \Eprint
  {http://arxiv.org/abs/2104.10559} {arXiv:2104.10559 [gr-qc]} \BibitemShut
  {NoStop}%
\bibitem [{\citenamefont {Placidi}\ \emph {et~al.}(2022)\citenamefont
  {Placidi}, \citenamefont {Albanesi}, \citenamefont {Nagar}, \citenamefont
  {Orselli}, \citenamefont {Bernuzzi},\ and\ \citenamefont
  {Grignani}}]{Placidi:2021rkh}%
  \BibitemOpen
  \bibfield  {author} {\bibinfo {author} {\bibfnamefont {Andrea}\ \bibnamefont
  {Placidi}}, \bibinfo {author} {\bibfnamefont {Simone}\ \bibnamefont
  {Albanesi}}, \bibinfo {author} {\bibfnamefont {Alessandro}\ \bibnamefont
  {Nagar}}, \bibinfo {author} {\bibfnamefont {Marta}\ \bibnamefont {Orselli}},
  \bibinfo {author} {\bibfnamefont {Sebastiano}\ \bibnamefont {Bernuzzi}}, \
  and\ \bibinfo {author} {\bibfnamefont {Gianluca}\ \bibnamefont {Grignani}},\
  }\bibfield  {title} {\enquote {\bibinfo {title} {{Exploiting
  Newton-factorized, 2PN-accurate waveform multipoles in effective-one-body
  models for spin-aligned noncircularized binaries}},}\ }\href {\doibase
  10.1103/PhysRevD.105.104030} {\bibfield  {journal} {\bibinfo  {journal}
  {Phys. Rev. D}\ }\textbf {\bibinfo {volume} {105}},\ \bibinfo {pages}
  {104030} (\bibinfo {year} {2022})},\ \Eprint
  {http://arxiv.org/abs/2112.05448} {arXiv:2112.05448 [gr-qc]} \BibitemShut
  {NoStop}%
\bibitem [{\citenamefont {Albanesi}\ \emph {et~al.}(2022)\citenamefont
  {Albanesi}, \citenamefont {Placidi}, \citenamefont {Nagar}, \citenamefont
  {Orselli},\ and\ \citenamefont {Bernuzzi}}]{Albanesi:2022xge}%
  \BibitemOpen
  \bibfield  {author} {\bibinfo {author} {\bibfnamefont {Simone}\ \bibnamefont
  {Albanesi}}, \bibinfo {author} {\bibfnamefont {Andrea}\ \bibnamefont
  {Placidi}}, \bibinfo {author} {\bibfnamefont {Alessandro}\ \bibnamefont
  {Nagar}}, \bibinfo {author} {\bibfnamefont {Marta}\ \bibnamefont {Orselli}},
  \ and\ \bibinfo {author} {\bibfnamefont {Sebastiano}\ \bibnamefont
  {Bernuzzi}},\ }\bibfield  {title} {\enquote {\bibinfo {title} {{New avenue
  for accurate analytical waveforms and fluxes for eccentric compact
  binaries}},}\ }\href {\doibase 10.1103/PhysRevD.105.L121503} {\bibfield
  {journal} {\bibinfo  {journal} {Phys. Rev. D}\ }\textbf {\bibinfo {volume}
  {105}},\ \bibinfo {pages} {L121503} (\bibinfo {year} {2022})},\ \Eprint
  {http://arxiv.org/abs/2203.16286} {arXiv:2203.16286 [gr-qc]} \BibitemShut
  {NoStop}%
\bibitem [{\citenamefont {Placidi}\ \emph {et~al.}(2023)\citenamefont
  {Placidi}, \citenamefont {Grignani}, \citenamefont {Harmark}, \citenamefont
  {Orselli}, \citenamefont {Gliorio},\ and\ \citenamefont
  {Nagar}}]{Placidi:2023ofj}%
  \BibitemOpen
  \bibfield  {author} {\bibinfo {author} {\bibfnamefont {Andrea}\ \bibnamefont
  {Placidi}}, \bibinfo {author} {\bibfnamefont {Gianluca}\ \bibnamefont
  {Grignani}}, \bibinfo {author} {\bibfnamefont {Troels}\ \bibnamefont
  {Harmark}}, \bibinfo {author} {\bibfnamefont {Marta}\ \bibnamefont
  {Orselli}}, \bibinfo {author} {\bibfnamefont {Sara}\ \bibnamefont {Gliorio}},
  \ and\ \bibinfo {author} {\bibfnamefont {Alessandro}\ \bibnamefont {Nagar}},\
  }\bibfield  {title} {\enquote {\bibinfo {title} {{2.5PN accurate waveform
  information for generic-planar-orbit binaries in effective one-body
  models}},}\ }\href {\doibase 10.1103/PhysRevD.108.024068} {\bibfield
  {journal} {\bibinfo  {journal} {Phys. Rev. D}\ }\textbf {\bibinfo {volume}
  {108}},\ \bibinfo {pages} {024068} (\bibinfo {year} {2023})},\ \Eprint
  {http://arxiv.org/abs/2305.14440} {arXiv:2305.14440 [gr-qc]} \BibitemShut
  {NoStop}%
\bibitem [{\citenamefont {Faggioli}\ \emph {et~al.}(2025)\citenamefont
  {Faggioli}, \citenamefont {van~de Meent}, \citenamefont {Buonanno},
  \citenamefont {Gamboa}, \citenamefont {Khalil},\ and\ \citenamefont
  {Khanna}}]{Faggioli:2024ugn}%
  \BibitemOpen
  \bibfield  {author} {\bibinfo {author} {\bibfnamefont {Guglielmo}\
  \bibnamefont {Faggioli}}, \bibinfo {author} {\bibfnamefont {Maarten}\
  \bibnamefont {van~de Meent}}, \bibinfo {author} {\bibfnamefont {Alessandra}\
  \bibnamefont {Buonanno}}, \bibinfo {author} {\bibfnamefont {Aldo}\
  \bibnamefont {Gamboa}}, \bibinfo {author} {\bibfnamefont {Mohammed}\
  \bibnamefont {Khalil}}, \ and\ \bibinfo {author} {\bibfnamefont {Gaurav}\
  \bibnamefont {Khanna}},\ }\bibfield  {title} {\enquote {\bibinfo {title}
  {{Testing eccentric corrections to the radiation-reaction force in the
  test-mass limit of effective-one-body models}},}\ }\href {\doibase
  10.1103/PhysRevD.111.044036} {\bibfield  {journal} {\bibinfo  {journal}
  {Phys. Rev. D}\ }\textbf {\bibinfo {volume} {111}},\ \bibinfo {pages}
  {044036} (\bibinfo {year} {2025})},\ \Eprint
  {http://arxiv.org/abs/2405.19006} {arXiv:2405.19006 [gr-qc]} \BibitemShut
  {NoStop}%
\bibitem [{\citenamefont {Gamboa}\ \emph {et~al.}(2025)\citenamefont {Gamboa},
  \citenamefont {Khalil},\ and\ \citenamefont {Buonanno}}]{Gamboa:2024imd}%
  \BibitemOpen
  \bibfield  {author} {\bibinfo {author} {\bibfnamefont {Aldo}\ \bibnamefont
  {Gamboa}}, \bibinfo {author} {\bibfnamefont {Mohammed}\ \bibnamefont
  {Khalil}}, \ and\ \bibinfo {author} {\bibfnamefont {Alessandra}\ \bibnamefont
  {Buonanno}},\ }\bibfield  {title} {\enquote {\bibinfo {title} {{Third
  post-Newtonian dynamics for eccentric orbits and aligned spins in the
  effective-one-body waveform model seobnrv5ehm}},}\ }\href {\doibase
  10.1103/rb1c-nx5f} {\bibfield  {journal} {\bibinfo  {journal} {Phys. Rev. D}\
  }\textbf {\bibinfo {volume} {112}},\ \bibinfo {pages} {044037} (\bibinfo
  {year} {2025})},\ \Eprint {http://arxiv.org/abs/2412.12831} {arXiv:2412.12831
  [gr-qc]} \BibitemShut {NoStop}%
\bibitem [{\citenamefont {Blanchet}(2014)}]{Blanchet:2013haa}%
  \BibitemOpen
  \bibfield  {author} {\bibinfo {author} {\bibfnamefont {Luc}\ \bibnamefont
  {Blanchet}},\ }\bibfield  {title} {\enquote {\bibinfo {title}
  {{Post-Newtonian Theory for Gravitational Waves}},}\ }\href {\doibase
  10.12942/lrr-2014-2} {\bibfield  {journal} {\bibinfo  {journal} {Living Rev.
  Rel.}\ }\textbf {\bibinfo {volume} {17}},\ \bibinfo {pages} {2} (\bibinfo
  {year} {2014})},\ \Eprint {http://arxiv.org/abs/1310.1528} {arXiv:1310.1528
  [gr-qc]} \BibitemShut {NoStop}%
\bibitem [{\citenamefont {{Chandrasekhar}}(1983)}]{1983mtbh.book.....C}%
  \BibitemOpen
  \bibfield  {author} {\bibinfo {author} {\bibfnamefont {S.}~\bibnamefont
  {{Chandrasekhar}}},\ }\href@noop {} {\emph {\bibinfo {title} {{The
  mathematical theory of black holes}}}}\ (\bibinfo {year} {1983})\BibitemShut
  {NoStop}%
\bibitem [{\citenamefont {Damour}\ and\ \citenamefont
  {Nagar}(2007)}]{Damour:2007xr}%
  \BibitemOpen
  \bibfield  {author} {\bibinfo {author} {\bibfnamefont {Thibault}\
  \bibnamefont {Damour}}\ and\ \bibinfo {author} {\bibfnamefont {Alessandro}\
  \bibnamefont {Nagar}},\ }\bibfield  {title} {\enquote {\bibinfo {title}
  {{Faithful effective-one-body waveforms of small-mass-ratio coalescing
  black-hole binaries}},}\ }\href {\doibase 10.1103/PhysRevD.76.064028}
  {\bibfield  {journal} {\bibinfo  {journal} {Phys. Rev. D}\ }\textbf {\bibinfo
  {volume} {76}},\ \bibinfo {pages} {064028} (\bibinfo {year} {2007})},\
  \Eprint {http://arxiv.org/abs/0705.2519} {arXiv:0705.2519 [gr-qc]}
  \BibitemShut {NoStop}%
\bibitem [{\citenamefont {Damour}\ and\ \citenamefont
  {Nagar}(2008)}]{Damour:2007yf}%
  \BibitemOpen
  \bibfield  {author} {\bibinfo {author} {\bibfnamefont {Thibault}\
  \bibnamefont {Damour}}\ and\ \bibinfo {author} {\bibfnamefont {Alessandro}\
  \bibnamefont {Nagar}},\ }\bibfield  {title} {\enquote {\bibinfo {title}
  {{Comparing Effective-One-Body gravitational waveforms to accurate numerical
  data}},}\ }\href {\doibase 10.1103/PhysRevD.77.024043} {\bibfield  {journal}
  {\bibinfo  {journal} {Phys. Rev. D}\ }\textbf {\bibinfo {volume} {77}},\
  \bibinfo {pages} {024043} (\bibinfo {year} {2008})},\ \Eprint
  {http://arxiv.org/abs/0711.2628} {arXiv:0711.2628 [gr-qc]} \BibitemShut
  {NoStop}%
\bibitem [{\citenamefont {Nagar}\ and\ \citenamefont
  {Akcay}(2012)}]{Nagar:2011aa}%
  \BibitemOpen
  \bibfield  {author} {\bibinfo {author} {\bibfnamefont {Alessandro}\
  \bibnamefont {Nagar}}\ and\ \bibinfo {author} {\bibfnamefont {Sarp}\
  \bibnamefont {Akcay}},\ }\bibfield  {title} {\enquote {\bibinfo {title}
  {{Horizon-absorbed energy flux in circularized, nonspinning black-hole
  binaries and its effective-one-body representation}},}\ }\href {\doibase
  10.1103/PhysRevD.85.044025} {\bibfield  {journal} {\bibinfo  {journal} {Phys.
  Rev. D}\ }\textbf {\bibinfo {volume} {85}},\ \bibinfo {pages} {044025}
  (\bibinfo {year} {2012})},\ \Eprint {http://arxiv.org/abs/1112.2840}
  {arXiv:1112.2840 [gr-qc]} \BibitemShut {NoStop}%
\bibitem [{\citenamefont {{Kojima}}\ and\ \citenamefont
  {{Nakamura}}(1984)}]{1984PThPh..72..494K}%
  \BibitemOpen
  \bibfield  {author} {\bibinfo {author} {\bibfnamefont {Y.}~\bibnamefont
  {{Kojima}}}\ and\ \bibinfo {author} {\bibfnamefont {T.}~\bibnamefont
  {{Nakamura}}},\ }\bibfield  {title} {\enquote {\bibinfo {title}
  {{Gravitational Radiation from a Particle Scattered by a Kerr Black Hole}},}\
  }\href {\doibase 10.1143/PTP.72.494} {\bibfield  {journal} {\bibinfo
  {journal} {Progress of Theoretical Physics}\ }\textbf {\bibinfo {volume}
  {72}},\ \bibinfo {pages} {494--504} (\bibinfo {year} {1984})}\BibitemShut
  {NoStop}%
\bibitem [{\citenamefont {Rifat}\ \emph {et~al.}(2019)\citenamefont {Rifat},
  \citenamefont {Khanna},\ and\ \citenamefont {Burko}}]{Rifat:2019fkt}%
  \BibitemOpen
  \bibfield  {author} {\bibinfo {author} {\bibfnamefont {Nur E.~M.}\
  \bibnamefont {Rifat}}, \bibinfo {author} {\bibfnamefont {Gaurav}\
  \bibnamefont {Khanna}}, \ and\ \bibinfo {author} {\bibfnamefont {Lior~M.}\
  \bibnamefont {Burko}},\ }\bibfield  {title} {\enquote {\bibinfo {title}
  {{Repeated Ringing of the Black Hole's Bell: Quasi-Normal Bursts from Highly
  Eccentric, Extreme Mass-Ratio Binaries}},}\ }\href {\doibase
  10.1103/PhysRevResearch.1.033150} {\bibfield  {journal} {\bibinfo  {journal}
  {Phys. Rev. Research.}\ }\textbf {\bibinfo {volume} {1}},\ \bibinfo {pages}
  {033150} (\bibinfo {year} {2019})},\ \Eprint
  {http://arxiv.org/abs/1910.03462} {arXiv:1910.03462 [gr-qc]} \BibitemShut
  {NoStop}%
\bibitem [{\citenamefont {Thornburg}\ \emph {et~al.}(2020)\citenamefont
  {Thornburg}, \citenamefont {Wardell},\ and\ \citenamefont {van~de
  Meent}}]{Thornburg:2019ukt}%
  \BibitemOpen
  \bibfield  {author} {\bibinfo {author} {\bibfnamefont {Jonathan}\
  \bibnamefont {Thornburg}}, \bibinfo {author} {\bibfnamefont {Barry}\
  \bibnamefont {Wardell}}, \ and\ \bibinfo {author} {\bibfnamefont {Maarten}\
  \bibnamefont {van~de Meent}},\ }\bibfield  {title} {\enquote {\bibinfo
  {title} {{Excitation of Kerr quasinormal modes in extreme--mass-ratio
  inspirals}},}\ }\href {\doibase 10.1103/PhysRevResearch.2.013365} {\bibfield
  {journal} {\bibinfo  {journal} {Phys. Rev. Res.}\ }\textbf {\bibinfo {volume}
  {2}},\ \bibinfo {pages} {013365} (\bibinfo {year} {2020})},\ \Eprint
  {http://arxiv.org/abs/1906.06791} {arXiv:1906.06791 [gr-qc]} \BibitemShut
  {NoStop}%
\bibitem [{\citenamefont {Spearman}(1904)}]{spearman1904}%
  \BibitemOpen
  \bibfield  {author} {\bibinfo {author} {\bibfnamefont {Charles}\ \bibnamefont
  {Spearman}},\ }\bibfield  {title} {\enquote {\bibinfo {title} {The proof and
  measurement of association between two things},}\ }\href {\doibase
  10.2307/1412159} {\ \textbf {\bibinfo {volume} {15}},\ \bibinfo {pages}
  {72--101} (\bibinfo {year} {1904})}\BibitemShut {NoStop}%
\bibitem [{\citenamefont {Zwillinger}\ and\ \citenamefont
  {Kokoska}(2000)}]{zwillinger2000}%
  \BibitemOpen
  \bibfield  {author} {\bibinfo {author} {\bibfnamefont {Daniel}\ \bibnamefont
  {Zwillinger}}\ and\ \bibinfo {author} {\bibfnamefont {Stephen}\ \bibnamefont
  {Kokoska}},\ }\href@noop {} {\emph {\bibinfo {title} {CRC Standard
  Probability and Statistics Tables and Formulae}}}\ (\bibinfo  {publisher}
  {Chapman \& Hall},\ \bibinfo {address} {New York},\ \bibinfo {year}
  {2000})\BibitemShut {NoStop}%
\bibitem [{\citenamefont {Kendall}\ and\ \citenamefont
  {Stuart}(1973)}]{kendall1973}%
  \BibitemOpen
  \bibfield  {author} {\bibinfo {author} {\bibfnamefont {Maurice~G.}\
  \bibnamefont {Kendall}}\ and\ \bibinfo {author} {\bibfnamefont {Alan}\
  \bibnamefont {Stuart}},\ }\href@noop {} {\emph {\bibinfo {title} {The
  Advanced Theory of Statistics, Volume 2: Inference and Relationship}}},\
  \bibinfo {edition} {3rd}\ ed.\ (\bibinfo  {publisher} {Griffin},\ \bibinfo
  {year} {1973})\BibitemShut {NoStop}%
\bibitem [{\citenamefont {Albertini}\ \emph
  {et~al.}(2022{\natexlab{a}})\citenamefont {Albertini}, \citenamefont {Nagar},
  \citenamefont {Pound}, \citenamefont {Warburton}, \citenamefont {Wardell},
  \citenamefont {Durkan},\ and\ \citenamefont {Miller}}]{Albertini:2022rfe}%
  \BibitemOpen
  \bibfield  {author} {\bibinfo {author} {\bibfnamefont {Angelica}\
  \bibnamefont {Albertini}}, \bibinfo {author} {\bibfnamefont {Alessandro}\
  \bibnamefont {Nagar}}, \bibinfo {author} {\bibfnamefont {Adam}\ \bibnamefont
  {Pound}}, \bibinfo {author} {\bibfnamefont {Niels}\ \bibnamefont
  {Warburton}}, \bibinfo {author} {\bibfnamefont {Barry}\ \bibnamefont
  {Wardell}}, \bibinfo {author} {\bibfnamefont {Leanne}\ \bibnamefont
  {Durkan}}, \ and\ \bibinfo {author} {\bibfnamefont {Jeremy}\ \bibnamefont
  {Miller}},\ }\bibfield  {title} {\enquote {\bibinfo {title} {{Comparing
  second-order gravitational self-force, numerical relativity, and effective
  one body waveforms from inspiralling, quasicircular, and nonspinning black
  hole binaries}},}\ }\href {\doibase 10.1103/PhysRevD.106.084061} {\bibfield
  {journal} {\bibinfo  {journal} {Phys. Rev. D}\ }\textbf {\bibinfo {volume}
  {106}},\ \bibinfo {pages} {084061} (\bibinfo {year} {2022}{\natexlab{a}})},\
  \Eprint {http://arxiv.org/abs/2208.01049} {arXiv:2208.01049 [gr-qc]}
  \BibitemShut {NoStop}%
\bibitem [{\citenamefont {Albertini}\ \emph
  {et~al.}(2022{\natexlab{b}})\citenamefont {Albertini}, \citenamefont {Nagar},
  \citenamefont {Pound}, \citenamefont {Warburton}, \citenamefont {Wardell},
  \citenamefont {Durkan},\ and\ \citenamefont {Miller}}]{Albertini:2022dmc}%
  \BibitemOpen
  \bibfield  {author} {\bibinfo {author} {\bibfnamefont {Angelica}\
  \bibnamefont {Albertini}}, \bibinfo {author} {\bibfnamefont {Alessandro}\
  \bibnamefont {Nagar}}, \bibinfo {author} {\bibfnamefont {Adam}\ \bibnamefont
  {Pound}}, \bibinfo {author} {\bibfnamefont {Niels}\ \bibnamefont
  {Warburton}}, \bibinfo {author} {\bibfnamefont {Barry}\ \bibnamefont
  {Wardell}}, \bibinfo {author} {\bibfnamefont {Leanne}\ \bibnamefont
  {Durkan}}, \ and\ \bibinfo {author} {\bibfnamefont {Jeremy}\ \bibnamefont
  {Miller}},\ }\bibfield  {title} {\enquote {\bibinfo {title} {{Comparing
  second-order gravitational self-force and effective one body waveforms from
  inspiralling, quasicircular and nonspinning black hole binaries. II. The
  large-mass-ratio case}},}\ }\href {\doibase 10.1103/PhysRevD.106.084062}
  {\bibfield  {journal} {\bibinfo  {journal} {Phys. Rev. D}\ }\textbf {\bibinfo
  {volume} {106}},\ \bibinfo {pages} {084062} (\bibinfo {year}
  {2022}{\natexlab{b}})},\ \Eprint {http://arxiv.org/abs/2208.02055}
  {arXiv:2208.02055 [gr-qc]} \BibitemShut {NoStop}%
\bibitem [{\citenamefont {van~de Meent}\ \emph {et~al.}(2023)\citenamefont
  {van~de Meent}, \citenamefont {Buonanno}, \citenamefont {Mihaylov},
  \citenamefont {Ossokine}, \citenamefont {Pompili}, \citenamefont {Warburton},
  \citenamefont {Pound}, \citenamefont {Wardell}, \citenamefont {Durkan},\ and\
  \citenamefont {Miller}}]{vandeMeent:2023ols}%
  \BibitemOpen
  \bibfield  {author} {\bibinfo {author} {\bibfnamefont {Maarten}\ \bibnamefont
  {van~de Meent}}, \bibinfo {author} {\bibfnamefont {Alessandra}\ \bibnamefont
  {Buonanno}}, \bibinfo {author} {\bibfnamefont {Deyan~P.}\ \bibnamefont
  {Mihaylov}}, \bibinfo {author} {\bibfnamefont {Serguei}\ \bibnamefont
  {Ossokine}}, \bibinfo {author} {\bibfnamefont {Lorenzo}\ \bibnamefont
  {Pompili}}, \bibinfo {author} {\bibfnamefont {Niels}\ \bibnamefont
  {Warburton}}, \bibinfo {author} {\bibfnamefont {Adam}\ \bibnamefont {Pound}},
  \bibinfo {author} {\bibfnamefont {Barry}\ \bibnamefont {Wardell}}, \bibinfo
  {author} {\bibfnamefont {Leanne}\ \bibnamefont {Durkan}}, \ and\ \bibinfo
  {author} {\bibfnamefont {Jeremy}\ \bibnamefont {Miller}},\ }\bibfield
  {title} {\enquote {\bibinfo {title} {{Enhancing the SEOBNRv5
  effective-one-body waveform model with second-order gravitational self-force
  fluxes}},}\ }\href {\doibase 10.1103/PhysRevD.108.124038} {\bibfield
  {journal} {\bibinfo  {journal} {Phys. Rev. D}\ }\textbf {\bibinfo {volume}
  {108}},\ \bibinfo {pages} {124038} (\bibinfo {year} {2023})},\ \Eprint
  {http://arxiv.org/abs/2303.18026} {arXiv:2303.18026 [gr-qc]} \BibitemShut
  {NoStop}%
\bibitem [{\citenamefont {Albertini}\ \emph
  {et~al.}(2024{\natexlab{a}})\citenamefont {Albertini}, \citenamefont {Gamba},
  \citenamefont {Nagar},\ and\ \citenamefont {Bernuzzi}}]{Albertini:2023aol}%
  \BibitemOpen
  \bibfield  {author} {\bibinfo {author} {\bibfnamefont {Angelica}\
  \bibnamefont {Albertini}}, \bibinfo {author} {\bibfnamefont {Rossella}\
  \bibnamefont {Gamba}}, \bibinfo {author} {\bibfnamefont {Alessandro}\
  \bibnamefont {Nagar}}, \ and\ \bibinfo {author} {\bibfnamefont {Sebastiano}\
  \bibnamefont {Bernuzzi}},\ }\bibfield  {title} {\enquote {\bibinfo {title}
  {{Effective-one-body waveforms for extreme-mass-ratio binaries: Consistency
  with second-order gravitational self-force quasicircular results and
  extension to nonprecessing spins and eccentricity}},}\ }\href {\doibase
  10.1103/PhysRevD.109.044022} {\bibfield  {journal} {\bibinfo  {journal}
  {Phys. Rev. D}\ }\textbf {\bibinfo {volume} {109}},\ \bibinfo {pages}
  {044022} (\bibinfo {year} {2024}{\natexlab{a}})},\ \Eprint
  {http://arxiv.org/abs/2310.13578} {arXiv:2310.13578 [gr-qc]} \BibitemShut
  {NoStop}%
\bibitem [{\citenamefont {Albertini}\ \emph
  {et~al.}(2024{\natexlab{b}})\citenamefont {Albertini}, \citenamefont {Nagar},
  \citenamefont {Mathews},\ and\ \citenamefont
  {Lukes-Gerakopoulos}}]{Albertini:2024rrs}%
  \BibitemOpen
  \bibfield  {author} {\bibinfo {author} {\bibfnamefont {Angelica}\
  \bibnamefont {Albertini}}, \bibinfo {author} {\bibfnamefont {Alessandro}\
  \bibnamefont {Nagar}}, \bibinfo {author} {\bibfnamefont {Josh}\ \bibnamefont
  {Mathews}}, \ and\ \bibinfo {author} {\bibfnamefont {Georgios}\ \bibnamefont
  {Lukes-Gerakopoulos}},\ }\bibfield  {title} {\enquote {\bibinfo {title}
  {{Comparing second-order gravitational self-force and effective-one-body
  waveforms from inspiralling, quasicircular black hole binaries with a
  nonspinning primary and a spinning secondary}},}\ }\href {\doibase
  10.1103/PhysRevD.110.044034} {\bibfield  {journal} {\bibinfo  {journal}
  {Phys. Rev. D}\ }\textbf {\bibinfo {volume} {110}},\ \bibinfo {pages}
  {044034} (\bibinfo {year} {2024}{\natexlab{b}})},\ \Eprint
  {http://arxiv.org/abs/2406.04108} {arXiv:2406.04108 [gr-qc]} \BibitemShut
  {NoStop}%
\bibitem [{\citenamefont {Albertini}\ \emph {et~al.}(2025)\citenamefont
  {Albertini}, \citenamefont {Skoup{\'y}}, \citenamefont {Lukes-Gerakopoulos},\
  and\ \citenamefont {Nagar}}]{Albertini:2024agg}%
  \BibitemOpen
  \bibfield  {author} {\bibinfo {author} {\bibfnamefont {Angelica}\
  \bibnamefont {Albertini}}, \bibinfo {author} {\bibfnamefont {Viktor}\
  \bibnamefont {Skoup{\'y}}}, \bibinfo {author} {\bibfnamefont {Georgios}\
  \bibnamefont {Lukes-Gerakopoulos}}, \ and\ \bibinfo {author} {\bibfnamefont
  {Alessandro}\ \bibnamefont {Nagar}},\ }\bibfield  {title} {\enquote {\bibinfo
  {title} {{Comparing effective-one-body and Mathisson-Papapetrou-Dixon results
  for a spinning test particle on circular equatorial orbits around a Kerr
  black hole}},}\ }\href {\doibase 10.1103/PhysRevD.111.064086} {\bibfield
  {journal} {\bibinfo  {journal} {Phys. Rev. D}\ }\textbf {\bibinfo {volume}
  {111}},\ \bibinfo {pages} {064086} (\bibinfo {year} {2025})},\ \Eprint
  {http://arxiv.org/abs/2412.16077} {arXiv:2412.16077 [gr-qc]} \BibitemShut
  {NoStop}%
\bibitem [{\citenamefont {Leather}\ \emph {et~al.}(2025)\citenamefont
  {Leather}, \citenamefont {Buonanno},\ and\ \citenamefont {van~de
  Meent}}]{Leather:2025nhu}%
  \BibitemOpen
  \bibfield  {author} {\bibinfo {author} {\bibfnamefont {Benjamin}\
  \bibnamefont {Leather}}, \bibinfo {author} {\bibfnamefont {Alessandra}\
  \bibnamefont {Buonanno}}, \ and\ \bibinfo {author} {\bibfnamefont {Maarten}\
  \bibnamefont {van~de Meent}},\ }\bibfield  {title} {\enquote {\bibinfo
  {title} {{Inspiral-merger-ringdown waveforms with gravitational self-force
  results within the effective-one-body formalism}},}\ }\href {\doibase
  10.1103/6qc3-xn17} {\bibfield  {journal} {\bibinfo  {journal} {Phys. Rev. D}\
  }\textbf {\bibinfo {volume} {112}},\ \bibinfo {pages} {044012} (\bibinfo
  {year} {2025})},\ \Eprint {http://arxiv.org/abs/2505.11242} {arXiv:2505.11242
  [gr-qc]} \BibitemShut {NoStop}%
\bibitem [{\citenamefont {Albanesi}\ \emph {et~al.}(2026)\citenamefont
  {Albanesi}, \citenamefont {Bernuzzi},\ and\ \citenamefont
  {Nagar}}]{Albanesi:2026qtx}%
  \BibitemOpen
  \bibfield  {author} {\bibinfo {author} {\bibfnamefont {Simone}\ \bibnamefont
  {Albanesi}}, \bibinfo {author} {\bibfnamefont {Sebastiano}\ \bibnamefont
  {Bernuzzi}}, \ and\ \bibinfo {author} {\bibfnamefont {Alessandro}\
  \bibnamefont {Nagar}},\ }\bibfield  {title} {\enquote {\bibinfo {title}
  {{Ringdown modeling for effective-one-body waveforms in the test-mass limit
  for eccentric equatorial orbits around a Kerr black hole}},}\ }\href@noop {}
  {\  (\bibinfo {year} {2026})},\ \Eprint {http://arxiv.org/abs/2603.19413}
  {arXiv:2603.19413 [gr-qc]} \BibitemShut {NoStop}%
\bibitem [{\citenamefont {Faggioli}\ \emph {et~al.}(2026)\citenamefont
  {Faggioli}, \citenamefont {Buonanno}, \citenamefont {van~de Meent},\ and\
  \citenamefont {Khanna}}]{Faggioli:2026alx}%
  \BibitemOpen
  \bibfield  {author} {\bibinfo {author} {\bibfnamefont {Guglielmo}\
  \bibnamefont {Faggioli}}, \bibinfo {author} {\bibfnamefont {Alessandra}\
  \bibnamefont {Buonanno}}, \bibinfo {author} {\bibfnamefont {Maarten}\
  \bibnamefont {van~de Meent}}, \ and\ \bibinfo {author} {\bibfnamefont
  {Gaurav}\ \bibnamefont {Khanna}},\ }\bibfield  {title} {\enquote {\bibinfo
  {title} {{Modeling the merger-ringdown of an eccentric test-mass inspiral
  into a Kerr black hole using the effective-one-body framework}},}\
  }\href@noop {} {\  (\bibinfo {year} {2026})},\ \Eprint
  {http://arxiv.org/abs/2603.19913} {arXiv:2603.19913 [gr-qc]} \BibitemShut
  {NoStop}%
\bibitem [{\citenamefont {Cipriani}\ \emph
  {et~al.}(2026{\natexlab{a}})\citenamefont {Cipriani}, \citenamefont {Fucito},
  \citenamefont {Heissenberg}, \citenamefont {Morales},\ and\ \citenamefont
  {Russo}}]{Cipriani:2026myb}%
  \BibitemOpen
  \bibfield  {author} {\bibinfo {author} {\bibfnamefont {Andrea}\ \bibnamefont
  {Cipriani}}, \bibinfo {author} {\bibfnamefont {Francesco}\ \bibnamefont
  {Fucito}}, \bibinfo {author} {\bibfnamefont {Carlo}\ \bibnamefont
  {Heissenberg}}, \bibinfo {author} {\bibfnamefont {Jose~Francisco}\
  \bibnamefont {Morales}}, \ and\ \bibinfo {author} {\bibfnamefont {Rodolfo}\
  \bibnamefont {Russo}},\ }\bibfield  {title} {\enquote {\bibinfo {title}
  {{''Waveforms'' at the Horizon}},}\ }\href@noop {} {\  (\bibinfo {year}
  {2026}{\natexlab{a}})},\ \Eprint {http://arxiv.org/abs/2602.05766}
  {arXiv:2602.05766 [gr-qc]} \BibitemShut {NoStop}%
\bibitem [{\citenamefont {Cipriani}\ \emph
  {et~al.}(2026{\natexlab{b}})\citenamefont {Cipriani}, \citenamefont {Nagar},
  \citenamefont {Fucito},\ and\ \citenamefont {Morales}}]{Cipriani:2026xmx}%
  \BibitemOpen
  \bibfield  {author} {\bibinfo {author} {\bibfnamefont {Andrea}\ \bibnamefont
  {Cipriani}}, \bibinfo {author} {\bibfnamefont {Alessandro}\ \bibnamefont
  {Nagar}}, \bibinfo {author} {\bibfnamefont {Francesco}\ \bibnamefont
  {Fucito}}, \ and\ \bibinfo {author} {\bibfnamefont {Jos{\'e}~Francisco}\
  \bibnamefont {Morales}},\ }\bibfield  {title} {\enquote {\bibinfo {title}
  {{From the confluent Heun equation to a new factorized and resummed
  gravitational waveform for circularized, nonspinning, compact binaries}},}\
  }\href@noop {} {\  (\bibinfo {year} {2026}{\natexlab{b}})},\ \Eprint
  {http://arxiv.org/abs/2602.08833} {arXiv:2602.08833 [gr-qc]} \BibitemShut
  {NoStop}%
\bibitem [{\citenamefont {Nishimura}\ \emph {et~al.}(2026)\citenamefont
  {Nishimura}, \citenamefont {Buonanno}, \citenamefont {Faggioli},
  \citenamefont {van~de Meent},\ and\ \citenamefont
  {Khanna}}]{Nishimura:2026nse}%
  \BibitemOpen
  \bibfield  {author} {\bibinfo {author} {\bibfnamefont {Nami}\ \bibnamefont
  {Nishimura}}, \bibinfo {author} {\bibfnamefont {Alessandra}\ \bibnamefont
  {Buonanno}}, \bibinfo {author} {\bibfnamefont {Guglielmo}\ \bibnamefont
  {Faggioli}}, \bibinfo {author} {\bibfnamefont {Maarten}\ \bibnamefont {van~de
  Meent}}, \ and\ \bibinfo {author} {\bibfnamefont {Gaurav}\ \bibnamefont
  {Khanna}},\ }\bibfield  {title} {\enquote {\bibinfo {title} {{Advancing the
  Effective-One-Body Framework in the Test-Mass Limit}},}\ }\href@noop {} {\
  (\bibinfo {year} {2026})},\ \Eprint {http://arxiv.org/abs/2603.05601}
  {arXiv:2603.05601 [gr-qc]} \BibitemShut {NoStop}%
\bibitem [{\citenamefont {Nagni}\ \emph {et~al.}(2026)\citenamefont {Nagni},
  \citenamefont {Nagar}, \citenamefont {Gamba}, \citenamefont {Albanesi},\ and\
  \citenamefont {Bernuzzi}}]{Nagni:2025cdw}%
  \BibitemOpen
  \bibfield  {author} {\bibinfo {author} {\bibfnamefont {Luca}\ \bibnamefont
  {Nagni}}, \bibinfo {author} {\bibfnamefont {Alessandro}\ \bibnamefont
  {Nagar}}, \bibinfo {author} {\bibfnamefont {Rossella}\ \bibnamefont {Gamba}},
  \bibinfo {author} {\bibfnamefont {Simone}\ \bibnamefont {Albanesi}}, \ and\
  \bibinfo {author} {\bibfnamefont {Sebastiano}\ \bibnamefont {Bernuzzi}},\
  }\bibfield  {title} {\enquote {\bibinfo {title} {{Binary black hole merger in
  the extreme mass ratio limit: A multipolar analysis of the inclined orbit
  case}},}\ }\href {\doibase 10.1103/11xp-8jtk} {\bibfield  {journal} {\bibinfo
   {journal} {Phys. Rev. D}\ }\textbf {\bibinfo {volume} {113}},\ \bibinfo
  {pages} {044052} (\bibinfo {year} {2026})},\ \Eprint
  {http://arxiv.org/abs/2509.17478} {arXiv:2509.17478 [gr-qc]} \BibitemShut
  {NoStop}%
\bibitem [{\citenamefont {Forseth}\ \emph {et~al.}(2016)\citenamefont
  {Forseth}, \citenamefont {Evans},\ and\ \citenamefont
  {Hopper}}]{Forseth:2015oua}%
  \BibitemOpen
  \bibfield  {author} {\bibinfo {author} {\bibfnamefont {Erik}\ \bibnamefont
  {Forseth}}, \bibinfo {author} {\bibfnamefont {Charles~R.}\ \bibnamefont
  {Evans}}, \ and\ \bibinfo {author} {\bibfnamefont {Seth}\ \bibnamefont
  {Hopper}},\ }\bibfield  {title} {\enquote {\bibinfo {title} {{Eccentric-orbit
  extreme-mass-ratio inspiral gravitational wave energy fluxes to 7PN
  order}},}\ }\href {\doibase 10.1103/PhysRevD.93.064058} {\bibfield  {journal}
  {\bibinfo  {journal} {Phys. Rev. D}\ }\textbf {\bibinfo {volume} {93}},\
  \bibinfo {pages} {064058} (\bibinfo {year} {2016})},\ \Eprint
  {http://arxiv.org/abs/1512.03051} {arXiv:1512.03051 [gr-qc]} \BibitemShut
  {NoStop}%
\bibitem [{\citenamefont {Munna}\ \emph {et~al.}(2020)\citenamefont {Munna},
  \citenamefont {Evans}, \citenamefont {Hopper},\ and\ \citenamefont
  {Forseth}}]{Munna:2020juq}%
  \BibitemOpen
  \bibfield  {author} {\bibinfo {author} {\bibfnamefont {Christopher}\
  \bibnamefont {Munna}}, \bibinfo {author} {\bibfnamefont {Charles~R.}\
  \bibnamefont {Evans}}, \bibinfo {author} {\bibfnamefont {Seth}\ \bibnamefont
  {Hopper}}, \ and\ \bibinfo {author} {\bibfnamefont {Erik}\ \bibnamefont
  {Forseth}},\ }\bibfield  {title} {\enquote {\bibinfo {title} {{Determination
  of new coefficients in the angular momentum and energy fluxes at infinity to
  9PN order for eccentric Schwarzschild extreme-mass-ratio inspirals using
  mode-by-mode fitting}},}\ }\href {\doibase 10.1103/PhysRevD.102.024047}
  {\bibfield  {journal} {\bibinfo  {journal} {Phys. Rev. D}\ }\textbf {\bibinfo
  {volume} {102}},\ \bibinfo {pages} {024047} (\bibinfo {year} {2020})},\
  \Eprint {http://arxiv.org/abs/2005.03044} {arXiv:2005.03044 [gr-qc]}
  \BibitemShut {NoStop}%
\bibitem [{\citenamefont {Munna}\ \emph {et~al.}(2023)\citenamefont {Munna},
  \citenamefont {Evans},\ and\ \citenamefont {Forseth}}]{Munna:2023vds}%
  \BibitemOpen
  \bibfield  {author} {\bibinfo {author} {\bibfnamefont {Christopher}\
  \bibnamefont {Munna}}, \bibinfo {author} {\bibfnamefont {Charles~R.}\
  \bibnamefont {Evans}}, \ and\ \bibinfo {author} {\bibfnamefont {Erik}\
  \bibnamefont {Forseth}},\ }\bibfield  {title} {\enquote {\bibinfo {title}
  {{Tidal heating and torquing of the primary black hole in eccentric-orbit,
  nonspinning, extreme-mass-ratio inspirals to 22PN order}},}\ }\href {\doibase
  10.1103/PhysRevD.108.044039} {\bibfield  {journal} {\bibinfo  {journal}
  {Phys. Rev. D}\ }\textbf {\bibinfo {volume} {108}},\ \bibinfo {pages}
  {044039} (\bibinfo {year} {2023})},\ \Eprint
  {http://arxiv.org/abs/2306.12481} {arXiv:2306.12481 [gr-qc]} \BibitemShut
  {NoStop}%
\bibitem [{\citenamefont {Prasad}\ \emph {et~al.}(2020)\citenamefont {Prasad},
  \citenamefont {Gupta}, \citenamefont {Bose}, \citenamefont {Krishnan},\ and\
  \citenamefont {Schnetter}}]{Prasad:2020xgr}%
  \BibitemOpen
  \bibfield  {author} {\bibinfo {author} {\bibfnamefont {Vaishak}\ \bibnamefont
  {Prasad}}, \bibinfo {author} {\bibfnamefont {Anshu}\ \bibnamefont {Gupta}},
  \bibinfo {author} {\bibfnamefont {Sukanta}\ \bibnamefont {Bose}}, \bibinfo
  {author} {\bibfnamefont {Badri}\ \bibnamefont {Krishnan}}, \ and\ \bibinfo
  {author} {\bibfnamefont {Erik}\ \bibnamefont {Schnetter}},\ }\bibfield
  {title} {\enquote {\bibinfo {title} {{News from horizons in binary black hole
  mergers}},}\ }\href {\doibase 10.1103/PhysRevLett.125.121101} {\bibfield
  {journal} {\bibinfo  {journal} {Phys. Rev. Lett.}\ }\textbf {\bibinfo
  {volume} {125}},\ \bibinfo {pages} {121101} (\bibinfo {year} {2020})},\
  \Eprint {http://arxiv.org/abs/2003.06215} {arXiv:2003.06215 [gr-qc]}
  \BibitemShut {NoStop}%
\bibitem [{\citenamefont {Prasad}(2025)}]{Prasad:2023bwa}%
  \BibitemOpen
  \bibfield  {author} {\bibinfo {author} {\bibfnamefont {Vaishak}\ \bibnamefont
  {Prasad}},\ }\bibfield  {title} {\enquote {\bibinfo {title} {{Shear at the
  common dynamical horizon in binary black hole mergers and its imprint in
  their gravitational radiation}},}\ }\href {\doibase
  10.1103/PhysRevD.111.084070} {\bibfield  {journal} {\bibinfo  {journal}
  {Phys. Rev. D}\ }\textbf {\bibinfo {volume} {111}},\ \bibinfo {pages}
  {084070} (\bibinfo {year} {2025})},\ \Eprint
  {http://arxiv.org/abs/2312.01136} {arXiv:2312.01136 [gr-qc]} \BibitemShut
  {NoStop}%
\bibitem [{\citenamefont {Prasad}(2024)}]{Prasad:2024vsz}%
  \BibitemOpen
  \bibfield  {author} {\bibinfo {author} {\bibfnamefont {Vaishak}\ \bibnamefont
  {Prasad}},\ }\bibfield  {title} {\enquote {\bibinfo {title} {{Tidal
  deformation of dynamical horizons in binary black hole mergers and its
  imprint on gravitational radiation}},}\ }\href {\doibase
  10.1103/PhysRevD.109.044033} {\bibfield  {journal} {\bibinfo  {journal}
  {Phys. Rev. D}\ }\textbf {\bibinfo {volume} {109}},\ \bibinfo {pages}
  {044033} (\bibinfo {year} {2024})}\BibitemShut {NoStop}%
\bibitem [{\citenamefont {Harris}\ \emph {et~al.}(2020)\citenamefont {Harris}
  \emph {et~al.}}]{Harris:2020xlr}%
  \BibitemOpen
  \bibfield  {author} {\bibinfo {author} {\bibfnamefont {Charles~R.}\
  \bibnamefont {Harris}} \emph {et~al.},\ }\bibfield  {title} {\enquote
  {\bibinfo {title} {{Array programming with NumPy}},}\ }\href {\doibase
  10.1038/s41586-020-2649-2} {\bibfield  {journal} {\bibinfo  {journal}
  {Nature}\ }\textbf {\bibinfo {volume} {585}},\ \bibinfo {pages} {357--362}
  (\bibinfo {year} {2020})},\ \Eprint {http://arxiv.org/abs/2006.10256}
  {arXiv:2006.10256 [cs.MS]} \BibitemShut {NoStop}%
\bibitem [{\citenamefont {Virtanen}\ \emph {et~al.}(2020)\citenamefont
  {Virtanen} \emph {et~al.}}]{Virtanen:2019joe}%
  \BibitemOpen
  \bibfield  {author} {\bibinfo {author} {\bibfnamefont {Pauli}\ \bibnamefont
  {Virtanen}} \emph {et~al.},\ }\bibfield  {title} {\enquote {\bibinfo {title}
  {{SciPy 1.0--Fundamental Algorithms for Scientific Computing in Python}},}\
  }\href {\doibase 10.1038/s41592-019-0686-2} {\bibfield  {journal} {\bibinfo
  {journal} {Nature Meth.}\ }\textbf {\bibinfo {volume} {17}},\ \bibinfo
  {pages} {261} (\bibinfo {year} {2020})},\ \Eprint
  {http://arxiv.org/abs/1907.10121} {arXiv:1907.10121 [cs.MS]} \BibitemShut
  {NoStop}%
\bibitem [{\citenamefont {Meurer}\ \emph {et~al.}(2017)\citenamefont {Meurer}
  \emph {et~al.}}]{Meurer:2017yhf}%
  \BibitemOpen
  \bibfield  {author} {\bibinfo {author} {\bibfnamefont {Aaron}\ \bibnamefont
  {Meurer}} \emph {et~al.},\ }\bibfield  {title} {\enquote {\bibinfo {title}
  {{SymPy: symbolic computing in Python}},}\ }\href {\doibase
  10.7717/peerj-cs.103} {\bibfield  {journal} {\bibinfo  {journal} {PeerJ
  Comput. Sci.}\ }\textbf {\bibinfo {volume} {3}},\ \bibinfo {pages} {e103}
  (\bibinfo {year} {2017})}\BibitemShut {NoStop}%
\bibitem [{\citenamefont {Johansson}\ \emph {et~al.}(2023)\citenamefont
  {Johansson} \emph {et~al.}}]{mpmath}%
  \BibitemOpen
  \bibfield  {author} {\bibinfo {author} {\bibfnamefont {Fredrik}\ \bibnamefont
  {Johansson}} \emph {et~al.},\ }\href {http://mpmath.org/} {\enquote {\bibinfo
  {title} {{mpmath: a Python library for arbitrary-precision floating-point
  arithmetic (version 1.3.0)}},}\ } (\bibinfo {year} {2023})\BibitemShut
  {NoStop}%
\bibitem [{\citenamefont {Hunter}(2007)}]{Hunter:2007}%
  \BibitemOpen
  \bibfield  {author} {\bibinfo {author} {\bibfnamefont {John~D.}\ \bibnamefont
  {Hunter}},\ }\bibfield  {title} {\enquote {\bibinfo {title} {Matplotlib: A 2d
  graphics environment},}\ }\href {\doibase 10.1109/MCSE.2007.55} {\bibfield
  {journal} {\bibinfo  {journal} {Computing in Science \& Engineering}\
  }\textbf {\bibinfo {volume} {9}},\ \bibinfo {pages} {90--95} (\bibinfo {year}
  {2007})}\BibitemShut {NoStop}%
\bibitem [{\citenamefont {Waskom}(2021)}]{Waskom:2021}%
  \BibitemOpen
  \bibfield  {author} {\bibinfo {author} {\bibfnamefont {Michael}\ \bibnamefont
  {Waskom}},\ }\bibfield  {title} {\enquote {\bibinfo {title} {{seaborn:
  statistical data visualization}},}\ }\href {\doibase 10.21105/joss.03021}
  {\bibfield  {journal} {\bibinfo  {journal} {J. Open Source Softw.}\ }\textbf
  {\bibinfo {volume} {6}} (\bibinfo {year} {2021}),\
  10.21105/joss.03021}\BibitemShut {NoStop}%
\bibitem [{\citenamefont {Breuer}\ \emph {et~al.}(1973)\citenamefont {Breuer},
  \citenamefont {Ruffini}, \citenamefont {Tiomno},\ and\ \citenamefont
  {Vishveshwara}}]{Breuer:1973kt}%
  \BibitemOpen
  \bibfield  {author} {\bibinfo {author} {\bibfnamefont {R.~A.}\ \bibnamefont
  {Breuer}}, \bibinfo {author} {\bibfnamefont {R.}~\bibnamefont {Ruffini}},
  \bibinfo {author} {\bibfnamefont {J.}~\bibnamefont {Tiomno}}, \ and\ \bibinfo
  {author} {\bibfnamefont {C.~V.}\ \bibnamefont {Vishveshwara}},\ }\bibfield
  {title} {\enquote {\bibinfo {title} {{Vector and tensor radiation from
  schwarzschild relativistic circular geodesics}},}\ }\href {\doibase
  10.1103/PhysRevD.7.1002} {\bibfield  {journal} {\bibinfo  {journal} {Phys.
  Rev. D}\ }\textbf {\bibinfo {volume} {7}},\ \bibinfo {pages} {1002--1007}
  (\bibinfo {year} {1973})}\BibitemShut {NoStop}%
\bibitem [{\citenamefont {Chrzanowski}\ and\ \citenamefont
  {Misner}(1974)}]{Chrzanowski:1974nr}%
  \BibitemOpen
  \bibfield  {author} {\bibinfo {author} {\bibfnamefont {P.~L.}\ \bibnamefont
  {Chrzanowski}}\ and\ \bibinfo {author} {\bibfnamefont {Charles~W.}\
  \bibnamefont {Misner}},\ }\bibfield  {title} {\enquote {\bibinfo {title}
  {{Geodesic synchrotron radiation in the Kerr geometry by the method of
  asymptotically factorized Green's functions}},}\ }\href {\doibase
  10.1103/PhysRevD.10.1701} {\bibfield  {journal} {\bibinfo  {journal} {Phys.
  Rev. D}\ }\textbf {\bibinfo {volume} {10}},\ \bibinfo {pages} {1701--1721}
  (\bibinfo {year} {1974})}\BibitemShut {NoStop}%
\bibitem [{\citenamefont {Sundararajan}\ \emph {et~al.}(2007)\citenamefont
  {Sundararajan}, \citenamefont {Khanna},\ and\ \citenamefont
  {Hughes}}]{Sundararajan:2007jg}%
  \BibitemOpen
  \bibfield  {author} {\bibinfo {author} {\bibfnamefont {Pranesh~A.}\
  \bibnamefont {Sundararajan}}, \bibinfo {author} {\bibfnamefont {Gaurav}\
  \bibnamefont {Khanna}}, \ and\ \bibinfo {author} {\bibfnamefont {Scott~A.}\
  \bibnamefont {Hughes}},\ }\bibfield  {title} {\enquote {\bibinfo {title}
  {{Towards adiabatic waveforms for inspiral into Kerr black holes. I. A New
  model of the source for the time domain perturbation equation}},}\ }\href
  {\doibase 10.1103/PhysRevD.76.104005} {\bibfield  {journal} {\bibinfo
  {journal} {Phys. Rev. D}\ }\textbf {\bibinfo {volume} {76}},\ \bibinfo
  {pages} {104005} (\bibinfo {year} {2007})},\ \Eprint
  {http://arxiv.org/abs/gr-qc/0703028} {arXiv:gr-qc/0703028} \BibitemShut
  {NoStop}%
\end{thebibliography}
\end{document}